\documentclass[12pt]{article}
\usepackage{amsmath}
\usepackage{graphicx}
\usepackage{natbib}
\usepackage[title]{appendix}
\usepackage{float}
\usepackage[colorinlistoftodos]{todonotes}
\usepackage[colorlinks=true, allcolors=blue]{hyperref}
\usepackage{listings}
\usepackage{url}
\usepackage{bm}
\usepackage{setspace}
\usepackage{subfig}
\usepackage{algorithm2e}
\usepackage{amssymb}
\usepackage{csquotes}
\usepackage{multirow}
\usepackage{cancel}
\usepackage{dsfont}

\setcitestyle{notesep={ }}

\newcommand{\blind}{0}

\usepackage{xr-hyper}
\makeatletter
\newcommand*{\addFileDependency}[1]{
 \typeout{(#1)}
 \@addtofilelist{#1}
 \IfFileExists{#1}{}{\typeout{No file #1.}}
}
\makeatother

\newcommand*{\myexternaldocument}[1]{
    \externaldocument{#1}
    \addFileDependency{#1.tex}
    \addFileDependency{#1.aux}
}

\myexternaldocument{JCGS_Rev-supp}

\begin{document}

\def\spacingset#1{\renewcommand{\baselinestretch}%
{#1}\small\normalsize} \spacingset{1}

\newcommand{\Z}{Z}
\newcommand{\z}{\bm{z}}
\newcommand{\Q}{Q}
\newcommand{\tildebw}{\tilde{\bm{w}}}
\newcommand{\tildew}{\tilde{w}}
\newcommand{\checkw}{\check{w}}
\newcommand{\V}{V}
\newcommand{\X}{X}
\newcommand{\x}{\bm{x}}

\newcommand{\rPpdf}{\lambda_r}
\newcommand{\excReg}{\mathcal{A}_r}
\newcommand{\estimand}{\bm{\psi}, \bm{f}}

\if0\blind
{
  \title{\bf Modeling Nonstationary Extremal Dependence via Deep Spatial Deformations}
  \author{Xuanjie Shao$^1$, Jordan Richards$^2$, and Rapha\"el Huser$^1$\hspace{.2cm}\\
    $^1$Statistics Program, Computer, Electrical and Mathematical Sciences and\\
     Engineering (CEMSE) Division, King Abdullah University of Science \\ and Technology (KAUST), Thuwal 23955-6900, Saudi Arabia\\
    $^2$School of Mathematics and Maxwell Institute for Mathematical Sciences, \\ University of Edinburgh, Edinburgh, EH9 3FD, UK
    }
    \date{}
  \maketitle
} \fi


\if1\blind
{
  \bigskip
  \bigskip
  \bigskip
  \begin{center}
    {\LARGE\bf Modeling Nonstationary Extremal Dependence via Deep Spatial Deformations}
\end{center}
  \medskip
} \fi


\bigskip
\begin{abstract}
Modeling nonstationarity that often prevails in extremal dependence of spatial data can be challenging, and typically requires bespoke or complex spatial models that are difficult to estimate. Inference for stationary and isotropic models is considerably easier, but the assumptions that underpin these models are rarely met by data observed over large or topographically complex domains. A possible approach for accommodating nonstationarity in a spatial model is to warp the spatial domain to a latent space where stationarity and isotropy can be reasonably assumed. Although this approach is very flexible, estimating the warping function can be computationally expensive, and the transformation is not always guaranteed to be bijective, which may lead to physically unrealistic transformations when the domain folds onto itself. We overcome these challenges by developing deep compositional spatial models to capture nonstationarity in extremal dependence. Specifically, we focus on modeling high threshold exceedances of process functionals by leveraging efficient inference methods for limiting $r$-Pareto processes. A detailed high-dimensional simulation study demonstrates the superior performance of our model in estimating the warped space. We illustrate our method by modeling UK precipitation extremes and show that we can efficiently estimate the extremal dependence structure of data observed at thousands of locations.



\end{abstract}

\noindent%
{\it Keywords:} Deep Learning, Nonstationarity, $r$-Pareto Process, Spatial Extremes

\spacingset{1.5} 

\section{Introduction}



The most impactful extreme events, such as Storm Dennis, which brought record-breaking rainfall leading to severe flooding across the UK \citep{sefton20212019}, are often spatial in nature; therefore, there is a critical need to account for spatial extremal dependence when assessing the risk of unprecedented environmental extremes.
Natural processes observed over large geographical domains are often inhomogeneous due to orography or proximity to the coast, among other factors, and capturing such (potentially complex) nonstationary behavior prevailing in both margins and dependence can be challenging. Failing to adequately model this nonstationarity can lead to poor risk assessment and suboptimal mitigation strategies. Herein, we focus on modeling nonstationarity in spatial dependence. Various modeling approaches have been proposed to account for nonstationarity in the extremal dependence structure of spatial processes; they can be generally categorized into two major classes: i) those that directly impose nonstationarity in the extremal dependence structure itself \citep{huser2016non, saunders2021regionalisation, forster2025non, shao2025flexible}, and ii) those that rely on a deformation of the spatial domain to capture nonstationarity \citep{youngman2020flexible, chevalier2021modeling, richards2021spatial}.


For the first class of models, nonstationarity is often incorporated into the covariance function (or variogram) of an underlying Gaussian process used in the model construction, which induces nonstationarity in the resulting model's extremal dependence structure. An important consideration, therefore, is to create flexible nonstationary covariance functions that retain positive definiteness. One possible approach is to exploit the \cite{paciorek2006spatial} covariance model, which defines a globally nonstationary model through simpler local (spatially-varying) covariance kernels. Covariates can then easily be incorporated into local range and anisotropy parameters \citep{huser2016non}. However, a prediction-driven approach often lacks the flexibility to capture complex nonstationary behaviors. Balancing model flexibility and parsimony is indeed pivotal, and various approaches have been proposed to deal with this trade-off. A popular approach involves regionalization of the study domain into multiple homogeneous subregions, with some subregions enforced to share similar extremal dependence behavior, through equal dependence parameters. \cite{buhl2016anisotropic} and \cite{saunders2021regionalisation} proposed to identify the subregions by clustering sites using information about their extremal dependence structure. \cite{shao2025flexible} instead advocated a spatially-motivated penalized estimation procedure to identify homogeneous subregions; similar methods have also been proposed in classical geostatistics \citep[see, e.g.,][]{parker2016fused}. For modeling stationary but anisotropic dependence, a common choice is to use a parametric anisotropic covariance function or variogram, where the notion of distance considers the relative orientation of stations \citep[see, e.g.,][]{huser2016non, de2018high}. 


Instead of directly modeling nonstationary dependence via complex parametrizations of covariance/variogram functions, the spatial deformation approach, proposed by \cite{sampson1992nonparametric}, warps the original spatial domain into a latent space where stationarity and isotropy of data can be reasonably assumed. Estimation of the warping function is challenging: this task may be computationally expensive; the chosen class of warping functions may lack flexibility (especially in cases when there is strong nonstationarity); or, conversely, the model may be too flexible and overfit the data. The estimated warping may also display the notorious problem of ``space-folding'' if not appropriately constrained, which is undesirable and implausible for environmental data \citep{schmidt2003bayesian}. To estimate the warping function $\bm{f}: \mathcal{S} \rightarrow \mathcal{W}$ from the original spatial domain $\mathcal{S}\subseteq\mathbb{R}^K$ to the latent space $\mathcal{W}\subseteq\mathbb{R}^K$ (typically with $K=2$ or $K=3$), \cite{sampson1992nonparametric} exploited multidimensional scaling; this approach was extended by \cite{chevalier2021modeling} for the modeling of nonstationary extremal dependence. \cite{smith1996estimating} instead adopted a spline-based approach for modeling the warping function in the Gaussian setting, followed by \cite{youngman2020flexible} and \cite{richards2021spatial} in the extremal setting, with practical techniques advocated to reduce the chance of estimating non-bijective warpings (i.e., ``space-folding''). Bayesian hierarchical models were also considered by \cite{schmidt2003bayesian}, where the warping function $\bm{f}(\cdot)$ was modeled as a bivariate Gaussian process, thus precluding non-bijective warpings via specification of suitable priors.

To fully preclude space-folding, \cite{perrin1999modelling} and \cite{iovleff2004estimating} used a composition of bijective radial basis functions to model $\bm{f}(\cdot)$; \cite{zammit2022deep} further developed the deep compositional spatial model (DCSM), a bijective warping model constructed from a composition of simple bijective warping units. 
Specifically, suppose $X(\cdot)$ is a spatial process on some spatial domain $\mathcal{S} \subseteq \mathbb{R}^K$. The spatial deformation approach assumes that on the latent warped space $\mathcal{W}\subseteq\mathbb{R}^K$, the dependence structure solely depends on the Euclidean distance, i.e., ${\|\bm{f}(\bm{s}_1) - \bm{f}(\bm{s}_2)\|}$, but not on the individual sites $\bm{s}_1, \bm{s}_2\in\mathcal{S}$. Deep compositional spatial models further construct the warping function $\bm{f}(\cdot)$ using a composition $\bm{f}(\cdot) = 
\bm{f}_n\circ \bm{f}_{n-1}\circ\cdots \circ \bm{f}_1(\cdot)$ of simple bijective warping functions $\bm{f}_i:\mathcal{W}_{i-1}\to \mathcal{W}_{i}$, where $\mathcal{W}_{i-1}\subseteq\mathbb{R}^K$ is the domain of $\bm{f}_i(\cdot)$ and $\mathcal{W}_i\subseteq\mathbb{R}^K$ is its image; here, $\mathcal{W}_0:=\mathcal{S}$ is the original space, $\{\mathcal{W}_i,i=1,\ldots,n-1\}$ are intermediate warped spaces, and $\mathcal{W}_{n}:=\mathcal{W}$ is the final warped space. The simple functions $\bm{f}_i(\cdot)$ may be specified, for example, as axial warping units, radial basis functions, or based on the M\"obius transform (see Section~\ref{Methodology:DCrPP} for details), and they may differ across $i=1,\ldots,n$. 


With efficient optimization software, the modeling capability of the DCSM has been extended to much higher-dimensional settings than was previously possible using alternative approaches. While \cite{zammit2022deep} originally developed the DCSM for univariate spatial Gaussian processes, \cite{vu2022modeling, vu2023constructing} further extended this framework to multivariate and space-time Gaussian processes, but they did not consider non-Gaussian settings. The number of parameters describing the warping function in DCSMs is typically quite large; as a result, they can flexibly capture a wide range of nonstationary features. A sufficiently flexible mapping from $\mathbb{R}^K$ to itself makes realistic and comprehensible kriging possible and enhances the interpretability of the estimated space \citep[see][for details]{zammit2022deep}, relative to models based on dimensional expansion \citep[see, e.g.,][]{bornn2012modeling, chevalier2021modeling}. To mitigate the risk of overfitting, one can either increase the model's flexibility progressively by adding warping units step-by-step or use suitable regularization techniques. In this work, we use both strategies, namely progressive enrichment of the warping architecture and regularization of more flexible models. 
\textcolor{black}{To free oneself from parametric warping units, normalizing flows have recently been used to warp the spatial domain \citep{nag2025modeling}; see also \cite{de2025generative} and \cite{hu2025gpdflow} who also employ normalizing flows to flexibly capture the (tail) dependence structure in models for multivariate extremes.}
To model spatial extremal dependence, a popular approach has been to use parametric max-stable processes \citep[see, e.g.,][]{brown1977extreme, smith1990max, schlather2002models, opitz2013extremal}, amongst which the Brown--Resnick class is one of the most popular choices. 
Although they have proven a useful tool for modeling nonstationary and anisotropic spatial extremal dependence \citep[see, e.g.,][]{huser2016non, chevalier2021modeling, shao2025flexible, zhong2025fast}, the use of max-stable processes in environmental applications has been criticized, because they are not designed to describe the spatial extreme events that actually occurred but, rather, spatial maxima representing composite spatial extreme events that occurred at different times (see \citealp{huser2025modeling}, for an extensive review of their limitations). \textcolor{black}{Moving beyond max-stability, \cite{zhang2026fast} proposed a variational autoencoder for high-dimensional spatial extremes based on a flexible nonstationary max-infinitely divisible model that accommodates both asymptotic dependence and independence, though the framework is still intended for block maxima.} 

On the other hand, peaks-over-threshold models based on Pareto processes offer a more suitable alternative since they allow modeling the original extreme events directly. \cite{ferreira2014generalized} introduced the generalized Pareto process, with spatial exceedances defined using the sup-norm; that is, a process is termed extreme if it is large at at least one site. This class of models was extended by \cite{dombry2015functional} to model high threshold exceedances defined in terms of a general nonnegative and homogeneous risk functional (e.g., the spatial average), here denoted by $r(\cdot)$, applied to the spatial process. These so-called $r$-Pareto processes offer more flexibility in defining spatial extreme events. \cite{de2018high}, with extension to generalized $r$-Pareto processes by \cite{de2022functional}, proposed computationally-efficient inference methods for $r$-Pareto processes constructed from log-Gaussian random functions, which share similar properties with Brown--Resnick max-stable processes. Their proposed estimation approach leverages the gradient scoring rule \citep{hyvarinen2005estimation}, which bypasses the computational bottlenecks of likelihood-based inference methods and thus allows $r$-Pareto models to be fitted to high-dimensional data; \textcolor{black}{see also \citet{koh2025using}, who fitted a Brown--Resnick $r$-Pareto process within a machine-learning pipeline by minimizing the gradient score, allowing predictor-driven temporal nonstationarity, and \citet{Zhong02025spatialmodeling}, who similarly incorporated a physics-based temporal covariate to make future projections using $r$-Pareto models. Related work also includes \citet{Healy2024inference}, who used a semiparametric Brown--Resnick $r$-Pareto process with temporally varying parameters to study extreme spatial temperature events in a changing climate.} 
Due to the flexibility of the $r$-Pareto processes in defining spatial extremes with various risk functionals and the efficient algorithm available for their inference, we thus follow this line of work and here use $r$-Pareto processes combined with the DCSM to capture spatially nonstationary extremal dependence. 

\textcolor{black}{In contrast to \citet{chevalier2021modeling} and \citet{richards2021spatial}, who captured nonstationary extremal dependence through injective deformations and max-stable processes, our goal here is to learn a flexible bijective deformation in combination with a peaks-over-threshold $r$-Pareto model and scalable loss-based inference.}
\textcolor{black}{These two strands of literature remain largely separate: existing deformation or dimension-expansion approaches for modeling nonstationary spatial extremes have been developed in max-stable or related settings, whereas recent work on $r$-Pareto processes has primarily focused on model construction, scalable inference, or predictor-driven nonstationarity. We bridge this gap by combining $r$-Pareto processes with deep bijective warpings into a unified framework for modeling high-dimensional nonstationary spatial extremes.}
This yields a coherent workflow in which the warping function $\bm{f}(\cdot)$ and all other dependence parameters are jointly estimated within a deep learning framework. \textcolor{black}{The computational aspect is central here: by leveraging \texttt{TensorFlow}~v2, automatic differentiation, and stochastic gradient descent, the proposed framework makes joint estimation practically feasible at scales that would be difficult to handle with standard implementations for spatial extremal dependence models.} Precisely, inference proceeds either via a weighted least-squares criterion based on conditional exceedance probabilities \citep[similar to][]{richards2021spatial} or via gradient score matching \citep{de2018high}. Regularization strategies to control deformation complexity are also explored, and the effect of different risk functionals, $r(\cdot)$, on the estimated extremal dependence pattern is investigated. \textcolor{black}{Our implementation of the proposed methodology is incorporated into an open-source \texttt{R} package \citep{vu2025deepspat}, thereby facilitating adoption and extension in real data analyses. To our knowledge, we provide the first \texttt{TensorFlow}-based modeling framework for spatial extremes, via modernization of the original DCSM codebase, paving the way for future high-dimensional applications to spatial extremes.}

The paper is organized as follows. In Section \ref{Methodology}, we develop deep compositional $r$-Pareto processes to model nonstationary extremal dependence. In Section~\ref{Inference}, we detail our joint inference approach for simultaneously estimating the warping function and model parameters. In Section~\ref{SimulationStudy}, we conduct a detailed simulation study to validate the efficacy of our proposed methodology. In Section~\ref{RealDataApplication}, we analyze United Kingdom (UK) precipitation extremes and investigate nonstationary extremal dependence patterns of rainfall anomalies around Wales. \textcolor{black}{The proposed modeling approach is found to capture the nonstationarity in the extremal dependence structure well, while making joint inference practically feasible at spatial scales that would be difficult to handle with standard implementations. In Section~\ref{Concluding}, we conclude with some discussion and avenues for future research.}
\section{Modeling with deep compositional $r$-Pareto processes} \label{Methodology}


\subsection{$r$-Pareto process construction}\label{Methodology:Construction}

Let $\{Y(\bm{s}), \bm{s}\in\mathcal{S}\}$ be our process of interest, defined on the compact spatial domain $\mathcal{S} \subseteq \mathbb{R}^K$, and let $F_{\bm{s}}(y)$ denote its marginal distribution (assumed continuous) at site $\bm{s}\in\mathcal{S}$. Since we focus on modeling the extremal dependence structure of $Y$, we can consider, without loss of generality, the standardized process $\{X(\bm{s}), \bm{s}\in \mathcal{S}\}$, obtained as $X(\bm{s})=1 /\left[1-F_{\bm{s}}\{Y(\bm{s})\}\right]$, which has standard Pareto margins, i.e., $\mathbb{P}(\X(\bm{s}) > x) = 1/x$, $x\geq 1$, for all $\bm{s}\in\mathcal{S}$.
There are multiple ways to define spatial extreme events of $\X$. From the peaks-over-threshold perspective, \cite{ferreira2014generalized} defined $\X$ to be extreme if $\sup_{\bm{s}\in\mathcal{S}}\{X(\bm{s})\}\geq u$ for some large threshold $u>1$. They showed that, under mild assumptions, the only possible limit of the rescaled process $X / u$ given $\sup_{\bm{s} \in \mathcal{S}} \{X(\bm{s})\}\geq u$, as $u$ tends to infinity, is the so-called simple Pareto process, $\{Z(\bm{s}), \bm{s}\in\mathcal{S}\}$, which admits the following construction:
\begin{equation} \label{eq:Pareto}
    Z(\bm{s}) = UV(\bm{s}),  \quad \bm{s}\in\mathcal{S},
\end{equation}
where $U$ is a standard Pareto random variable and $\{V(\bm{s}), \bm{s}\in\mathcal{S}\}$ is a nonnegative stochastic process independent of $U$ and satisfying $\sup_{\bm{s}\in\mathcal{S}}V(\bm{s}) = 1$. This definition of a spatial extreme event using the sup-norm is quite restrictive in practice, as the process $Y$ (or $X$) is generally not observed throughout the domain $\mathcal{S}$. To mitigate this issue, \cite{dombry2015functional} studied limits of high threshold exceedances defined in terms of a more general nonnegative $1$-homogeneous risk functional $r(\cdot)$ (i.e., $r(c\times\cdot) = cr(\cdot)$, for all $c>0$). A process $\X$ is now considered to be extreme if $r(\X)\geq u$ for some large threshold $u>1$, and we say that $\X\mid r(\X)\geq u$ is an $r$-exceedance above the threshold $u$. Under mild assumptions, \cite{dombry2015functional} showed that the only possible limit of the rescaled process $X/u$ given that $r(\X)\geq u$, as $u$ tends to infinity, is the so-called simple $r$-Pareto process, i.e., 
\begin{equation}\label{eq:limiting}
\mathbb{P}\left\{u^{-1} X \in \cdot \mid r(X) \geq u\right\} \rightarrow \mathbb{P}\left(Z \in \cdot\right) \text{ as } u\rightarrow \infty,
\end{equation}
where the limiting process $Z$ has a similar construction as in (\ref{eq:Pareto}) but with the constraint that $r(V)=1$ instead. To satisfy the latter condition, one may take $V=\Q/r(\Q)$ for some nonnegative stochastic process $\{Q(\bm{s}), \bm{s}\in\mathcal{S}\}$. An essential property of $r$-Pareto processes is peaks-over-threshold stability \citep[see][]{dombry2015functional}, also called threshold invariance by \cite{de2018high}. This means that for any measurable set $\mathcal{A}\subset \{\Z : r(\Z) \geq 1\}$ and $u > 1$, a simple $r$-Pareto process satisfies $\mathbb{P}\left\{u^{-1} \Z \in \mathcal{A} \mid r(\Z) \geq u\right\} = \mathbb{P}\left(\Z \in \mathcal{A}\right)$, which can be exploited for joint tail extrapolation.





Risk functionals of practical interest include, among others: i) the max-functional, $r_{\text{max}}(\X):= \max_{\bm{s}\in\mathcal{D}} X(\bm{s})$, which defines extremes such that the process is large at at least one site within a finite subset $\mathcal{D}\subset\mathcal{S}$---in practice, $\mathcal{D}$ is often chosen as the set of observed sampling locations or a subset thereof; ii) the sum-functional, $r_{\text{sum}}(\X):= \sum_{\bm{s}\in\mathcal{D}} X(\bm{s})$, which defines extremes as events such that the process is large on average over $\mathcal{D}$; and iii) the site-functional, $r_{\text{site}}(\X):= X(\bm{s}_0)$, which is a special case of the max- and sum-functionals when $\mathcal{D}=\{\bm{s}_0\}$ contains a single specific site of interest. For practical applications of the sum-functional, \cite{de2018high} argued that a direct summation of the rescaled process $\X$ on the standard Pareto scale may not be physically meaningful, and proposed using a modified sum-functional that first transforms the tail of $\X$ to match that of the original process $Y$ before computing the sum. Precisely, for some parameter $\beta>0$, they defined $r_{\text{sum};\beta}(\X):= \left( \sum_{\bm{s}\in\mathcal{D}} |X(\bm{s})|^{\beta} \right)^{1/\beta}$. As $\X$ is on the standard Pareto scale, $\beta$ can be selected as an estimate of the marginal tail index of $Y$, assuming that $Y$ is heavy-tailed with $\mathbb{P}(Y(\bm{s})>y)\sim c y^{-1/\beta}$, $c>0$, as $y\rightarrow\infty$, for all $\bm{s}\in\mathcal{S}$; see Section~\ref{RealDataApplication:Modeling} for details. 

A particular class of $r$-Pareto processes that we consider in this paper are the so-called Brown--Resnick $r$-Pareto processes; these are $r$-Pareto processes associated with Brown--Resnick max-stable processes, constructed as $\tilde{\Z}(\bm{s}) = \max_{i\geq 1}\Gamma_{i}^{-1}\tilde{V}_{i}(\bm{s})$, $\bm{s}\in\mathcal{S}$, where $\{\Gamma_{i}\}_{i\geq 1}$ are the points of a unit rate Poisson point process and, independently, $\{\tilde{V}_{i}\}_{i\geq 1}$ are independent and identically distributed log-Gaussian processes. Specifically, each $\tilde{V}_{i}$ is a copy of the process $\{\tilde{V}(\bm{s}), \bm{s}\in\mathcal{S}\}$ where $\tilde{V}(\bm{s}) = \exp\left\{ \epsilon(\bm{s}) - \sigma^2(\bm{s})/2\right\}$, $\bm{s}\in\mathcal{S}$, and $\{\epsilon(\bm{s}),\bm{s}\in\mathcal{S}\}$ is a zero-mean Gaussian process with marginal variance $\sigma^2(\bm{s})$ at site $\bm{s}$ and semivariogram $\gamma(\cdot, \cdot)$. The process $V$ arising, as in \eqref{eq:Pareto}, to construct the corresponding $r$-Pareto process counterpart  
has distribution characterized by $\mathbb{P}(V\in\cdot) = \mathbb{E}[r(\tilde{V})\mathds{1}\{\tilde{V}/r(\tilde{V})\in\cdot\}] / \mathbb{E}[r(\tilde{V})]$, where $\mathds{1}(\cdot)$ is the indicator function; see \cite{dombry2024pareto} for further details.

To measure the strength of extremal dependence between the process observed at two sites $\bm{s}_i, \bm{s}_j\in \mathcal{S}$, \cite{de2018high} proposed using the conditional exceedance probability (CEP), defined as
\begin{equation} \label{eq:cep}
\pi_{ij}(u, u')=\mathbb{P}\left[X\left(\bm{s}_j\right) \geqslant u' \mid\left\{X\left(\bm{s}_i\right) \geqslant u'\right\} \cap\left\{r\left(\X / u\right) \geqslant 1\right\}\right],
\end{equation}
for some high thresholds $u, u' > 1$. One can show that when both $u\to\infty$ and $u'\to\infty$ such that $u'/u\to\infty$, the CEP of the aforementioned Brown--Resnick $r$-Pareto process takes the limiting form $\pi_{ij} = \lim_{u,u'\to\infty} \pi_{ij}(u,u') = 2\left[1-\Phi\left\{\sqrt{\gamma(\bm{s}_i, \bm{s}_j)/2}\right\}\right]$. We use this extremal dependence measure for both inference and model checking. 



\subsection{Modeling nonstationarity}\label{Methodology:DCrPP}


The deep compositional spatial modeling approach of \cite{zammit2022deep} constructs a nonstationary covariance function using a composition of bijective warping functions, and we here extend it to the (non-Gaussian) extremal setting. Assume that $r$-exceedances of the process $\{X(\bm{s}), \bm{s}\in \mathcal{S}\}$ converge, when appropriately rescaled, to a Brown--Resnick $r$-Pareto process, characterized by a nonstationary semivariogram denoted here by $\gamma_{\mathcal{S}}: \mathcal{S}\times\mathcal{S}\rightarrow\mathbb{R}_{+}$. We construct a spatial deformation, also called ``warping'', $\bm{f}: \mathcal{S}\rightarrow\mathcal{W}$, such that the process $\X$ is approximately stationary and isotropic on the warped domain $\mathcal{W}$. In other words, we assume that the semivariogram can be expressed as $\gamma_{\mathcal{S}}(\bm{s}_1, \bm{s}_2) = \gamma_{\mathcal{W}}(\|\bm{f}(\bm{s}_1) - \bm{f}(\bm{s}_2)\|)$ for $\bm{s}_1, \bm{s}_2 \in \mathcal{S}$, where $\gamma_{\mathcal{W}}(\cdot)$ is a semivariogram function that only depends on the distance $\|\bm{f}(\bm{s}_1) - \bm{f}(\bm{s}_2)\|$ between $\bm{f}(\bm{s}_1),\bm{f}(\bm{s}_2)\in\mathcal{W}$. 
In this paper, we take the (common) power variogram $\gamma_{\mathcal{W}}(\tilde{\bm{s}}_1, \tilde{\bm{s}}_2) = (\|\tilde{\bm{s}}_1 - \tilde{\bm{s}}_2\|/\varphi)^{\kappa}$, $\tilde{\bm{s}}_1, \tilde{\bm{s}}_2 \in \mathcal{W}$, with range parameter $\varphi>0$ and smoothness parameter $\kappa\in(0,2)$, and let $\bm{\psi}=(\varphi, \kappa)'$. 
Therefore, on the original space, the semivariogram $\gamma_{\mathcal{S}}$ is parametrized by both the parameter vector $\bm\psi$ and the warping function $\bm f$ and has the form $\gamma_{\mathcal{S}}(\bm s_1,\bm s_2) := \gamma_{\mathcal{S}}(\bm s_1,\bm s_2;\bm\psi,\bm f)=(\|\bm{f}(\bm{s}_1) - \bm{f}(\bm{s}_2)\|/\varphi)^{\kappa}$.

Constructing a flexible family of bijective warpings that can be estimated efficiently is challenging. Following \cite{zammit2022deep}, we build the warping function $\bm{f}(\cdot)$ as a composition of $n$ warping units, i.e., $\bm{f}(\cdot) = \bm{f}_n \circ \bm{f}_{n-1} \circ \cdots \circ \bm{f}_1(\cdot)$, where each $\bm{f}_i(\cdot)$ is taken to be a simple bijective function.
Each warping unit $\bm{f}_i: \mathcal{W}_{i-1} \rightarrow \mathcal{W}_i$ maps $\mathcal{W}_{i-1} \subset \mathbb{R}^K$ to $\mathcal{W}_{i} \subset \mathbb{R}^K$, where $\mathcal{W}_{0} := \mathcal{S}$ is the original spatial domain, and $\mathcal{W}_{n}:= \mathcal{W}$ denotes the target warped space $\mathcal{W}$. For all $i=1,\dots,n$, we parametrize each warping unit $\bm{f}_i(\cdot)$ in terms of basis functions $\bm{\phi}_i(\cdot; \bm{\Theta}_i)$, which depend on a parameter vector $\bm{\Theta}_i$ to be estimated, and a weight matrix $\bm{W}_i$ also to be estimated. 
\cite{zammit2022deep} proposed three types of warping units, which we also consider in this paper and review below: axial warping, radial basis function, and M\"obius transformation units. 

\paragraph{Axial Warping (AW) Unit.} An AW unit is a nonlinear mapping of one input dimension of the spatial ordinate $\bm{s}\in\mathcal{S}$. The map is constrained to be strictly monotonic, and hence bijective. For some fixed $k = 1, \ldots, K$, the AW unit, $\bm{f}_i(\cdot) = \left\{f_{i 1}(\cdot), \ldots, f_{i K}(\cdot) \right\}$, warps the $k$-th input dimension and is defined such that $f_{i k}(\bm{s}) = \bm{W}_i^{(k)} \bm{\phi}_i\left(\bm{s} ; \bm{\Theta}_i\right)$ and $f_{i k^{\prime}}(\bm{s}) = s_{k^{\prime}}$ for $k^{\prime} \neq k$ (that is, only the $k$-th component of $\bm{s}$ is affected), where $\bm{\phi}_i(\cdot) = (\phi_{i 1}(\cdot), \ldots, \phi_{i m_i}(\cdot))'$ comprises $m_i$ basis functions with $\phi_{i 1}\left(\bm{s}\right)=s_k$ and $\phi_{i j}\left(\bm{s} ; \bm{\theta}_{ij}\right)=\left[1+\exp \left\{-\theta_{ij1}\left(s_k-\theta_{ij2}\right)\right\}\right]^{-1}$, $\bm{\theta}_{ij}=(\theta_{ij1},\theta_{ij2})'$, for $j= 2, \ldots, m_i$. Weights $\bm{W}_i^{(k)} = (w_{i1}, \ldots, w_{im_i})$ are required to be nonnegative to ensure the bijectivity of $\bm{f}_i(\cdot)$, and $\bm{\Theta}_i = (\theta_{i21}, \theta_{i22}, \ldots, \theta_{im_i1}, \theta_{im_i2})'$ are fixed to produce smooth warping functions over the entire domain of the $k$-th input dimension of $\mathcal{W}_{i-1}$. Among all $m_i$ basis functions, the first basis function $\phi_{i 1}(\cdot)$ controls linear scaling of the $k$-th input dimension, and the remaining basis functions, $\phi_{i 2}(\cdot), \ldots, \phi_{i m_i}(\cdot)$, control its nonlinear scaling with sigmoid functions.

\paragraph{Radial Basis Function (RBF) Unit.} The RBF unit provides a localized expansion or contraction of the space around a centroid. An RBF unit $\bm{f}_i(\cdot)$, for $\bm{s} \in \mathcal{W}_{i-1}$, is defined as
$$
\bm{f}_i(\bm{s}) = \bm{s}+w_i \left(\bm{s}-\bm{c}_{i}\right) \exp \left\{-b_i\left\|\bm{s}-\bm{c}_{i}\right\|^2\right\},
$$
where $b_i>0$ is a learnable rate parameter, $\bm{c}_i = (c_{i1}, \ldots, c_{iK})' \in \mathcal{W}_{i-1}$ denotes the warping centroid, and $w_i$ controls the intensity of stretching ($w_i>0$) or shrinking ($w_i<0$); when $w_i=0$, $\bm{f}_i(\cdot)$ is the identity function. For $K = 2$, $\bm{f}_i(\cdot)$ can be equivalently expressed as $\bm{f}_i(\cdot) = \bm{W}_i\bm{\phi}_i(\cdot; \bm{\Theta}_i)$ with
$$
\begin{aligned}
\bm{\phi}_i(\cdot) &= \left( \phi_{i1}(\cdot), \phi_{i2}(\cdot), \phi_{i3}(\cdot), \phi_{i4}(\cdot) \right)' \text{ with } \begin{cases}
\phi_{i 1}(\bm{s})=s_1 \\
\phi_{i 2}(\bm{s})=s_2 \\
\phi_{i 3}\left(\bm{s} ; \bm{\Theta}_i\right) = \left(s_1-c_{i1}\right) \exp \left(-b_i\left\|\bm{s}-\bm{c}_i\right\|^2\right) \\
\phi_{i 4}\left(\bm{s} ; \bm{\Theta}_i\right) = \left(s_2-c_{i2}\right) \exp \left(-b_i\left\|\bm{s}-\bm{c}_i\right\|^2\right)
\end{cases};
\end{aligned}
$$
here $\bm{\Theta}_i = \left(\bm{c}_i', b_i\right)'$ and $\bm{W}_i = \left[\bm{I}_2 \; w_i \bm{I}_2\right]$, with $\bm{I}_2$ the $2\times2$ identity matrix. The weight $w_i$ must satisfy $-1 < w_i<\exp (3 / 2) / 2$ to enforce injectivity \citep{perrin1999modelling}.

In order to simplify computations, \cite{zammit2022deep} utilized the single-resolution RBF unit, which is parameterized by a resolution $l \in \mathbb{N}_+$; we denote this by SR-RBF($l$). The SR-RBF($l$) unit is formed as a composition of RBF units, with centroids $\mathbf{c}_i$ fixed on a $3^l \times 3^l$ grid in $\mathcal{W}_{i-1}$ and $b_i$ set to $b_i=2 \left(3^l-1\right)^2$ for all layers. As a result, the SR-RBF($l$) unit has $3^{2 l}$ layers and only one weight, $w_i$, needs to be estimated per layer. 


\paragraph{M\"obius Transformation (MT) Unit.} The M\"obius transformation is a bijective, conformal map from the Riemann sphere onto itself; it rotates and moves the sphere to a new location and orientation in space, and then performs stereographic projection back to the plane. Thus, it can only be applied when $\mathcal{S}\subseteq\mathbb{R}^2$, i.e., $K=2$; we hereafter assume this to hold. The MT unit is defined through the function 
$$
\phi_i\left(\bm{s} ; \bm{\Theta}_i\right)=\frac{a_{1i} z(\bm{s})+a_{2i}}{a_{3i} z(\bm{s})+a_{4i}},
$$ 
where $z(\bm{s}) = s_1+s_2 \sqrt{-1} \in \mathbb{C}$, $\bm{s} = (s_1, s_2)'$, $\bm{\Theta}_i =\left(a_{1i}, \ldots, a_{4i}\right)'\in \mathbb{C}^4$. This warping unit contains eight unknown parameters (the real and imaginary components of $a_{1i},\ldots,a_{4i}$) and all weights $\bm{W}_i$ are fixed to one. Finally, $f_{i 1}(\bm{s})=\operatorname{Re}\left(\phi_i\left(\bm{s} ; \bm{\Theta}_i\right)\right)$ and $f_{i 2}(\bm{s})=\operatorname{Im}\left(\phi_i\left(\bm{s} ; \bm{\Theta}_i\right)\right)$, $\bm{s} \in \mathcal{W}_{i-1}$, where $\operatorname{Re}(\cdot)$ and $\operatorname{Im}(\cdot)$ return the real and imaginary components, respectively, of their arguments. Note that M\"obius transformations form a group under composition, so it is unnecessary to apply successive MT units in the compositional warping.

\section{Inference} \label{Inference}

\subsection{Model fitting by loss minimization} \label{Inference:Joint}


Suppose that $\x = \{x(\bm{s}_i)\}_{i=1}^D\in\mathbb{R}^{D}_{+}$ is a realization of the normalized process $\X = \{X(\bm{s}), \bm{s}\in\mathcal{S}\}$ (with standard Pareto margins) at $D$ locations in $\mathcal{D}_0 = \{\bm{s}_1, \ldots, \bm{s}_D\}\subset\mathcal{S}$, and assume that the rescaled $r$-exceedances of $\X$, i.e., $\X/u$ given $r(\X)>u$, converge weakly to a nonstationary $r$-Pareto process $\Z$ as $u$ goes to infinity. 
In the sequel, we only consider risk functionals that involve variables observed at locations within the finite set $\mathcal{D}_0$ and thus, by abuse of notation, we write $r(\bm{x})$ to denote the risk functional evaluated at the realization $\bm{x}$. 
An approximate realization of the limit process $\Z$ at the locations in $\mathcal{D}_0$ is then given by $\z = \bm{x}/u$ such that $r(\bm{x})>u$, for some high threshold $u>1$. In practice, for $N$ independent replicates $\{\x_t\}_{t=1}^{N}$, the threshold $u$ is taken to be a high empirical quantile of $\{r(\x_t)\}^N_{t=1}$. The (approximate) negative log-likelihood function associated with the limiting $r$-Pareto process may thus be written as
\begin{equation} \label{eq:likelihood}
\ell(\estimand) = -\sum^N_{t=1} \mathds{1}\left\{r\left(\frac{\x_t}{u}\right) \geq 1\right\} \log\rPpdf\left(\frac{\x_t}{u}; \estimand\right), 
\end{equation}
where $\rPpdf(\z; \estimand)$ denotes the probability density function of the $r$-Pareto process, which is of the form $\rPpdf(\z; \estimand) := \lambda(\z; \estimand) / \Lambda\left(\excReg;\estimand\right)$ defined over the support $\excReg := \left\{\z \in \mathbb{R}_{+}^D: r(\z) \geqslant 1\right\}$, and with $\bm{\psi}$ and $\bm{f}$ the vector of extremal dependence parameters and warping function, respectively, to be estimated; here, $\lambda(\bm{z}; \estimand)$ is a suitable parametric intensity function and $\Lambda\left(\excReg;\estimand\right)$ is a normalizing factor given by integrating $\lambda(\z; \estimand)$ over the exceedance region $\excReg$, i.e., $\Lambda\left(\excReg;\estimand\right) := \int_{\excReg} \lambda(\z; \estimand) \mathrm{d} \z$. Note that $\rPpdf(\z; \estimand)$ is dependent on the chosen risk functional $r(\cdot)$. Specifically, when $\Z$ is the Brown--Resnick $r$-Pareto process defined in terms of log-Gaussian functions with semivariogram $\gamma_{\mathcal{S}}(\cdot,\cdot; \estimand)$, the intensity function is \citep{engelke2015estimation}
\begin{equation}\label{eq:intensity}
\lambda(\z; \estimand)=\frac{\left|\bm{\Sigma}_{\estimand}\right|^{-1 / 2}}{z_1^2 z_2 \cdots z_D(2 \pi)^{(D-1) / 2}} \exp \left(-\frac{1}{2} \tilde{\z}' \bm{\Sigma}_{\estimand}^{-1} \tilde{\z}\right),
\end{equation}
with $(D-1)$-dimensional vector $\tilde{\z} = \left\{\log \left(z_i / z_1\right)+\gamma_{i, 1}\right\}_{i=2}^D$, and $(D-1)$-dimensional square matrix $\bm{\Sigma}_{\estimand} = \left(\gamma_{i, 1}+\gamma_{j, 1}-\gamma_{i, j}\right)_{2 \leq i, j \leq D}$ for $z_i = z(\bm{s}_i)$ and $\gamma_{i,j} = \gamma_{\mathcal{S}}(\bm{s}_i,\bm{s}_j;\estimand)$, $i, j =1,\ldots, D$. 
The computational bottleneck when minimizing (\ref{eq:likelihood}) comes from the evaluation of the multivariate $D$-dimensional integral $\Lambda\left(\excReg; \estimand\right)$ for a general risk functional $r(\cdot)$, which makes full likelihood inference for $r$-Pareto processes intractable when $D$ is large. Furthermore, \cite{wadsworth2014efficient} and \cite{de2018high} developed a censored likelihood based on \eqref{eq:likelihood} 
that incorporates both fully observed extreme values and partially-observed extreme values, where any component $z_i$ lying below a marginal threshold is treated as censored; evaluating a censored likelihood further requires multiple partial integrals of the density $\rPpdf(\z; \estimand)$. Although general expressions exist for censored likelihood contributions when considering the Brown--Resnick $r$-Pareto process, they involve multivariate Gaussian distribution functions that are costly to evaluate in high dimensions. \cite{de2018high} advocated quasi-Monte Carlo methods to accelerate their evaluation, but these remain intensive when estimating model parameters with hundreds of locations. We here make use of the censored likelihood to assess model performance in subsequent studies (see Sections~\ref{SimulationStudy} and \ref{RealDataApplication}), but adopt a more computationally efficient loss function for estimation; for brevity, we do not detail the censored likelihood and instead direct the reader to \cite{de2018high} for full mathematical details.


Inference for the deep compositional $r$-Pareto process $\X$ thus involves jointly optimizing the dependence parameters $\bm{\psi}$ along with the warping function weights and parameters, denoted by $\bm{W} = \{\bm{W}_1, \ldots, \bm{W}_n\}$ and $\bm{\Theta} = \{\bm{\Theta}_1, \ldots, \bm{\Theta}_n\}$, with respect to a chosen loss function. Each warping unit $\bm{f}_i(\cdot)$, parameterized by $\bm{W}_i$ and $\bm{\Theta}_i$, maps sampling locations $\mathcal{D}_{i-1} \subset \mathcal{W}_{i-1}$ to $\mathcal{D}_i \subset \mathcal{W}_i$, $i=1,\ldots,n$, where $\mathcal{D}_0 \subset \mathcal{S}$ denotes the original sampling locations. All location sets, $\mathcal{D}_0, \ldots, \mathcal{D}_n$, are rescaled to lie in $[-0.5, 0.5]^2$ to ensure numerical stability of model training. Only the relative distance between locations in the warped space affects the extremal dependence structure; rescaling may influence the range parameter $\varphi$, but preserves the underlying geometry of the warped space.


We consider two loss functions for fitting deep compositional $r$-Pareto processes:
i) a weighted square error loss on conditional exceedance probabilities with data-adaptive weights, which is computationally efficient and only uses information from pairs of sites; and
ii) the gradient score matching approach of \cite{de2018high}, which exploits information contained in the entire multivariate vector.

\paragraph{Weighted least squares.} An intuitive approach for inference is to minimize the (weighted) square error between the theoretical model-based CEPs, i.e., $\pi_{ij}(u, u'; \estimand)$ in (\ref{eq:cep}), determined by the dependence parameters $\bm{\psi}$ and warping $\bm{f}$, and their corresponding empirical counterparts. We pre-estimate the empirical counterparts, denoted as $\hat{\pi}_{ij}(u, u')$ for sites $\bm{s}_i$ and $\bm{s}_j$, using the formula:
\begin{equation*}
\hat{\pi}_{ij}(u, u') = \frac{\sum^N_{t=1}\mathds{1}\left\{x_t(\bm{s}_i)\geq u', x_t(\bm{s}_j)\geq u', r(\x_t)\geq u\right\}}{ \frac{1}{2}\left[ \sum^N_{t=1}\mathds{1}\left\{x_t(\bm{s}_i)\geq u', r(\x_t)\geq u\right\} + \sum^N_{t=1}\mathds{1}\left\{x_t(\bm{s}_j)\geq u', r(\x_t)\geq u\right\} \right] },
\end{equation*}
where the denominator achieves symmetry, i.e., $\hat{\pi}_{ij}(u, u') = \hat{\pi}_{ji}(u, u')$. In practice, $u$ is chosen as described above, while $u'$, a common marginal threshold, is chosen to be some (common) high empirical marginal quantile of $\{x_t(\bm{s}), \bm{s}\in\mathcal{D}_0\}^N_{t=1}$.
We write the (weighted) least squares loss function $\ell_{\text{WLS}}$ as 
\begin{equation}\label{eq:WLS_loss}
\ell_{\text{WLS}}(\estimand) = \sum_{1 \leq i<j \leq D} \checkw_{ij}\left\{\pi_{ij}(u, u'; \estimand) - \hat{\pi}_{ij}(u, u')\right\}^2,
\end{equation}
with weights $\checkw_{ij}\geq 0, i,j=1,\ldots,D$. The estimated dependence parameter vector $\hat{\bm{\psi}}_{\text{WLS}}$ and warping $\hat{\bm{f}}_{\text{WLS}}$, based on the warping weights $\hat{\bm{W}}_{\text{WLS}}$ and parameters $\hat{\bm{\Theta}}_{\text{WLS}}$, are obtained by minimizing $\ell_{\text{WLS}}(\estimand)$. \textcolor{black}{Under certain regularity conditions and assumptions (see Section~\ref{Appendix:WLS} of the Supplementary Material), the weighted least squares estimator $\hat{\bm{\psi}}_{\text{WLS}}$ is consistent and asymptotically normal as $N \to \infty$ \citep{newey1994large}.}


For the choice of weights in (\ref{eq:WLS_loss}), a natural approach is to assign more weight $\checkw_{ij}$ to closer location pairs and to downweight distant ones, but it may not be appropriate for nonstationary models. \cite{huser2016non} indicate that observation pairs with stronger extremal dependence contribute more to the fit of stationary models; based on this observation, we choose a data-driven scheme to assign more weight to pairs $(\bm{s}_i,\bm{s}_j)$ with stronger extremal dependence, where two simple choices include i) $\checkw_{ij} = \hat{\pi}_{ij}(u, u') \in [0,1]$ or ii) $\checkw_{ij} = 1/\{2-\hat{\pi}_{ij}(u, u')\} \in [1/2,1]$; we find the latter preferable in our experiments, as estimates of $\hat{\pi}_{ij}(u, u')$ often vanish when $u'$ is large, such that using the former may eliminate too many pairs that are weakly dependent in the tail, leading to information loss.


\paragraph{Gradient score matching.} \cite{de2018high} studied a computationally efficient inference method for Brown--Resnick $r$-Pareto processes that utilizes gradient score matching (GSM). The GSM method was developed by \cite{hyvarinen2005estimation} to estimate models for which the density cannot be easily normalized. The pivotal idea is that the normalizing constant in (\ref{eq:likelihood}), i.e., the multivariate integral $\Lambda\left(\excReg; \estimand\right)$, disappears when taking the gradient of the log-density, i.e., $\nabla_{\z} \log \rPpdf(\z; \estimand) = \nabla_{\z} \log \lambda(\z; \estimand)$, which eliminates the need to compute the multivariate integral $\Lambda\left(\excReg; \estimand\right)$. Hence, \cite{de2018high} proposed minimizing a divergence measure, defined by the expected (weighted) squared distance between the model score function and the true data generating score function, i.e., 
\begin{equation}\label{eq:GSM_theo}
\delta\left(\z; \estimand\right)=\int_{\excReg}\left\|\nabla_{\z} \log \lambda(\z; \estimand) \otimes \tildebw(\z)-\nabla_{\z} \log \lambda(\z; \bm{\psi}_0, \bm{f}_0) \otimes \tildebw(\z)\right\|_2^2 \lambda(\z; \bm{\psi}_0, \bm{f}_0) \mathrm{d} \z,
\end{equation}
where $\bm{\psi}_0$ and $\bm{f}_0$ denote the true dependence parameters and true warping, respectively, $\tildebw: \excReg \rightarrow \mathbb{R}_{+}^D$ is a positive weight function, and $\otimes$ denotes the componentwise product. Under some boundary and differentiability conditions imposed on the weight function, \cite{de2018high} showed that the empirical version of (\ref{eq:GSM_theo}) is, up to an additive constant, equal to
\begin{equation*}
\begin{aligned}
\hat{\delta}\left(\z; \estimand\right) = \sum_{i=1}^D&\left(2 \tildew_i(\z) \frac{\partial \tildew_i(\z)}{\partial z_i} \frac{\partial \log \lambda(\z; \estimand)}{\partial z_i} \right. \\
&\left.+\tildew_i^2(\z)\left[\frac{\partial^2 \log \lambda(\z; \estimand)}{\partial z_i^2}+\frac{1}{2}\left\{\frac{\partial \log \lambda(\z; \estimand)}{\partial z_i}\right\}^2\right]\right),
\end{aligned}
\end{equation*}
where $z_i\equiv z(\bm{s}_i)$ and $\tildew_i(\cdot)$ is the $i$-th component of $\tildebw(\cdot)$. 
To further adapt to extremes, $\tildebw(\cdot)$ can be designed to penalize low $r$-exceedances when $r(\z)$ is close to $u$ with low weight.
\cite{de2018high} proposed two weight functions. While they advocated one which is designed to mimic marginal censoring by downweighting components of $\z$ near $u$, in practice we found that such a design has the tendency to assign zero weight to too many low components; this removes substantial information, which leads to estimation difficulties in the presence of high-dimensional data.
Based on our experiments, we thus use the alternative weight function, defined as $\tildew_i(\z)=z_i[1-\exp\{1-r(\z)\}]$, $i=1,\ldots,D$, as it is simpler and performs better in our setting. 
The computational complexity of the loss function is now determined only by the inversion of the $(D-1)$-dimensional matrix $\bm{\Sigma}_{\estimand}$ in (\ref{eq:intensity}); the gradient score matching loss $\ell_{\text{GSM}}(\estimand)$ is
\begin{equation}\label{eq:GSM_loss}
\ell_{\text{GSM}}(\estimand)=\sum^N_{t=1}\mathds{1}\left\{r\left(\frac{\x_t}{u}\right) \geq 1\right\} \hat{\delta}\left(\frac{\x_t}{u}; \estimand\right).
\end{equation}
The estimated dependence parameter vector $\hat{\bm{\psi}}_{\text{GSM}}$ and warping $\hat{\bm{f}}_{\text{GSM}}$, based on the warping weights $\hat{\bm{W}}_{\text{GSM}}$ and parameters $\hat{\bm{\Theta}}_{\text{GSM}}$, are obtained by minimizing $\ell_{\text{GSM}}(\estimand)$.
\textcolor{black}{Under certain regularity conditions (e.g., parameter identifiability and assumptions on the smoothness of $\tilde{\bm{w}}(\cdot)$), the estimator $\hat{\bm{\psi}}_{\text{GSM}}$ minimizing \eqref{eq:GSM_loss}, with a fixed warping, is asymptotically normal, i.e., $\sqrt{N_u}\left( \hat{\bm{\psi}}_{\text{GSM}} - \bm{\psi}_{0} \right) \xrightarrow{d} \mathcal{N}\left\{\bm{0}, \bm{J}(\bm{\psi}_0)^{-1}\bm{K}(\bm{\psi}_0)\bm{J}(\bm{\psi}_0)^{-1} \right\}$, where ${N_u/N \rightarrow 0}$ as $N \rightarrow \infty$ with $N_u=\sum_{t=1}^N\mathds{1}\left\{r\left({\x_t/u}\right) \geq 1\right\}$ the number of functional excesses over threshold $u$, and $\bm{J}(\bm{\psi})=\mathbb{E}\left\{\partial^2 \delta(\Z; \bm{\psi})/(\partial \bm{\psi} \partial \bm{\psi}')\right\}$, $\bm{K}(\bm{\psi})=\mathbb{E}\left\{ ({\partial \delta(\Z; \bm{\psi})}/{\partial \bm{\psi}})({\partial \delta(\Z; \bm{\psi})}/{\partial \bm{\psi}'})\right\}$; see \citet{hyvarinen2007some} and \citet{de2018high} for details.}



An advantage of the gradient score matching method is that it exploits the full intensity $\lambda(\cdot)$, and hence information from the entire multivariate vector, rather than just information contained in pairwise extremal dependence estimates (as with the CEP in (\ref{eq:WLS_loss})). As it is based on the (unnormalized) likelihood function, we can expect it to be more efficient than a moment-based approach (like least squares) based on CEPs. However, the gradient score matching approach is potentially less robust to outliers and model misspecification, and has a heavier computational burden than the weighted least squares method.

\subsection{Regularization techniques} \label{Inference:Regularization}
To prevent overfitting when the warping $\bm{f}(\cdot)$ includes ``overly-flexible'' units, we add to the loss function a regularization term penalizing their weights. In our implementation, we only penalize the weights of SR-RBF($l$) units with $l \geq 2$ (recall Section~\ref{Methodology:DCrPP}), when such units are included in the model's architecture. The resulting regularized loss function, denoted $\tilde{\ell}(\cdot)$, modifies the original loss $\ell(\cdot)$---either $\ell_{\text{WLS}}(\cdot)$ in (\ref{eq:WLS_loss}) or $\ell_{\text{GSM}}(\cdot)$ in (\ref{eq:GSM_loss})---by adding a penalty equal to the sum of squared weights from the SR-RBF($l$) unit, i.e., 
\begin{equation}\label{eq:regularizedLoss}
    \tilde{\ell}(\cdot) = \ell(\cdot) + \alpha\sum^{n_l}_{i=1}w_i^2,
\end{equation}
where $\alpha>0$ is a penalty parameter, $n_l$ denotes the depth of the SR-RBF($l$) unit, and $w_i, i=1,\ldots,n_l$, denotes the weight of each layer. 
\textcolor{black}{The coefficient $\alpha$ can in principle be selected via a grid search; in the Supplementary Material (Section~\ref{Appendix:alpha_sensitivity}; Figures~\ref{pic:alpha_sens_800} and~\ref{pic:alpha_sens_2000}), we find that $\alpha=3$ performs well for the simulation study with $D=800$, while $\alpha=100$ is a reasonable choice for an idealized setting similar to the data application with $D=2000$. Due to the excessive computational burden, we have not performed the full grid search for the real-data application, and thus view the choice of $\alpha=100$ as practical guidance from an idealized simulation setting with a similar training size. We thus set $\alpha=3$ in Section~\ref{SimulationStudy} and use $\alpha=100$ in Section~\ref{RealDataApplication} as a practical default.}
Models with complex architectures comprising the SR-RBF($l$) unit with $l\geq 2$ are thus penalized by the above ridge regularizer, and we set $\tilde{\ell}(\cdot) = \ell(\cdot)$ if these units are not included.

An alternative approach to assess and guard against overfitting is to exploit cross-validation, which evaluates performance by assessing the loss over a carefully chosen out-of-sample validation set; see, e.g., \cite{wang2023spatial+}. Better performance over the validation locations usually indicates better generality of the estimated model. \textcolor{black}{In our setting, however, cross-validation can be difficult to interpret because withholding locations removes non-negligible spatial information and may distort the inferred deformation, since dependence is learned primarily from inter-site structure.} In practice, we found that the results can be very sensitive to the choice of validation locations, especially in a nonstationary context where estimating the spatial warping is a key goal, and this sensitivity may \textcolor{black}{artificially favor overly simple warpings} (i.e., underestimate the required warping complexity). \textcolor{black}{For this reason, we adopt the simple ridge regularization in \eqref{eq:regularizedLoss} to stabilize estimation and improve generalization, viewing cross-validation as a possible complement.}

\subsection{Implementation details and uncertainty quantification} \label{Inference:Implementation}

We minimize the regularized loss function $\tilde{\ell}(\cdot)$ in (\ref{eq:regularizedLoss}) using the Adam algorithm \citep{kingma2015adam}, which adaptively scales each parameter’s learning rate by keeping exponentially decaying averages of past gradients (first moments) and squared gradients (second moments).
\textcolor{black}{Adam can converge to different local optima depending on initialization for a nonconvex objective; in practice, however, repeated fits (including within our bootstrap) yield qualitatively stable deformations (up to rotation/scaling) and similar dependence.}
\textcolor{black}{As our objective function is nonconvex, Adam is sensitive to the initialization of the parameters. We thus fit an initial stationary model under the identity warping and use its estimates to initialize a subsequent nonstationary fit; this step is built into the optimization pipeline and yields stable estimation across repeated runs. Although local minima cannot be ruled out, the recovered warped geometry and estimated extremal dependence structure are typically stable across repeated fits, up to rotation and scaling.}
\textcolor{black}{Section~\ref{Appendix:Simulation_computation} of the Supplementary Material reports both single loss-evaluation times and full optimization times as functions of the dimension $D$ for WLS and GSM. The results confirm gentler scaling for WLS and steeper growth for GSM at larger $D$, while indicating that large spatial problems with tens of thousands of locations remain computationally manageable in practice.}

In contrast with the Gaussian case, providing suitable theoretical uncertainty quantification for extremal dependence estimates is challenging. We jointly assess the uncertainty of both dependence parameters, $\hat{\bm{\psi}}$, and the warping, $\hat{\bm{f}}(\cdot)$, as well as pairwise CEP estimates over the fitted warped space through the use of a nonparametric bootstrap. 
\textcolor{black}{For our simulation study in Section~\ref{SimulationStudy}, this is implemented at the replicate level by resampling $\{\bm{x}_t\}_{t=1}^n$ with replacement and re-computing the functional threshold $u$ (as a high empirical quantile of the bootstrap risk series) before selecting functional exceedances and re-estimating the extremal dependence model and warping. For the application in Section~\ref{RealDataApplication}, however, the bootstrap starts from the original data rather than from pre-standardized replicates, so we first resample the original observations and re-estimate the sitewise marginal standardization within each bootstrap replicate.}
Whilst the asymptotic normality of $\hat{\bm{\psi}}$ and the multivariate Delta method may provide another approach to quantify uncertainty, they are unable to account for the variability due to the estimation of the warping. Theoretical results for the asymptotic behavior of $\hat{\bm{\psi}}$ that account for the highly flexible, highly parametrized warping $\bm{f}(\cdot)$ are beyond the scope of this work.
\textcolor{black}{To further probe finite-sample calibration in the presence of a flexible warping, we additionally report, in the Section~\ref{Appendix:Bootstrap_design} of the Supplementary Material, results from a nested (two-level) resampling scheme that separates site-selection uncertainty from replicate/location resampling variability, together with an empirical coverage assessment of nominal 95\% intervals under several bootstrap variants in the simulation study (full refit versus conditional on the warping). The results are most encouraging for the more practically relevant pairwise CEP summaries, especially under WLS, indicating that the bootstrap is better calibrated for the practically relevant extremal dependence summaries than for the individual model parameters.} 

\textcolor{black}{The proposed model is implemented via the \texttt{Tensorflow} interface in \texttt{R} (TensorFlow~2.11.0; Python~3.7.11; \texttt{R}~4.3.2) and all experiments are run on a Linux workstation (Ubuntu kernel 5.4.0) equipped with an AMD Ryzen Threadripper PRO 3995WX CPU (64 cores) and 251~GB RAM (available memory $\approx$ 198~GB at run time).}

\section{Simulation study} \label{SimulationStudy}

\subsection{Overview} \label{SimulationStudy:Overview}

We illustrate the efficacy of our proposed model by simulating $r$-Pareto processes over a pre-defined warped space and estimating their nonstationary extremal dependence structure. We use a true compositional warping $\bm{f}(\cdot)$ consisting of three warping units (see Section~\ref{Methodology:DCrPP}), and consider four different architectures (see Table~\ref{tab:sim_archi}) of the deep model to capture nonstationarity. For the choice of architectures, we start with two general scenarios comprising all three types of warping units introduced in Section~\ref{Methodology:DCrPP}: axial warping (AW) units, single resolution radial basis functions (denoted as SR-RBF($l$) with resolution $l=1$ or $l=2$), and a M\"obius transformation (MT) unit. Specifically, Architecture~1 corresponds to a $12$-layered compositional warping $\bm{f}(\cdot)$, where $\bm{f}_1(\cdot)$ and $\bm{f}_2(\cdot)$ are axial warping units (one for each spatial dimension), $\bm{f}_3(\cdot), \ldots, \bm{f}_{11}(\cdot)$ comprise a $9$-layered SR-RBF(1) unit, and $\bm{f}_{12}(\cdot)$ is a MT unit. Architecture~2 has an additional 81-layered SR-RBF(2) unit after the SR-RBF(1) unit in Architecture~1. More complicated warping units, e.g., a 729-layered SR-RBF(3) unit, could also be used, but here we stop at resolution $l=2$ due to the additional computational burden and to reduce the risk of overfitting. In spite of this, we still find that our models are sufficiently flexible, and capture well the extremal dependence of the data in both our simulation study and subsequent data application. Architectures~3 and~4 are the same as Architectures~1 and~2, respectively, but without the MT unit. Models with Architecture~$k$ are hereafter termed as ``nonstationary model $k$'', $k=1,\ldots,4$. Architecture~0 denotes the ``null'' stationary and isotropic model, where no deformation is performed. 

\begin{table}[t!]
    \centering
    \caption{Architectures of deep compositional $r$-Pareto processes. The depth $n$ of each compositional model is also provided. All spatial locations are rescaled to $[-0.5,0.5]^2$ before fitting. Architecture~0 corresponds to the identity mapping on this rescaled domain, and hence to the stationary model.}
    \begin{tabular}{l|c|c|c|c|c}
    \hline \hline
    Architecture & $n$ & AW & SR-RBF(1) & SR-RBF(2) & MT \\
    \hline
    0 & 0 &  &  &  &  \\
    1 & 12 & $\surd$ & $\surd$ &  & $\surd$ \\
    2 & 93 & $\surd$ & $\surd$ & $\surd$ & $\surd$ \\
    3 & 11 & $\surd$ & $\surd$ &  &  \\
    4 & 92 & $\surd$ & $\surd$ & $\surd$ &  \\
    \hline
    \end{tabular} 
    \label{tab:sim_archi}
\end{table}

\begin{figure}[t!]
\begin{center}
\begin{tabular}{c}
\includegraphics[width=0.24\linewidth]{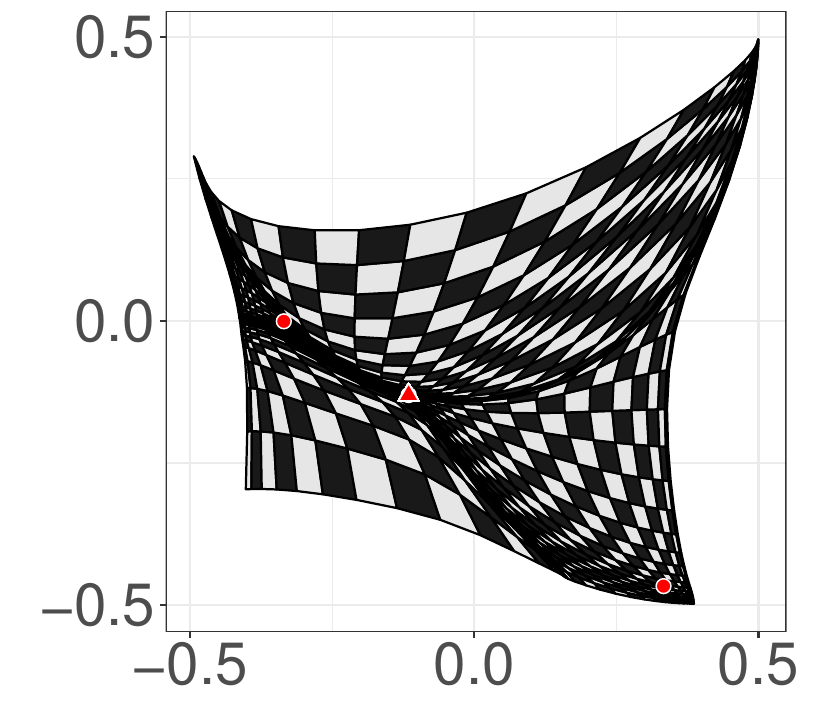}  \\
\includegraphics[width=0.24\linewidth]{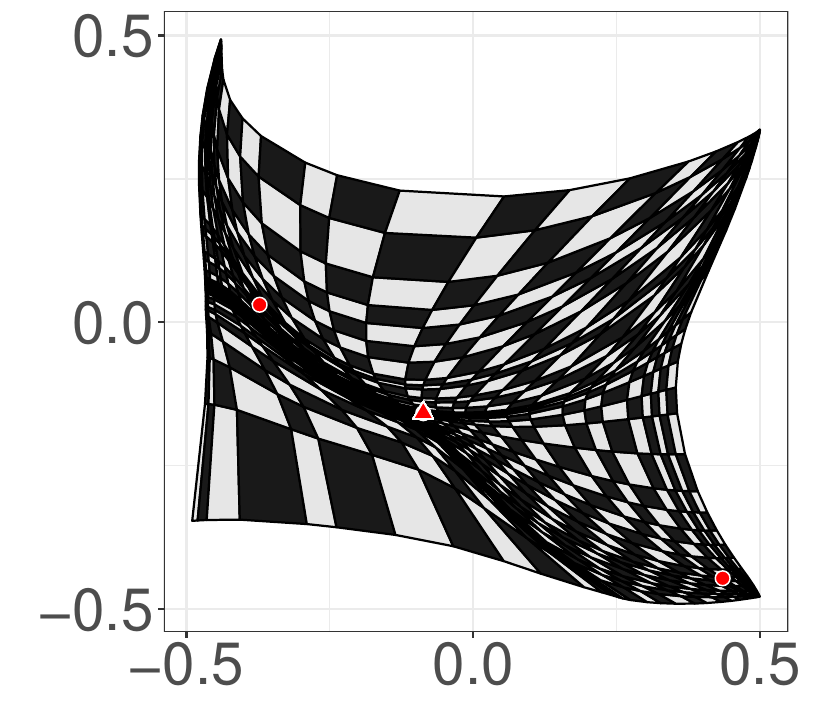} 
\includegraphics[width=0.24\linewidth]{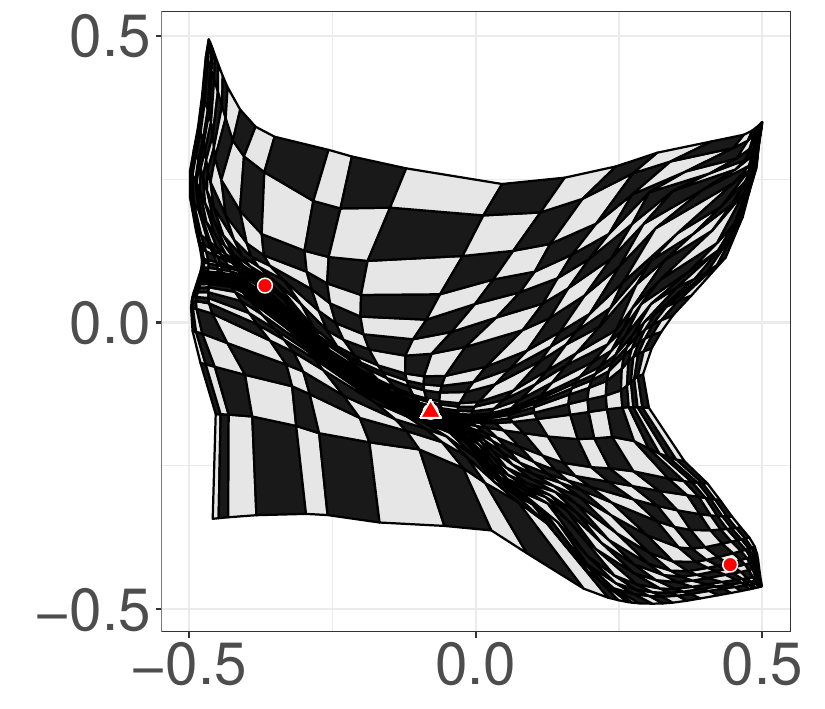}
\includegraphics[width=0.24\linewidth]{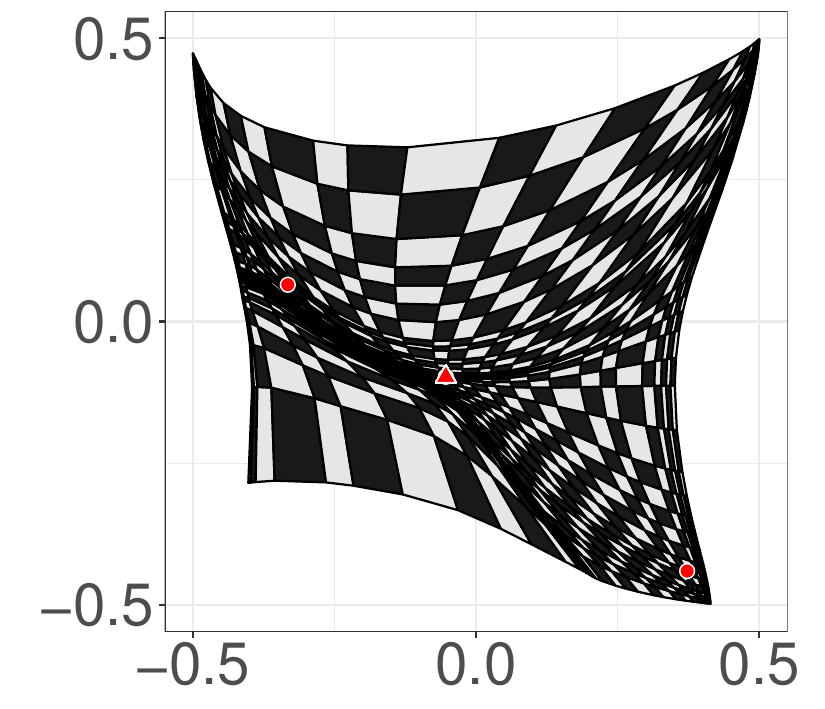}
\includegraphics[width=0.24\linewidth]{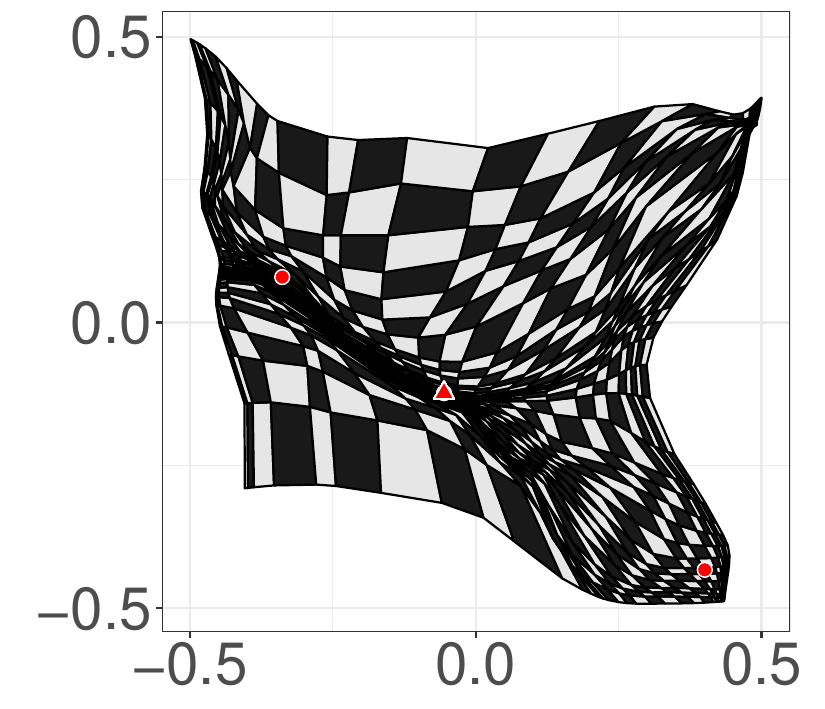} \\
\includegraphics[width=0.24\linewidth]{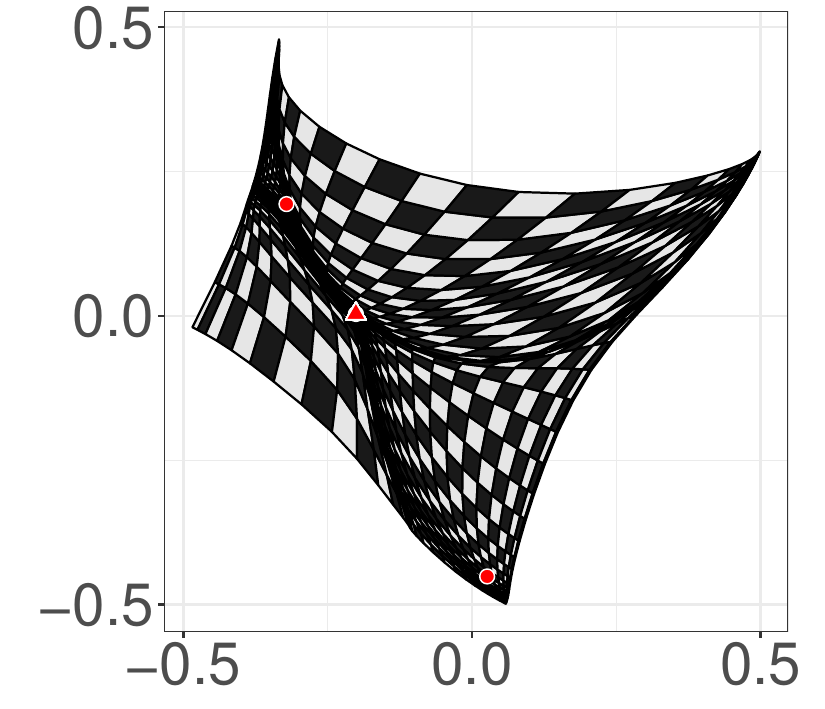} 
\includegraphics[width=0.24\linewidth]{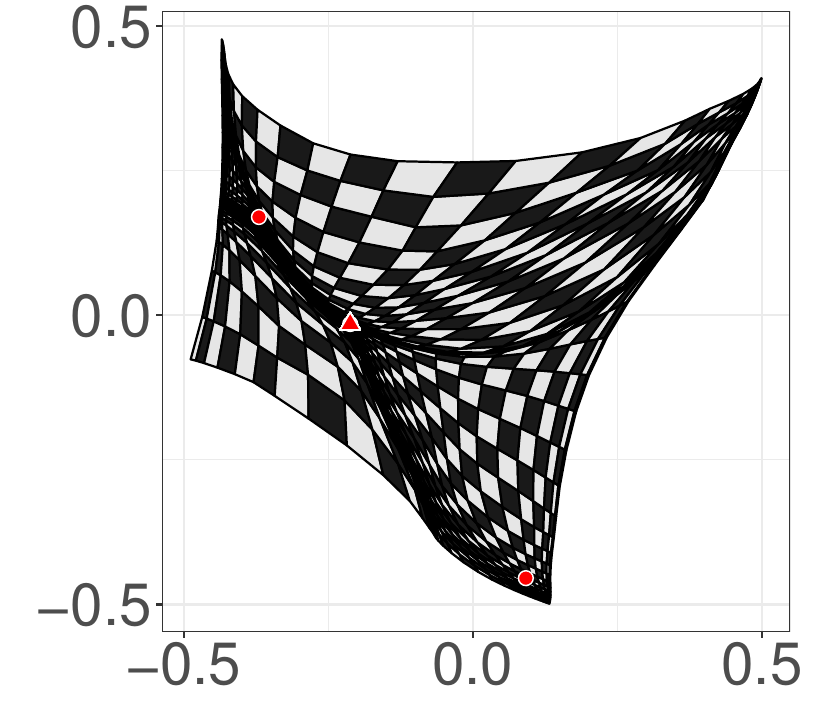}
\includegraphics[width=0.24\linewidth]{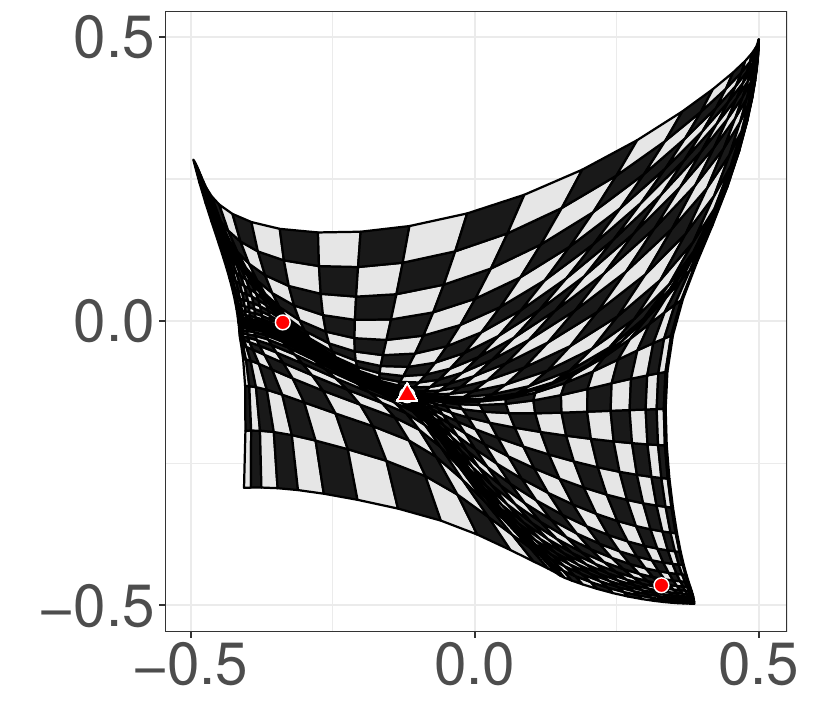}
\includegraphics[width=0.24\linewidth]{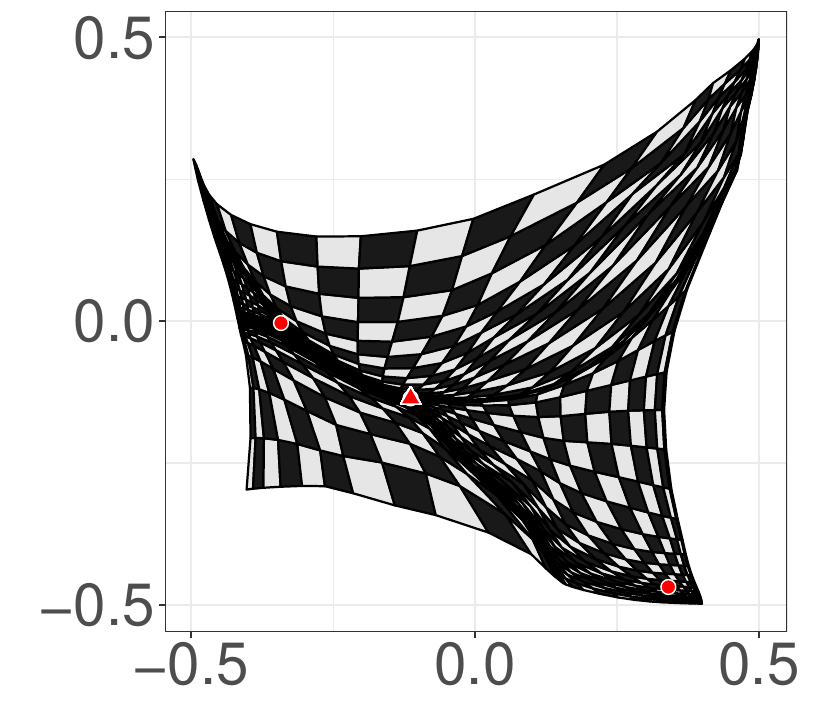}\\
\end{tabular}
\end{center}
\caption{Top row: true warped domain $\mathcal{W}$ generated using Architecture 3. Middle and bottom rows: estimated warped space $\hat{\mathcal{W}}$ using nonstationary models 1--4 (column 1--4) with WLS and GSM inference methods, respectively. The data are generated with the risk functional $r_{\text{max}}(\cdot)$, and three reference points are displayed as triangles and circles.} 
\label{pic:sim_space}
\end{figure}

In our simulation study, the original spatial locations $\mathcal{D}_0$ are chosen to be on a regular $101\times 101$ grid within the unit square $[-0.5, 0.5]^2$, and $\mathcal{D}_{n} = \bm{f}(\mathcal{D}_0)$ is created by randomly generating the weights $\bm{W} = \{\bm{W}_1, \ldots, \bm{W}_{n}\}$ and parameters $\bm{\Theta} = \{\bm{\Theta}_1, \ldots, \bm{\Theta}_{n}\}$ of the compositional model. Each architecture has its own ``true'' warped space $\mathcal{W}$, as shown in Figure~\ref{pic:sim_space} for Architecture 3. To facilitate comparisons, we maintain the same weights and parameters for all warping units, i.e., AW, SR-RBF(1), SR-RBF(2), and MT units in each architecture, which enables warped spaces generated using these four architectures to share some similarities and allows us to study the impact of each warping unit.
To assess the effect of architecture misspecification, which is likely to occur in real data applications, we perform a factorial simulation experiment. Specifically, we generate, for each architecture, its own true warped space. Then, we simulate a Brown--Resnick $r$-Pareto process with power semivariogram over each of the four true warped domains, i.e., we use the semivariogram  $\gamma_{\mathcal{S}}(\bm{s}_1, \bm{s}_2; \estimand) = (\|\bm{f}(\bm{s}_1) - \bm{f}(\bm{s}_2)\|/\varphi)^{\kappa}, \bm{s}_1, \bm{s}_2\in\mathcal{S}$ (with range parameter $\varphi = 0.2$ and smoothness parameter $\kappa = 1$). Finally, we fit the deep compositional $r$-Pareto processes, with each of the four different architectures, to each of the four datasets; this results in a total of 16 models to fit. Figure~\ref{pic:sim_space} provides the true warped domain for Architecture~3; see Figures~\ref{pic:sim_space1}, \ref{pic:sim_space2}, and \ref{pic:sim_space4} in the Supplementary Material for the true warped domains generated using Architectures 1, 2, and 4.

For each true warped domain $\mathcal{W}$, we generate $N = 5000$ replicates of a nonstationary $r$-Pareto process $\Z$, denoted as $\{\z_t\}_{t=1}^N$, with three considered risk functionals:
\begin{equation} \label{eq:sim_risks}
r_{\text{site}}\left(\Z\right) = Z(\bm{s}_0), \quad 
r_{\text{max}}\left(\Z\right)= \max_{\bm{s}\in\mathcal{D}_0} Z(\bm{s}), \quad \text{and } 
r_{\text{sum}}\left(\Z\right) = \sum_{\bm{s}\in\mathcal{D}_0}Z(\bm{s}),
\end{equation}
where $\mathcal{D}_0 = \{\bm{s}_1, \ldots, \bm{s}_D\}\in\mathcal{S}$. As $r_{\text{max}}(\cdot)$ is not differentiable, an approximation $r_{\text{max}}\left(\Z\right) \approx \left\{\sum_{\bm{s}\in\mathcal{D}_0} Z\left(\bm{s}\right)^{20}\right\}^{1 / 20}$ is used instead. 
Direct simulation of $r$-Pareto processes is only available for some risk functionals, e.g., $r_{\text{site}}(\cdot)$, but simulation is feasible for a general functional $r(\cdot)$ using the rejection sampling method of \cite{de2018high} and \cite{dombry2024pareto}. 
We implement the simulation of such processes in \texttt{R}, where parallel computing is exploited to relieve the low acceptance rate problem for some scenarios. The threshold $u$ is set to the empirical $95\%$ quantile of the observed risks $\{r(\z_1), \ldots, r(\z_N)\}$, giving rise to $N_u = 250$ $r$-exceedances, and $u'$ is set to the empirical $95\%$ marginal quantile. Among the $D = 101^2$ locations, \textcolor{black}{$D_{\text{train}}= 800$} and $D_{\text{test}}= 100$ locations are randomly chosen as the training and test locations, respectively; we use the former to fit the model and the latter to assess out-of-sample accuracy of the estimated nonstationary extremal dependence structure. To quantify model performance, we compute the censored likelihood over the test locations, evaluated at the estimated dependence parameters, using the \texttt{mvPot} package \citep{de2018high}. 

As the risk functional used in model fitting differs from the risk functional used to simulate the data, fitting $r$-Pareto process models inevitably introduces some risk-based misspecification, especially when using the max-functional. However, our subsequent simulation study shows that this risk-based misspecification has a negligible effect on the model fits and warped space estimates.

\textcolor{black}{In the Supplementary Material, Section~\ref{Appendix:Simulation_misspecification}, we also assess robustness to extremal dependence class misspecification (asymptotic dependence vs. asymptotic independence) by simulating nonstationary asymptotically independent data via a Gaussian-copula construction with standard Pareto margins on the same true warped domain, and then fitting our asymptotically dependent $r$-Pareto model. The estimated warped geometry remains close to the truth across risk functionals and both WLS/GSM inference approaches.}

\subsection{Results}\label{SimulationStudy:Results}

We illustrate the efficacy of the proposed modeling approach by investigating the extremal dependence structure relative to three reference sites (see, e.g., Figure~\ref{pic:sim_space}), where an obvious nonstationary extremal dependence pattern can be observed. 
For $r$-Pareto processes generated from each architecture, we first fit the deep compositional $r$-Pareto model using the true risk functional with four different architectures. In all four scenarios, both the gradient score matching (GSM) and weighted least squares (WLS) inference methods, described in Section~\ref{Inference:Joint}, are tested. Using the estimated warping weights $\hat{\bm{W}}$ and parameters $\hat{\bm{\Theta}}$ of the compositional model, we visualize the estimated warped spaces in Figure~\ref{pic:sim_space} for Architecture~3, and Figures~\ref{pic:sim_space1}, \ref{pic:sim_space2}, and \ref{pic:sim_space4} in the Supplementary Material for Architectures 1, 2, and 4, respectively, where the data are generated with the risk functional $r_{\text{max}}(\cdot)$.


\textcolor{black}{When the data are generated using Architecture~3 (AW + SR-RBF(1)), Figure~\ref{pic:sim_space} shows that nonstationary models~1 and~3 both perform well in recovering the warped space, with model~3 corresponding to the true architecture and model~1 differing only by an additional MT unit.
Table~\ref{tab:sim_compare}, showing the quantification of model fit accuracy using the censored likelihood, agrees with this observation: nonstationary models~1 and~3 generally perform better than the other models, while model~1, which includes an additional MT unit, is more flexible.
For data generated under Architecture 3, adding the extra regularized SR-RBF(2) component does not provide a systematic gain in out-of-sample censored likelihood, and may slightly degrade performance relative to the simpler model fits.
As Figure~\ref{pic:sim_space} is primarily qualitative, we additionally report two location-wise quantitative warping-recovery diagnostics in the Supplementary Material (Section~\ref{Appendix:Simulation_recovery}; Figures~\ref{pic:SWR} and \ref{pic:SB}): the sitewise warping recovery and signed bias, which compare each site’s estimated distance profile to that under the oracle deformation. These diagnostics explicitly localize where different inference methods and architectures under- or over-estimate the deformation across space. In this example, the GSM approach shows smaller and more homogeneous sitewise warping recovery values than the WLS alternative, and signed bias estimates indicate that the WLS tends to locally contract the warped space while the GSM provides recovery that is closer to unbiasedness.}

\textcolor{black}{Estimates of the extremal dependence parameters, $\varphi$ and $\kappa$, are also summarized in Table~\ref{tab:sim_table} for each risk functional and both inference methods. We observe generally less bias in estimates of the smoothness parameter $\kappa$ and smaller standard deviations of the dependence-parameter estimates when using the GSM inference method. At the same time, however, the empirical coverage results based on the nonparametric bootstrap are also somewhat lower for GSM. This is consistent with the fact that GSM tends to yield more ``stable'' estimates, both for the dependence parameters and for the warped geometry, so that the resulting bootstrap intervals are relatively narrow. If some finite-sample bias remains, whether due to limited sample size or other sources, such narrower intervals are more likely to miss the truth, leading to lower empirical coverage; see Section~\ref{Appendix:Bootstrap_coverage} of the Supplementary Material for details. Thus, while GSM has the advantage of providing more stable warping recovery and dependence-parameter estimation, it may also make any remaining systematic finite-sample bias more visible in coverage.}

\begin{table}[t!]
    \centering
    \caption{Test censored log-likelihoods (larger is preferred) evaluated for all fitted models, using both the gradient score matching (GSM) and weighted least squares (WLS) inference methods, and with three different risk functionals. The best models are highlighted in bold for each risk functional and each inference method, and the oracle model based on Architecture~3 is labeled with an asterisk.}
    \begin{tabular}{l|l|c|c|c}
    \hline \hline
    \multirow{2}{*}{Architecture} & Inference & \multicolumn{3}{c}{Risk functional} \\
    \cline{3-5}
    & method & $r_{\text{site}}(\cdot)$ & $r_{\text{max}}(\cdot)$ & $r_{\text{sum}}(\cdot)$ \\
    \hline
    Stationary & \multirow{5}{*}{WLS} & $-55088$ & $-14280$ & $10092$ \\ 
Nonstationary 1 &  & $\mathbf{-51847}$ & $-13592$ & $\mathbf{11739}$ \\ 
Nonstationary 2 &  & $-52671$ & $-13661$ & $11681$ \\ 
Nonstationary 3* &  & $-52727$ & $\mathbf{-13338}$ & $11659$ \\ 
Nonstationary 4 &  & $-52772$ & $-13995$ & $11662$ \\ 
\hline 
Stationary & \multirow{5}{*}{GSM} & $-54485$ & $-14071$ & $9862$ \\ 
Nonstationary 1 &  & $-50363$ & $\mathbf{-12688}$ & $\mathbf{12204}$ \\ 
Nonstationary 2 &  & $-51194$ & $-12691$ & $12193$ \\ 
Nonstationary 3* &  & $-50518$ & $-12695$ & $12197$ \\ 
Nonstationary 4 &  & $\mathbf{-50361}$ & $-12696$ & $12152$ \\ 
    \hline      
    \end{tabular} 
    \label{tab:sim_compare}
\end{table}

\begin{table}[t!]
    \centering
    \caption{Estimates of the extremal dependence parameters, $\hat{\bm{\psi}}' = (\hat{\varphi}, \hat{\kappa})$, for data generated with Architecture 3, using both the gradient score matching (GSM) and weighted least squares (WLS) inference methods, and three different risk functionals. The true values of the extremal dependence parameters are $\varphi_0 = 0.2$ and $\kappa_0 = 1$. Standard deviations obtained using a nonparametric bootstrap are reported in brackets as subscripts of the corresponding parameter estimate. The oracle model based on Architecture~3 is labeled with an asterisk.}
    \resizebox{\columnwidth}{!}{%
    \begin{tabular}{l|l|c|c|c}
    \hline \hline
    \multirow{2}{*}{Architecture} & Inference & \multicolumn{3}{c}{Risk functional} \\
    \cline{3-5}
    & method & $r_{\text{site}}(\cdot)$ & $r_{\text{max}}(\cdot)$ & $r_{\text{sum}}(\cdot)$ \\
    \hline 
    Stationary & \multirow{5}{*}{WLS} & $0.267_{(0.035)}$, $1.150_{(0.092)}$ & $0.293_{(0.056)}$, $1.191_{(0.115)}$ & $0.259_{(0.051)}$, $1.031_{(0.075)}$ \\ 
Nonstationary 1 &  & $0.176_{(0.035)}$, $1.134_{(0.103)}$ & $0.260_{(0.050)}$, $1.216_{(0.160)}$ & $0.164_{(0.036)}$, $0.978_{(0.089)}$ \\ 
Nonstationary 2 &  & $0.177_{(0.035)}$, $1.137_{(0.103)}$ & $0.259_{(0.048)}$, $1.210_{(0.150)}$ & $0.179_{(0.031)}$, $0.981_{(0.089)}$ \\ 
Nonstationary 3* &  & $0.203_{(0.030)}$, $1.156_{(0.118)}$ & $0.254_{(0.052)}$, $1.204_{(0.163)}$ & $0.199_{(0.043)}$, $0.992_{(0.095)}$ \\ 
Nonstationary 4 &  & $0.193_{(0.030)}$, $1.138_{(0.111)}$ & $0.250_{(0.052)}$, $1.203_{(0.159)}$ & $0.194_{(0.040)}$, $0.981_{(0.092)}$ \\ 
\hline 
Stationary & \multirow{5}{*}{GSM} & $0.153_{(0.001)}$, $0.943_{(0.002)}$ & $0.155_{(0.001)}$, $0.948_{(0.003)}$ & $0.154_{(0.001)}$, $0.946_{(0.003)}$ \\ 
Nonstationary 1 &  & $0.178_{(0.003)}$, $1.005_{(0.016)}$ & $0.179_{(0.003)}$, $1.004_{(0.014)}$ & $0.176_{(0.003)}$, $0.991_{(0.013)}$ \\ 
Nonstationary 2 &  & $0.180_{(0.003)}$, $1.014_{(0.012)}$ & $0.179_{(0.003)}$, $1.007_{(0.016)}$ & $0.177_{(0.003)}$, $0.997_{(0.016)}$ \\ 
Nonstationary 3* &  & $0.178_{(0.003)}$, $1.007_{(0.016)}$ & $0.181_{(0.003)}$, $1.021_{(0.016)}$ & $0.180_{(0.004)}$, $1.017_{(0.018)}$ \\ 
Nonstationary 4 &  & $0.180_{(0.002)}$, $1.019_{(0.011)}$ & $0.180_{(0.002)}$, $1.017_{(0.012)}$ & $0.179_{(0.003)}$, $1.012_{(0.015)}$ \\ 
    \hline
    \end{tabular} }
    \label{tab:sim_table}
\end{table}

\begin{figure}[t!]
\begin{center}
\begin{tabular}{c}
\includegraphics[width=0.74\linewidth]{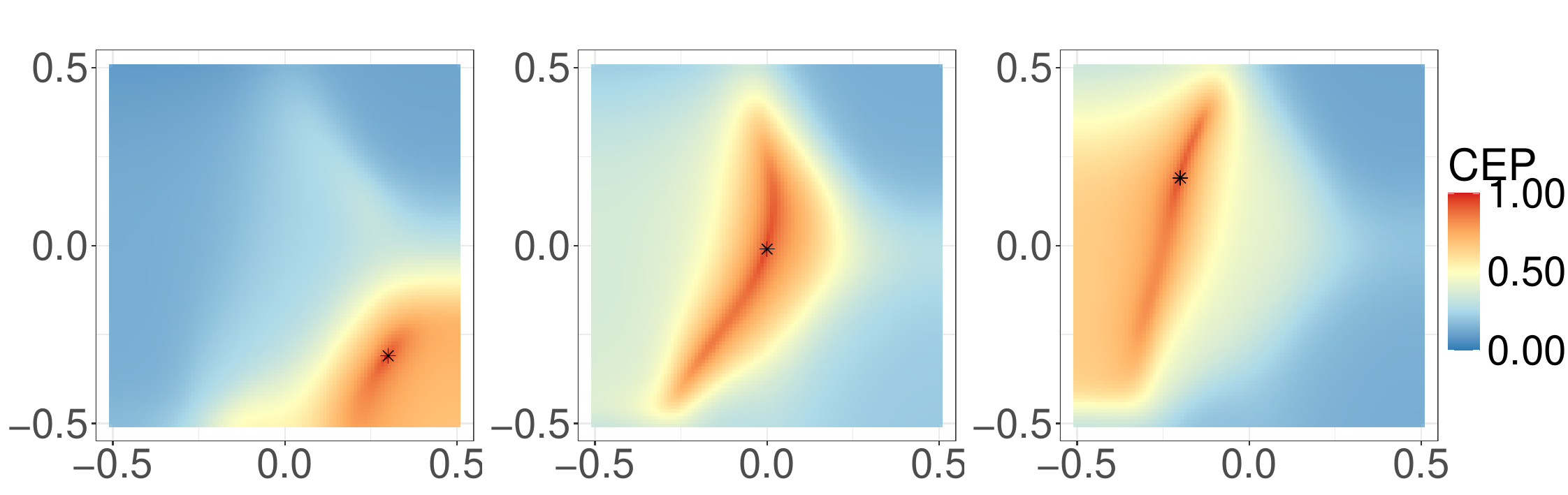}
\\
\includegraphics[width=0.74\linewidth]{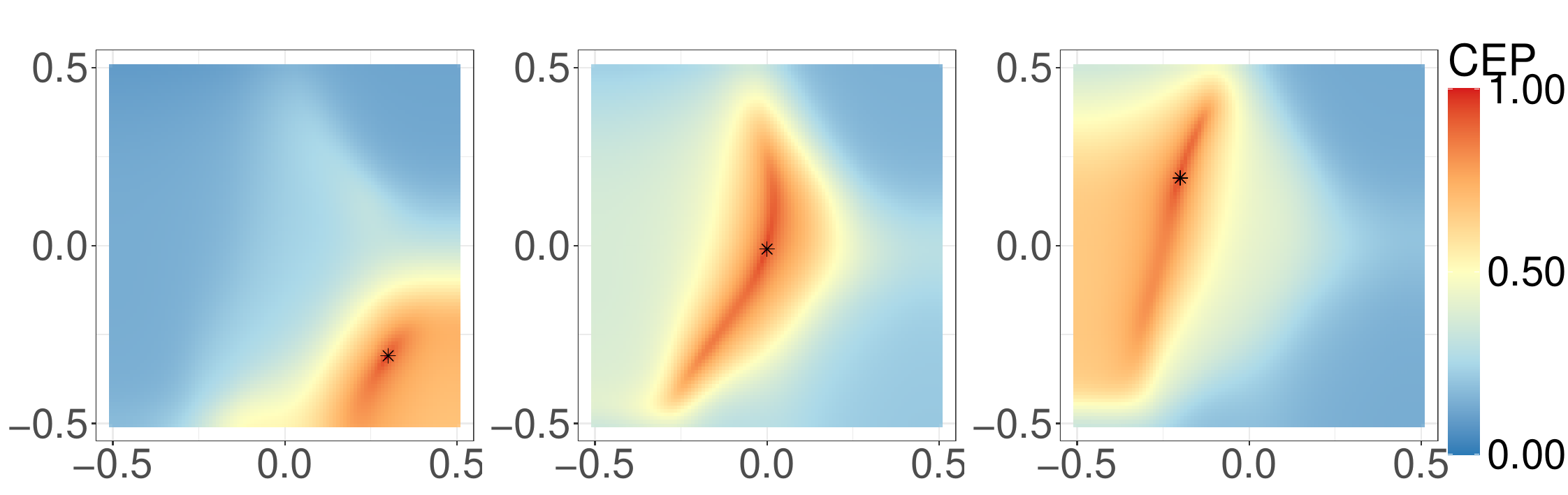}
\\
\includegraphics[width=0.74\linewidth]{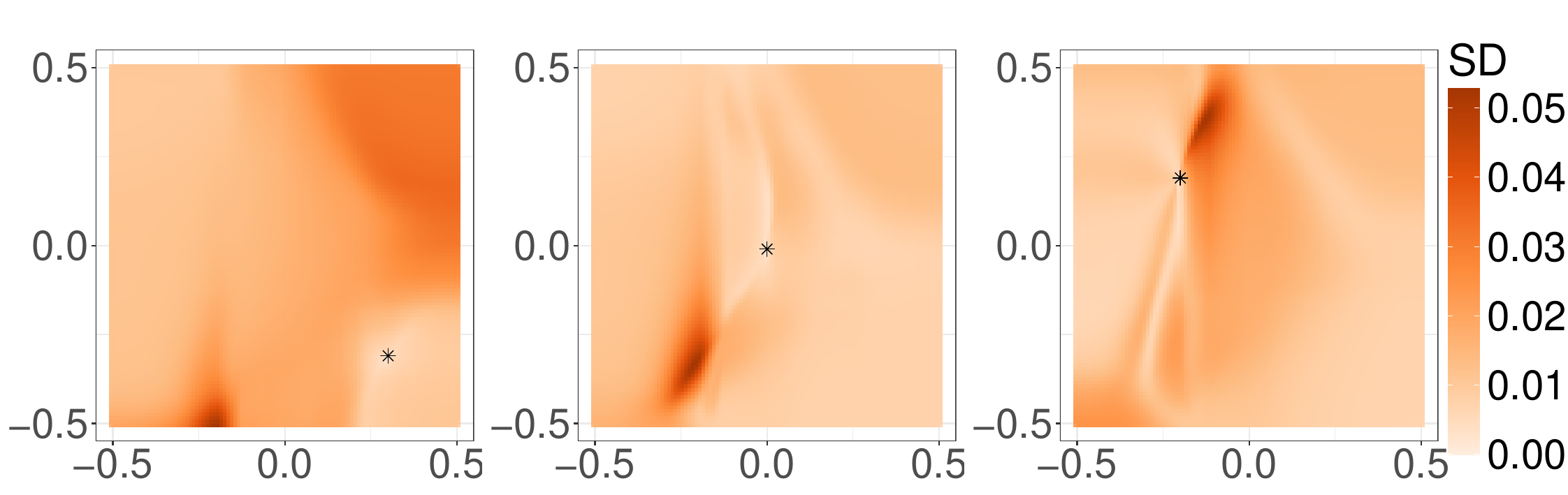}
\end{tabular}
\end{center}
\caption{Theoretical (top row) and estimated (middle row) pairwise CEPs in (\ref{eq:cep}) relative to site 1 (left column), site 2 (central column), and site 3 (right column) denoted by the asterisk. Third row: standard deviation (SD) of estimated pairwise CEPs. The data are generated with Architecture~3 with the risk functional $r_{\text{max}}(\cdot)$, and the model is fitted using nonstationary model~1 and the GSM method.}
\label{pic:sim_pairCEPs}
\end{figure}

\textcolor{black}{Figure~\ref{pic:sim_pairCEPs} provides a detailed look at the estimated extremal dependence structure, relative to three reference sites shown in Figure~\ref{pic:sim_space}; for each reference site, we evaluate the pairwise CEPs with respect to all other sites, where the data are generated with risk the functional $r_{\text{max}}(\cdot)$ and are fitted using nonstationary model~1 (with Architecture~1), and GSM inference method. Estimated pairwise CEPs generally match the oracle pattern very well. 
Furthermore, extreme events at reference sites exhibit strong yet nonstationary extremal dependence with nearby locations. For instance, extremes observed at reference sites 1 and 2 show pronounced spatial dependence, with the ``valley'' and ``edge'' regions varying dynamically across the direction and distance.
Uncertainty in the warped space is assessed via the standard deviation of fitted pairwise CEP estimates using a nonparametric bootstrap (with warping re-estimation and site resampling). These results reveal markedly lower variability in CEP estimates for pairs of sites exhibiting similar extremal behavior (indicative of strong dependence). In contrast, higher uncertainty emerges for pairs where one site lies within gradient regions of the warped space, i.e., areas characterized by abrupt transitions in extremal dependence structure.}


\textcolor{black}{Compared to Architecture 3, Architecture 1 includes an additional MT unit, with its warped-domain estimates shown in Figure \ref{pic:sim_space1}. We see pronounced differences in the estimated warped spaces between architectures with and without an MT unit (e.g., models 1 vs. 3, or 2 vs. 4). Table \ref{tab:sim_compare1} quantifies and agrees with this: nonstationary models incorporating an MT unit consistently achieve superior fit. Conversely, when fitting data generated without an MT unit, models that include an MT unit perform on par with those using the true (no MT unit) architecture. These results therefore support the inclusion of an MT unit in practice.}

\textcolor{black}{When the true extremal dependence structure is highly complex, as seen when data are generated under Architectures 2 and 4, visual estimates of the complex warped space favor the WLS inference method, as it better recovers the intricate dependence patterns (see Figures~\ref{pic:sim_space2} and \ref{pic:sim_space4} in the Supplementary Material). However, biased estimation of extremal dependence parameters, particularly the smoothness parameter $\kappa$, using WLS (see Tables~\ref{tab:sim_table2}, \ref{tab:sim_table4}) yields lower censored likelihood values compared to those obtained using GSM (see Tables~\ref{tab:sim_compare2}, \ref{tab:sim_compare4}). In these scenarios, oversimplified architectures do not suffice to estimate the complicated warped space, while architectures with the SR-RBF(2) unit generally perform better.} 

\textcolor{black}{These results provide practical guidance for choosing the warping architecture. In applications, we recommend starting with a parsimonious composition of AW and SR-RBF(1) units (Architectures 1 or 3 in Table~\ref{tab:sim_archi}), and only enriching it (e.g., by adding SR-RBF(2) units, Architectures 2 or 4) if diagnostics or out-of-sample scores indicate remaining lack of fit. In our experiments, including a MT unit tends to yield the best overall performance, and we therefore suggest considering it by default. The supplementary results (Section~\ref{Appendix:Simulation_124}) for the more complex Architectures 2 and 4 reinforce this complexity-dependent pattern: when the true deformation is richer, models that include an SR-RBF(2) unit become clearly more competitive when considering out-of-sample censored likelihood.}

\section{Data application} \label{RealDataApplication}

\subsection{UK precipitation extremes} \label{RealDataApplication:UKpr}

\begin{figure}[t!]
\begin{center}
\begin{tabular}{c}
\includegraphics[width=0.45\linewidth]{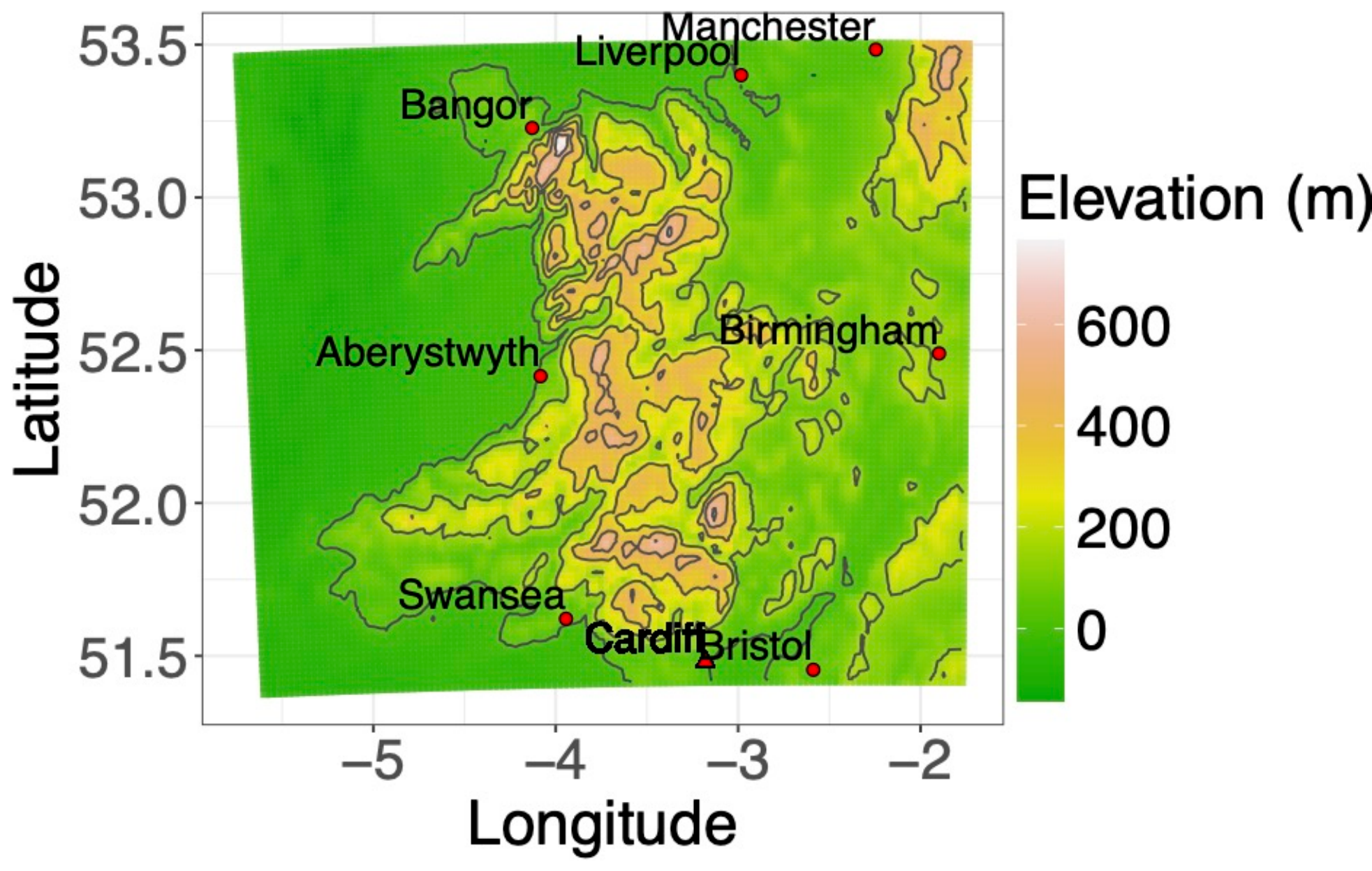}
\includegraphics[width=0.45\linewidth]{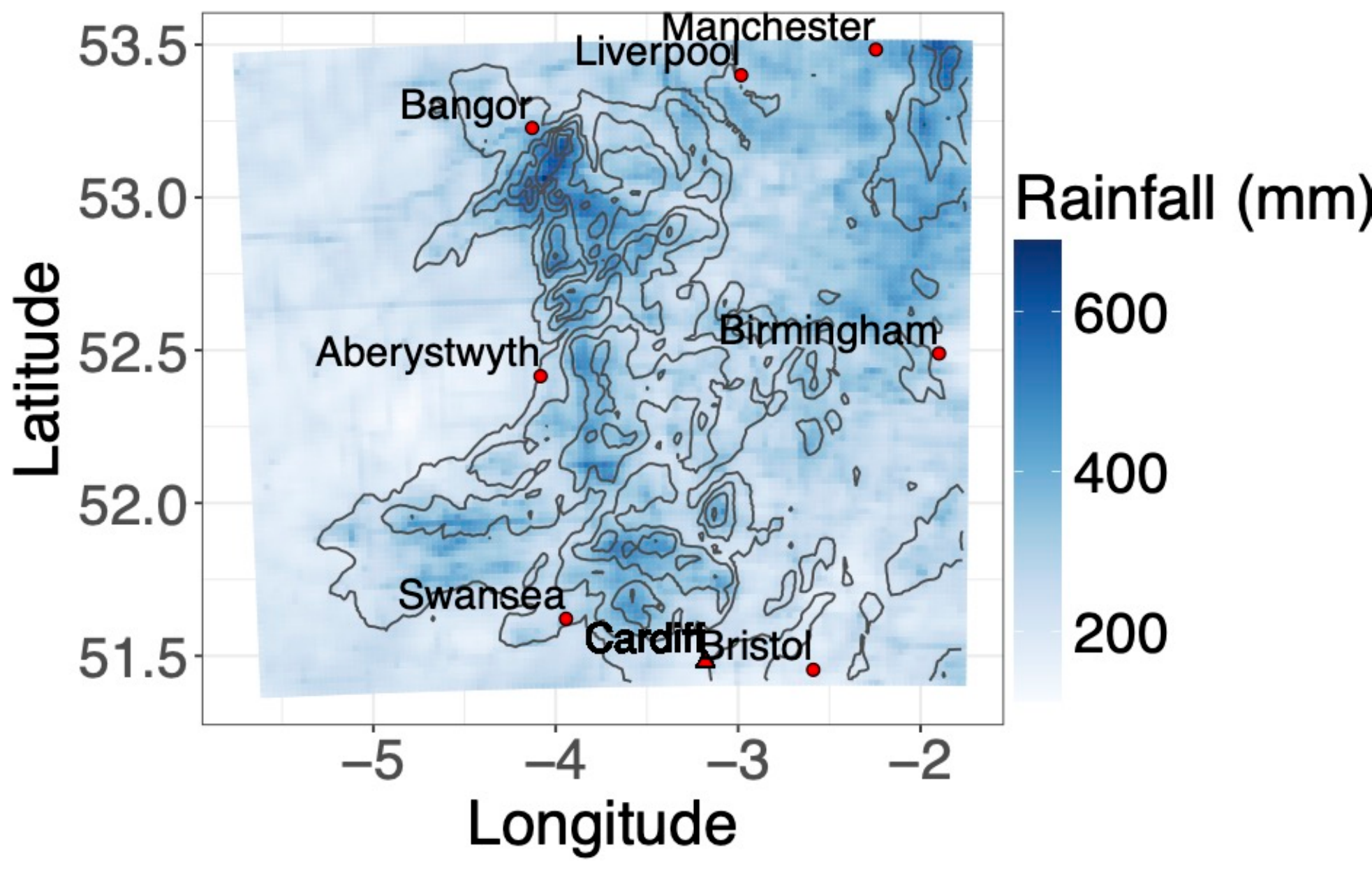} \\
\includegraphics[width=0.45\linewidth]{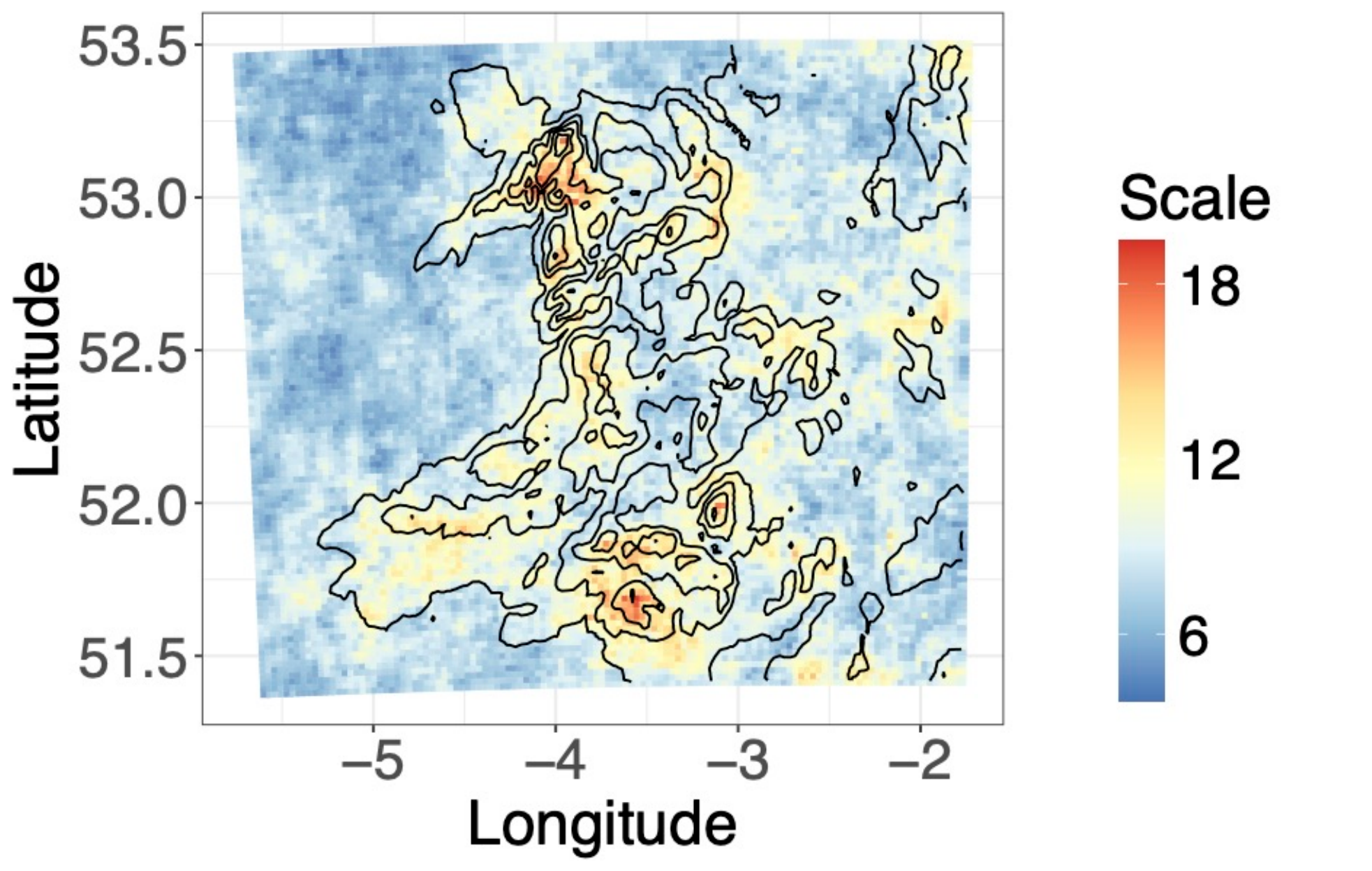}
\includegraphics[width=0.45\linewidth]{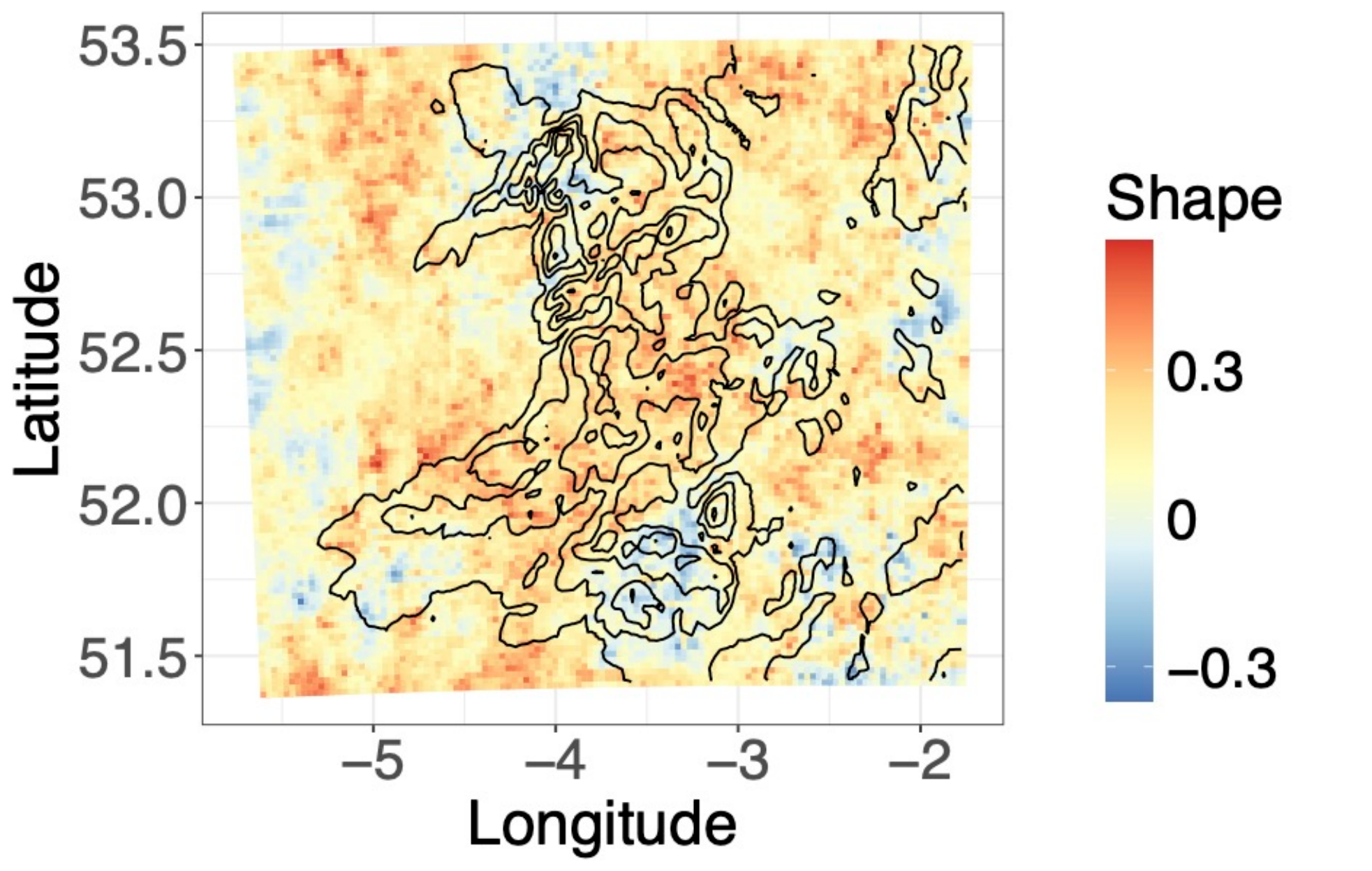}
\end{tabular}
\end{center}
\caption{Top row: topography of the study domain (left panel), and the cumulative rainfall for 2000 (mm; right panel). Bottom row: estimated scale parameter $\tau(\bm{s}_i)$ and shape parameter $\xi(\bm{s}_i)$ of (marginal) generalized Pareto distribution for $i \in 1,\ldots,D$. Elevation contour lines are plotted as solid curves in each panel for reference.}
\label{pic:UKpr_intro}
\end{figure}

We demonstrate the efficacy of our proposed model via an analysis of precipitation data from the 2018 UK climate projections \citep[UKCP18;][]{Lowe2019UKCP1S}, which contains values produced at hourly intervals on $2.2 \times 2.2 \mathrm{~km}^2$ grid boxes between the years 1980 and 2000; these data were also studied by \cite{richards2021spatial}, in the context of nonstationary extremal dependence modeling, and \cite{richards2022tail} and \cite{richards2023joint}, for inferring extremes of the risk functional $r_{\text{sum}}(\cdot)$. The data can be downloaded freely from the CEDA data catalogue; see \cite{datacite}. We focus on extreme rainfall over the west of Great Britain. As distance from the sea, proximity of mountains, and elevation are known drivers of extreme rainfall \citep{faulkner1998mapping}, we choose our study domain so that it includes complex topography which should induce nonstationarity in the spatial rainfall distribution: the coast of Wales and the Cambrian Mountains, and the West Midlands region to the west of the mountains; see Figure~\ref{pic:UKpr_intro}. 
We further visualize the cumulative rainfall over the year 2000 in Figure~\ref{pic:UKpr_intro}, which shows that the aforementioned factors influence the intensity of the precipitation distribution. The air carrying large amounts of water vapor from the Atlantic Ocean from west to east is lifted by the Cambrian Mountains, forming denser rainfall on the west side of the mountains. Therefore, rainfall anomalies are also concentrated in this area.

To reduce the number of zeros in the data, observations are aggregated to daily accumulated rainfall. Moreover, we take only the summer months, from June to August, to avoid seasonal effects; this leads to $N = 1800$ temporal replicates. 
\textcolor{black}{We randomly sample $D_{\text{train}} = 2000$ sites, from the total of $D=12600$ available sites, as the training set, and $D_{\text{test}}=100$ sites for testing.} 
Data margins are standardized to be standard Pareto using i) the empirical distribution function below a marginal threshold (95\% empirical quantile) and ii) a fitted generalized Pareto distribution (GPD) to excesses above the threshold (see, e.g., \citealp{coles2001introduction,davison2015statistics,deCarvalho2026handbook}). Specifically, the marginal model for rainfall exceedances $Y(\bm{s}_i)$ over a high threshold $u(\bm{s}_i)$ at site $\bm{s}_i \in \mathcal{S}$ is $\{Y(\bm{s}_i) - u(\bm{s}_i)\} \mid Y(\bm{s}_i) > u(\bm{s}_i) \overset{\mathrm{ind}}{\sim} \operatorname{GPD}\{\tau(\bm{s}_i), \xi(\bm{s}_i)\}$ for all $i = 1,\ldots,D$, which has the distribution function
\begin{equation}\label{eq:GPD}
G\{y; \tau(\bm{s}_i), \xi(\bm{s}_i)\}= \begin{cases}1-\left\{1+{\xi(\bm{s}_i) y/\tau(\bm{s}_i)}\right\}_{+}^{-1 / \xi(\bm{s}_i)}, & \xi(\bm{s}_i) \neq 0, \\ 1-\exp \left\{-{y/\tau(\bm{s}_i)}\right\}, & \xi(\bm{s}_i)=0,\end{cases}
\end{equation}
where the scale parameter $\tau(\bm{s}_i) > 0$ and shape parameter $\xi(\bm{s}_i)\in \mathbb{R}$ are spatially-varying, and $\{\cdot\}_+ = \max\{\cdot, 0\}$. Parameter estimates are presented in Figure~\ref{pic:UKpr_intro}. Larger scale and shape estimates to the west of the Cambrian Mountains suggest more extreme rainfall there, aligning with our explanatory analysis. Only five of 12600 locations violate the null hypothesis that the data at these sites do not follow a GPD. In the Supplementary Material, Figure~\ref{pic:UKpr_diag_map} shows model diagnostic plots for these sites, and Figure~\ref{pic:UKpr_diag} presents probability-probability (P-P) plots of the data for the total domain and at six representative sites. 
We then model the transformed data $\X(\bm{s}_i) = 1/[1-F_{\bm{s}_i}\{Y(\bm{s}_i)\}]$, for $i=1,\ldots,D$, using deep compositional Brown--Resnick $r$-Pareto processes, as described in Section~\ref{Methodology}.


\subsection{Extremal dependence modeling} \label{RealDataApplication:Modeling}
For defining extreme events, we consider the three risk functionals $r_{\text{site}}\left(\cdot\right)$, $r_{\text{max}}\left(\cdot\right)$, and $r_{\text{sum}}\left(\cdot\right)$ in (\ref{eq:app_risks}). For $r_{\text{site}}(\cdot)$, we choose the reference site $\bm{s}_0$ to be Cardiff, the capital city of Wales (see~Figure~\ref{pic:UKpr_intro}). 
We also consider the max-functional $r_{\text{max}}(\cdot)$ computed at all $D$ remaining sites, $\mathcal{D}_0$, and the modified sum-functional $r_{\text{sum}; \beta}(\cdot)$ with $\beta = D^{-1}\sum^D_{i=1}\hat{\xi}(\bm{s}_i) = 0.174$ taken to be the average of independent local estimates of the shape parameter from the fitted GPD marginal model.
We therefore consider the following risk functionals:
\begin{equation} \label{eq:app_risks}
r_{\text{site}}\left(\X\right) = X(\bm{s}_0), \quad 
r_{\text{max}}\left(\X\right)= \max_{\bm{s}\in\mathcal{D}_0} X(\bm{s}), \quad \text{and } 
r_{\text{sum}; \beta}\left(\X\right)=\left\{\sum_{\bm{s}\in\mathcal{D}_0} X\left(\bm{s}\right)^{\beta}\right\}^{1 / \beta}.
\end{equation}
\textcolor{black}{Following \cite{de2018high}, we guide the choice of functional threshold $u$ via a threshold-stability diagnostic by fitting a univariate GPD to excesses of the empirical risk series $\{r(\bm{x}_t)\}_{t=1}^N$ over a grid of candidate quantiles (see Figure~\ref{pic:stability} in the Supplementary Material). We therefore set $u$ to the empirical $0.9$-quantile of $\{r(\bm{x}_t)\}_{t=1}^N$ (yielding $N_u=180$ functional exceedances) and take $u'$ to be the empirical $0.95$-quantile.}
We conduct inference with i) the WLS method, where we pre-estimate the empirical CEPs using the training set, and ii) the GSM method.
Similar to the simulation study, we follow the guidance in Section~\ref{SimulationStudy:Results} and compare Architectures 0–4 for the warping $\bm{f}$ as shown in Table~\ref{tab:sim_archi} (which includes the stationary model for comparison). 

\paragraph{Model adequacy.}
\textcolor{black}{Our $r$-Pareto model assumes asymptotic dependence and peaks-over-threshold stability, which are idealized tail properties and may not hold exactly at observed levels \citep{huser2025modeling}. We therefore interpret this case study primarily at the chosen thresholds, without advocating extrapolation far into the tail. Empirical diagnostics suggest weakening tail dependence for most pairs. However, a deliberate extremal dependence misspecification experiment (see Section~\ref{Appendix:Simulation_misspecification} in the Supplementary Material) indicates that the learned warped geometry remains accurate even if the extremal dependence class is poorly specified; see, also, discussion by \cite{richards2021spatial}.}

\begin{table}[t!]
    \centering
    \caption{Model fit metrics including censored log-likelihoods (CL; larger is preferred), square errors (SE; smaller is preferred), and (weighted) gradient scores (GS; smaller is preferred) estimated on test data for fitted models with various architectures for $\bm{f}(\cdot)$, using both the gradient score matching (GSM) and weighted least squares (WLS) inference methods, and three different risk functionals. The best models, according to each metric, are highlighted in bold for each risk functional and inference method.}
    \resizebox{\columnwidth}{!}{%
    \begin{tabular}{l|l|c|c|c|c|c|c|c|c|c}
    \hline \hline
    \multirow{3}{*}{Architecture} & Inference & \multicolumn{9}{c}{Risk functional} \\
    \cline{3-11}
    & method & \multicolumn{3}{c|}{$r_{\text{site}}(\cdot)$} & \multicolumn{3}{c|}{$r_{\text{max}}(\cdot)$} & \multicolumn{3}{c}{$r_{\text{sum}}(\cdot)$} \\
    \cline{3-11}
    & & CL & SE & GS & CL & SE & GS & CL & SE & GS \\
    \hline
Stationary & \multirow{5}{*}{WLS} & $\mathbf{-37975}$ & $78.69$ & $-1.070$ & $-6375$ & $87.14$ & $\mathbf{-8.644}$ & $\mathbf{-27626}$ & $75.05$ & $-1.001$ \\
Nonstationary 1 &  & $-38294$ & $66.61$ & $-1.043$ & $-6384$ & $72.22$ & $-7.996$ & $-27854$ & $63.28$ & $-0.973$ \\
Nonstationary 2 &  & $-38324$ & $\mathbf{65.20}$ & $-1.049$ & $\mathbf{-6374}$ & $\mathbf{70.29}$ & $-8.433$ & $-27842$ & $\mathbf{61.97}$ & $-0.992$ \\
Nonstationary 3 &  & $-38272$ & $70.13$ & $-1.080$ & $-6405$ & $75.80$ & $-7.978$ & $-27930$ & $65.86$ & $-0.989$ \\
Nonstationary 4 &  & $-38264$ & $68.97$ & $\mathbf{-1.097}$ & $-6417$ & $74.34$ & $-7.725$ & $-27898$ & $63.66$ & $\mathbf{-1.006}$ \\
\hline
Stationary & \multirow{5}{*}{GSM} & $\mathbf{-38837}$ & $99.71$ & $-1.000$ & $-6577$ & $106.48$ & $1.000$ & $\mathbf{-28477}$ & $97.11$ & $\mathbf{-1.000}$ \\
Nonstationary 1 &  & $-38915$ & $99.97$ & $-1.026$ & $-6566$ & $98.25$ & $-1.052$ & $-28526$ & $94.48$ & $-0.967$ \\
Nonstationary 2 &  & $-38970$ & $99.60$ & $-1.005$ & $-6567$ & $98.86$ & $-1.480$ & $-28546$ & $95.86$ & $-0.965$ \\
Nonstationary 3 &  & $-38903$ & $\mathbf{97.49}$ & $\mathbf{-1.037}$ & $\mathbf{-6556}$ & $\mathbf{97.06}$ & $\mathbf{-1.971}$ & $-28531$ & $92.22$ & $-0.970$ \\
Nonstationary 4 &  & $-38947$ & $97.77$ & $-1.025$ & $-6563$ & $98.64$ & $-1.770$ & $-28542$ & $\mathbf{91.90}$ & $-0.970$ \\
    \hline
    \end{tabular} }
    \label{tab:app_compare}
\end{table}

\textcolor{black}{Table~\ref{tab:app_compare} provides model fit metrics over the test data for all fitted stationary and nonstationary models, in terms of censored likelihoods, as well as (weighted) gradient scores defined in Equation~\eqref{eq:GSM_loss} and square errors between empirical and fitted CEPs, using both inference methods and all risk functionals. Overall, the results give a mixed picture, and the improvement of the nonstationary models over their stationary counterparts does not appear substantial across all criteria. In particular, the nonstationary models are most clearly favored by the SE criterion, especially for the max-functional, whereas the CL criterion often favors the stationary model for the site- and modified sum-functionals, and the GS criterion gives more mixed conclusions. This pattern is nevertheless informative, as the censored likelihood mainly evaluates model performance in cases where all sites experience extreme events simultaneously. For partially observed extremes, the GS criterion provides mixed evidence across inference methods and risk functionals, whereas the SE criterion more consistently favors the nonstationary fits, with the clearest gains again occurring for the max-functional. Among the nonstationary models, the preferred architecture also appears to depend on the inference method rather than showing a uniform ranking: under WLS, models 1 and 2, which include the MT unit, are often more competitive, whereas under GSM, models 3 and 4 are often similarly competitive and sometimes perform better.}

\begin{figure}[t!]
\begin{center}
\begin{tabular}{c}
\includegraphics[width=0.9\linewidth]{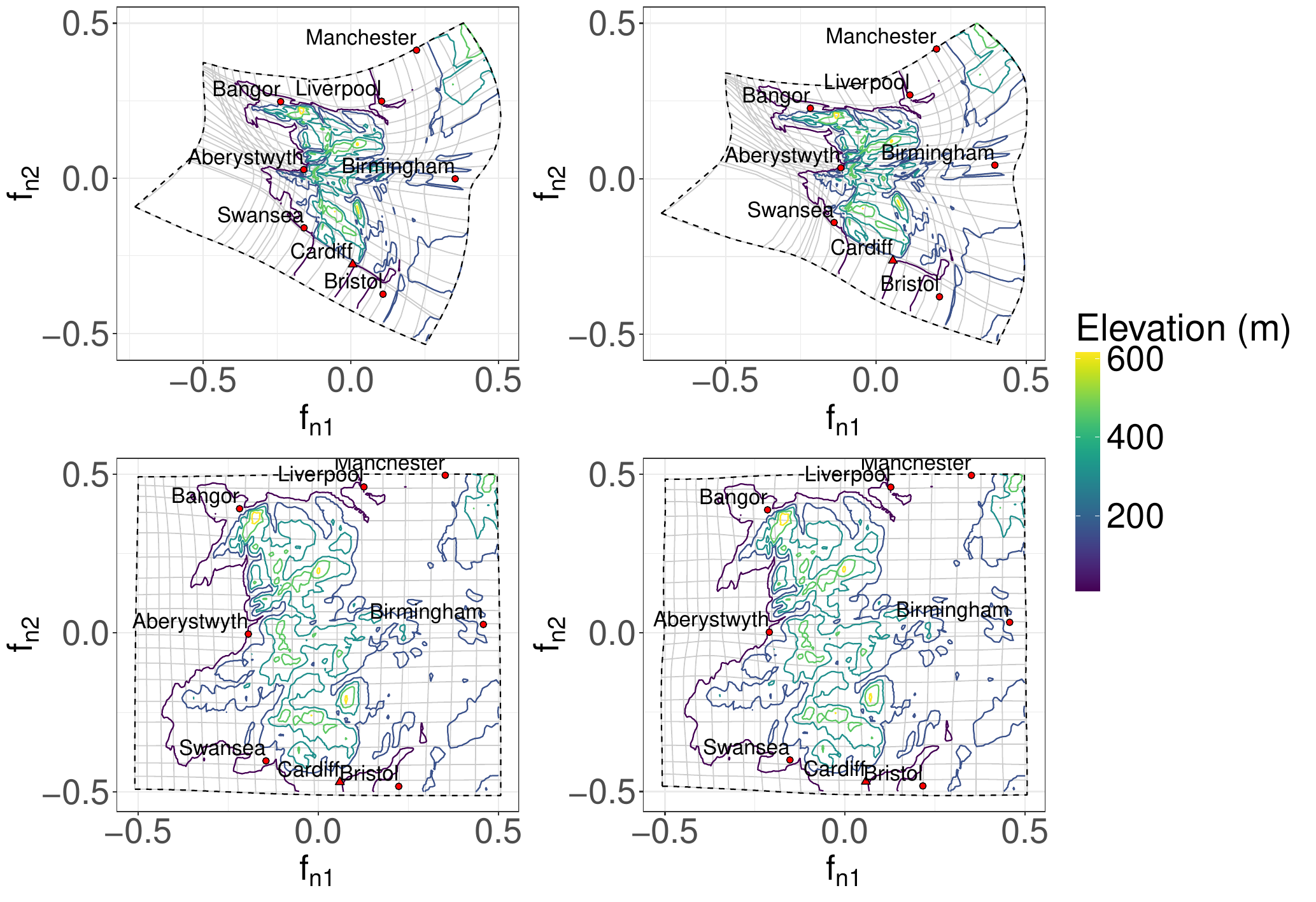}
\end{tabular}
\end{center}
\caption{Estimated warped space using the $r_{\text{max}}(\cdot)$ risk functional. The top row corresponds to WLS, with nonstationary model 1 (left) and model 2 (right), while the bottom row corresponds to GSM, with nonstationary model 3 (left) and model 4 (right). Warped longitude and latitude lines are shown, elevation contour lines are displayed according to their magnitude, and some reference cities are plotted as small dots.}
\label{pic:UKpr_space}
\end{figure}

\textcolor{black}{Figure~\ref{pic:UKpr_space} demonstrates the estimated warped spaces for the $r_{\text{max}}(\cdot)$ risk functional, with nonstationary models 1 and 2 under WLS (top row) and nonstationary models 3 and 4 under GSM (bottom row). The warped spaces estimated under the two inference methods are relatively different: WLS tends to produce more prominent deformation than GSM, aligning with the simulation results for more complex architectures (recall Section~\ref{SimulationStudy}). All warped spaces estimated with extremes defined via $r_{\text{max}}(\cdot)$ show a prominent contraction of the ocean and coastal area to the west of the mountains, together with an expansion of the eastern plain area, as seen from the warped longitude and latitude lines, indicating relatively stronger extremal dependence to the west of the mountains. Furthermore, the mountainous region is horizontally expanded, suggesting that rainfall extremes on the two sides of the mountains are less dependent because of the mountain barrier.}

\begin{table}[t!]
    \centering
    \caption{Estimates of the extremal dependence parameters, $\hat{\bm{\psi}}' = (\hat{\varphi}, \hat{\kappa})$, using both the gradient score matching (GSM) and weighted least squares (WLS) inference methods, and three different risk functionals. Standard deviations obtained using a nonparametric bootstrap are reported in brackets as subscripts of the corresponding parameter estimate.}
    \resizebox{\columnwidth}{!}{%
    \begin{tabular}{l|l|c|c|c}
    \hline \hline
    \multirow{2}{*}{Architecture} & Inference & \multicolumn{3}{c}{Risk functional} \\
    \cline{3-5}
    & method & $r_{\text{site}}(\cdot)$ & $r_{\text{max}}(\cdot)$ & $r_{\text{sum}}(\cdot)$ \\
    \hline 
Stationary & \multirow{5}{*}{WLS} & $0.074_{(0.037)}$, $0.583_{(0.081)}$ & $0.086_{(0.034)}$, $0.637_{(0.057)}$ & $0.090_{(0.036)}$, $0.670_{(0.055)}$ \\ 
Nonstationary 1 &  & $0.063_{(0.011)}$, $0.610_{(0.105)}$ & $0.064_{(0.010)}$, $0.669_{(0.065)}$ & $0.071_{(0.009)}$, $0.702_{(0.064)}$ \\ 
Nonstationary 2 &  & $0.062_{(0.011)}$, $0.616_{(0.096)}$ & $0.063_{(0.009)}$, $0.665_{(0.060)}$ & $0.071_{(0.009)}$, $0.707_{(0.059)}$ \\ 
Nonstationary 3 &  & $0.071_{(0.011)}$, $0.607_{(0.101)}$ & $0.070_{(0.010)}$, $0.665_{(0.064)}$ & $0.072_{(0.010)}$, $0.702_{(0.064)}$ \\ 
Nonstationary 4 &  & $0.071_{(0.011)}$, $0.611_{(0.099)}$ & $0.071_{(0.010)}$, $0.668_{(0.064)}$ & $0.075_{(0.010)}$, $0.703_{(0.065)}$ \\ 
\hline 
Stationary & \multirow{5}{*}{GSM} & $0.159_{(0.004)}$, $0.973_{(0.025)}$ & $0.128_{(0.002)}$, $0.832_{(0.007)}$ & $0.164_{(0.003)}$, $1.014_{(0.025)}$ \\ 
Nonstationary 1 &  & $0.165_{(0.004)}$, $0.960_{(0.021)}$ & $0.142_{(0.001)}$, $0.887_{(0.007)}$ & $0.168_{(0.004)}$, $0.996_{(0.024)}$ \\ 
Nonstationary 2 &  & $0.168_{(0.004)}$, $0.967_{(0.021)}$ & $0.144_{(0.001)}$, $0.894_{(0.007)}$ & $0.169_{(0.004)}$, $1.000_{(0.024)}$ \\ 
Nonstationary 3 &  & $0.163_{(0.004)}$, $0.978_{(0.021)}$ & $0.137_{(0.001)}$, $0.864_{(0.006)}$ & $0.167_{(0.004)}$, $1.011_{(0.022)}$ \\ 
Nonstationary 4 &  & $0.163_{(0.004)}$, $0.980_{(0.021)}$ & $0.137_{(0.001)}$, $0.866_{(0.006)}$ & $0.168_{(0.003)}$, $1.014_{(0.022)}$ \\ 
    \hline
    \end{tabular} }
    \label{tab:app_table}
\end{table}

\textcolor{black}{Dependence parameter estimates, $\hat{\bm{\psi}} = (\hat{\varphi}, \hat{\kappa})'$, are also reported in Table~\ref{tab:app_table}, together with their standard deviations obtained using a nonparametric bootstrap, in which the margins and warpings are re-estimated to account for the uncertainty brought by marginal estimation and by the deformation. 
Generally speaking, the effect of the deformation on the uncertainty of the dependence parameter estimates does not appear dramatic overall, although some differences are visible across models and inference methods. In particular, the bootstrap standard deviations of the smoothness parameter are of broadly similar magnitude for the stationary and nonstationary fits, whereas for the range parameter, the reduction in uncertainty under the nonstationary models is more noticeable for WLS than for GSM. 
Note that the range parameter may alter with the scale of the warped space, and rescaling of the warped space may be considered to stabilize the estimation of range parameters. Yet, the extremal dependence structure only depends on the relative positions among warped locations, and the range estimates are also adapted to different space scales, which would not affect the evaluated extremal dependence structure. 
Similar to the results of the simulation study, the standard deviations of the dependence parameter estimates for models fitted using the GSM inference method are generally smaller than those for models fitted using the WLS inference method.}

\begin{figure}[t!]
\begin{center}
\begin{tabular}{c}
\includegraphics[width=0.99\linewidth]{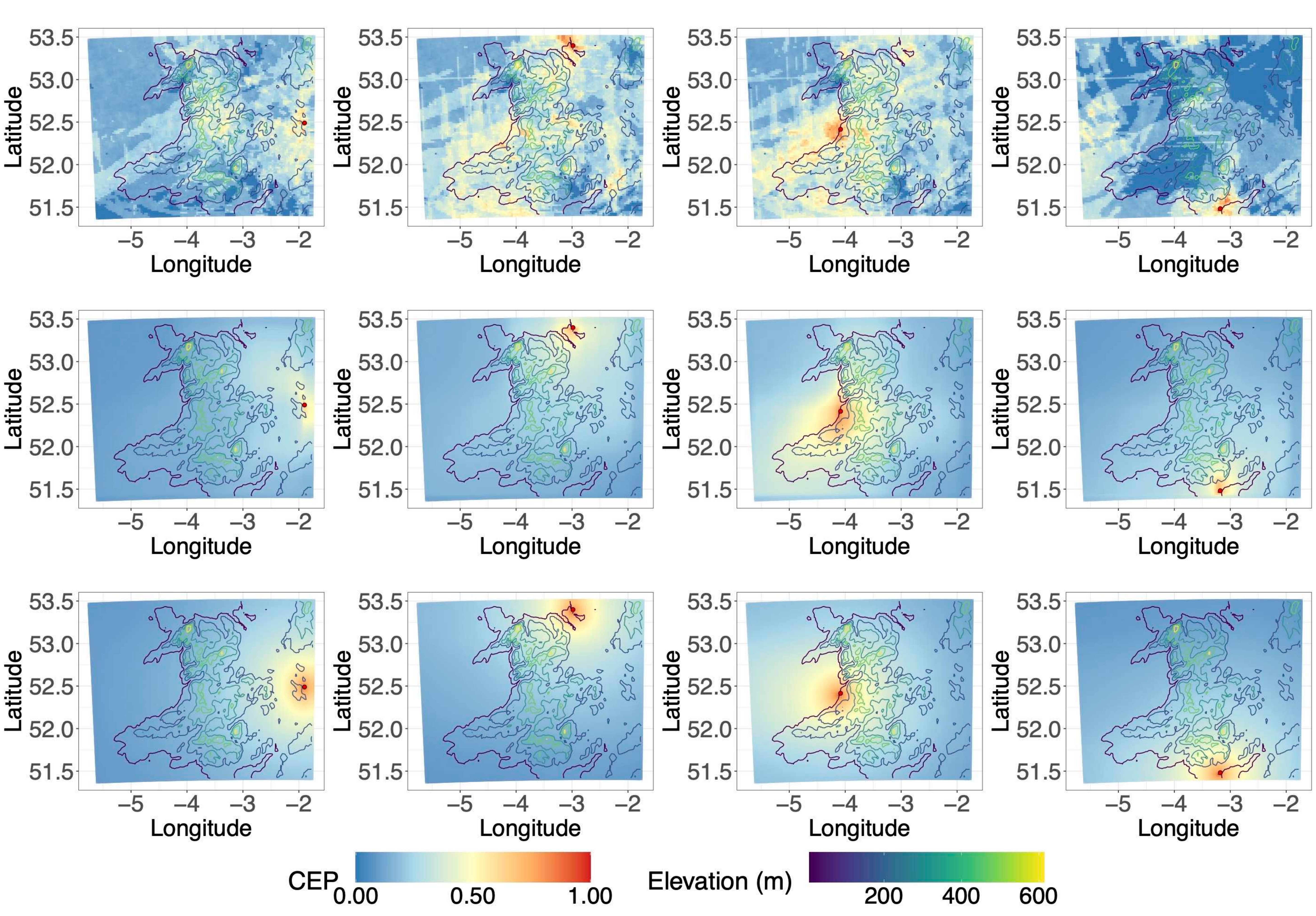}
\end{tabular}
\end{center}
\caption{Empirical (first row) and fitted pairwise CEPs relative to Birmingham (1st column), Liverpool (2nd column), Aberystwyth (3rd column), and Cardiff (4th column), where the fitted CEPs are obtained using nonstationary model 2 with WLS (middle row) and nonstationary model 3 with GSM (bottom row). Extremes are defined through $r_{\text{max}}(\cdot)$. Elevation contour lines are displayed according to their magnitude for reference.}
\label{pic:UKpr_pairCEPs}
\end{figure}


\textcolor{black}{In Figure~\ref{pic:UKpr_pairCEPs}, we further present the detailed extremal dependence structure estimated using nonstationary model~2 with WLS and nonstationary model~3 with GSM, under the risk functional $r_{\text{max}}(\cdot)$, with respect to the following reference cities: Birmingham, Liverpool, Aberystwyth, and Cardiff. The results are represented by maps of pairwise CEPs relative to the aforementioned reference cities, which show that the fitted nonstationary models capture the anisotropy and nonstationarity in the extremal dependence structure quite well. We can further observe the prominent effect of landform on extremal dependence; due to the mountain barrier, the rainfall anomaly in Aberystwyth is highly correlated in a large area to the west, while the rainfall anomaly in Birmingham is only highly correlated with a small adjacent terrestrial area in the east. 
Uncertainty assessment is reported, in terms of standard deviations computed using a nonparametric bootstrap (with warping re-estimation and site resampling), as discussed in Section~\ref{Inference:Implementation}, and these are shown for pairwise CEPs in Figure~\ref{pic:UKpr_unc}. We observe smaller standard deviations near the reference city, and larger uncertainty further away. The uncertainty pattern aligns with that of the estimated extremal dependence structure, which is influenced by the warped space. 
In contrast to the minimal impact observed on uncertainty for estimated dependence parameters, Figure~\ref{pic:UKpr_unc} indicates that the deformation plays a substantially larger role in inflating the uncertainty of pairwise CEPs: when taking the impact of the deformation into account, the standard deviations in estimated CEPs are markedly greater than those obtained when assuming the warping fixed.
Furthermore, the standard deviation is usually smaller when using the GSM inference method compared to the WLS method, which is consistent with our earlier observation that GSM tends to provide more ``stable'' warping recovery and dependence-parameter estimation. At the same time, such increased stability comes at the cost of a lower coverage probability of the nonparametric bootstrap.}

\begin{figure}[t!]
\begin{center}
\begin{tabular}{c}
\includegraphics[width=0.32\linewidth]{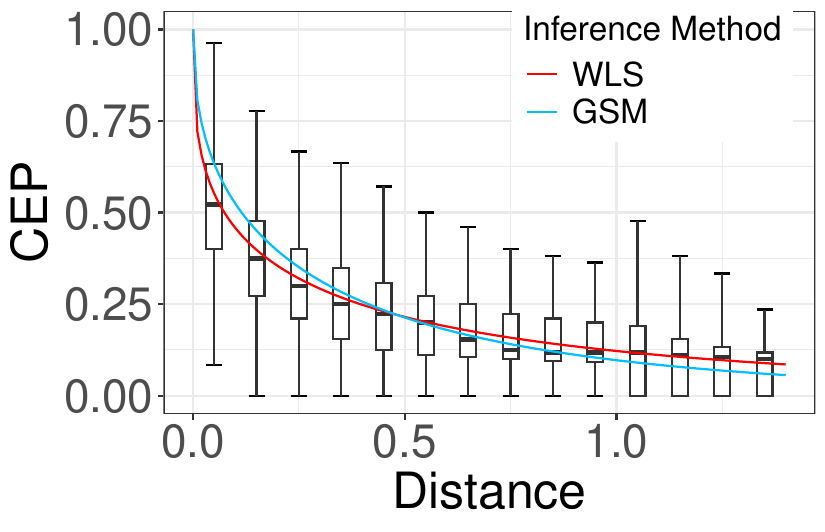}
\includegraphics[width=0.32\linewidth]{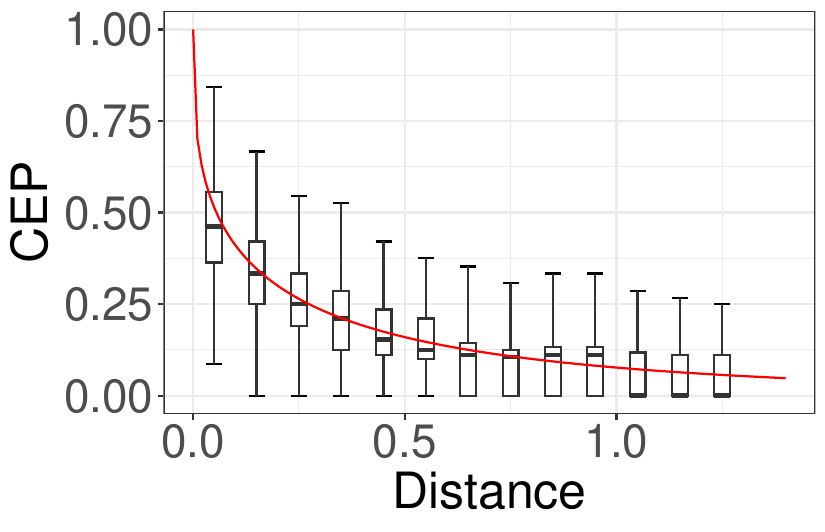}
\includegraphics[width=0.32\linewidth]{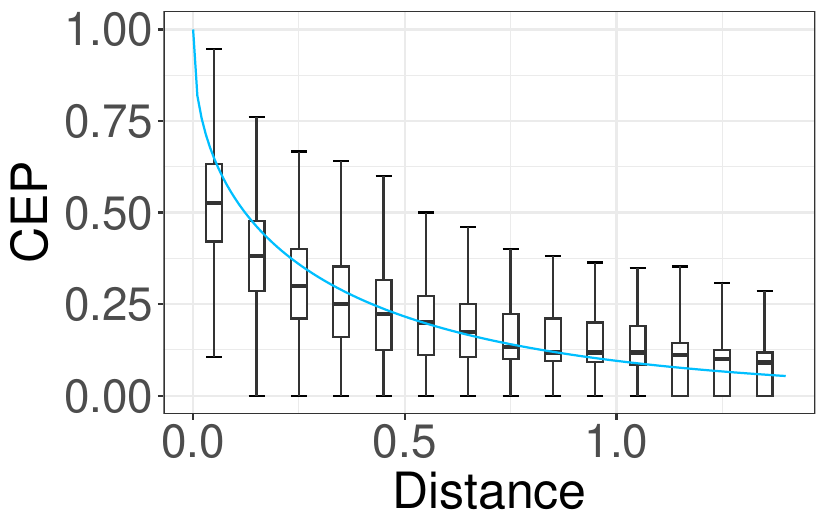}
\end{tabular}
\end{center}
\caption{Estimates of pairwise conditional exceedance probabilities $\hat{\pi}_{ij}(u, u')$ against relative distances in the (rescaled) original space (left panel) and in the estimated warped spaces, obtained using nonstationary model 2 for WLS (middle panel) and nonstationary model 3 for GSM (right panel). The reference lines show the corresponding theoretical pairwise CEPs as functions of distance under the fitted dependence parameters.}
\label{pic:UKpr_cloud}
\end{figure}

\textcolor{black}{To visualize the performance of our fitted models over the warped space, Figure~\ref{pic:UKpr_cloud} presents the empirical CEP estimates, $\hat{\pi}_{ij}(u,u')$, against distance computed both in the (rescaled) original and warped spaces, obtained using nonstationary model~2 for WLS and nonstationary model~3 for GSM under $r_{\text{max}}(\cdot)$. Estimates are grouped by distance and presented using box plots. The more concentrated the points around the reference line of the relative estimated CEP function, i.e., the reference line passing through the main box and error bars shrinking by distance, the more stationary the behavior of spatial extremes over that space. We see that the empirical CEPs-distance plots over the estimated warped space using both inference methods appear to be slightly more concentrated around the reference line, especially for the distant pairs. We further quantify the concentration by reporting the mean absolute difference between the empirical estimates and the theoretical one that comes from the reference curves: results show that the mean absolute differences for the nonstationary models are approximately $0.085$ and $0.100$ when using the WLS and GSM methods, respectively, compared with $0.097$ and $0.103$ for the stationary model under WLS and GSM. We hence conclude that the deformation has created processes that appear to be more nearly stationary with regard to their respective CEP estimates in the new warped domain, with the improvement being more pronounced for WLS than for GSM.}

\section{Concluding remarks}\label{Concluding}
In this work, we proposed a deformation approach to model nonstationarity in the extremal spatial dependence of high dimensional data based on the deep compositional spatial model (DCSM). We developed a flexible and computationally efficient method to estimate the nonstationary extremal dependence by warping the original spatial domain to a latent space where stationarity and isotropy in the extremal dependence structure can be reasonably assumed. By leveraging the DCSM framework, we efficiently estimated the warping function while ensuring bijectivity and hence overcame the space-folding issue prevailing in previous research. We proposed practical strategies to assess the uncertainty of the estimated extremal dependence structure via a bootstrapping method that accounts for multiple sources of uncertainty (training site selection, marginal fitting, dependence parameter and warping estimation). 
We leveraged Brown--Resnick $r$-Pareto processes for extremal dependence modeling, in order to characterize spatial peak-over-threshold events, and we performed inference by loss minimization, using either a weighted least squares error loss or a gradient score matching loss. The proposed framework is also compatible with other spatial extremal processes, e.g., max-stable or asymptotically independent processes (e.g., scale mixtures), subject to simple modifications of the loss function. In both our simulation study and real data application, we showed that our proposed methodology succeeds in providing good fits and capturing complex nonstationary extremal dependence structures, and is relatively robust to asymptotic dependence misspecification.

Future work includes an extension to incorporate temporal nonstationarity and jointly characterize nonstationary spatio-temporal extremal dependence patterns.
In the Gaussian setting, \cite{vu2023constructing} constructed nonstationary spatio-temporal covariance models via compositional warpings, which only allows rescaling time using axial warping units, yet non-separable space-time modeling is usually required for applications. 
It would also be interesting to extend our methodology to incorporate informative covariates, e.g., elevation, into the estimated warping for enhanced interpretability. 
Furthermore, current deep learning modeling approaches in spatial statistics require training a network over a specific study domain; extending it to other regions is not easy and retraining a different model for new data is time-consuming. \textcolor{black}{To speed up computations, it would be interesting to develop transfer-learning approaches \citep{pan2009survey} that can exploit pretrained models for inference in a new spatial domain, as well as general amortized inference methods \citep{richards2024neural,zammit2025neural} for nonstationary processes, and to extend the sparse $r$-Pareto process approach of \citet{bolin2025intrinsic} to the nonstationary context using deep compositional spatial models. 
Finally, our deep deformation approach is not tied to Euclidean domains and could be extended to general manifolds \citep[e.g.,][]{lindgren2011explicit} by constructing the warping as a composition of smooth bijections on the manifold and defining dependence using geodesic distances on the warped manifold.}

To promote reproducibility, our code and data are made fully available at the following GitHub repository: \url{https://github.com/shaox0a/DCSMExt/}.

\section*{Data Availability Statement}
The data that support the findings of this study are openly available in GitHub at \url{https://github.com/shaox0a/DCSMExt/}.

\setstretch{0.5}
\bibliographystyle{CUP}
\bibliography{Bibliography}

\spacingset{1.5}
\clearpage
\appendix

This supplementary material contains further methodological details and additional results on (i) the large-sample properties of the weighted least squares (WLS) estimator, (ii) the simulation study, including experiments to assess computational scaling and robustness to asymptotic dependence misspecification, as well as additional estimated warpings when the true deformation is generated from Architectures~1, 2, and~4, (iii) the real data application, including goodness-of-fit diagnostics for the marginal tail model and a threshold-stability check supporting the choice of the functional exceedance threshold, and (iv) additional uncertainty-quantification results for the fitted dependence parameters, including results for a nested bootstrap with site resampling and empirical coverage assessment.

{\color{black}
\section{Asymptotic behavior of the weighted least squares estimator $\hat{\bm{\psi}}_{\text{WLS}}$}  \label{Appendix:WLS}
This section provides details of the asymptotic behavior of the weighted least squares estimator $\hat{\bm{\psi}}_{\text{WLS}}$. We consider the spatial setting with $D$ fixed locations and $N \to \infty$ temporal observations. Collect the empirical conditional exceedance probabilities (CEPs) into a vector $\hat{\bm{\pi}} = \big(\hat\pi_{12}, \hat\pi_{13}, \ldots, \hat\pi_{D-1,D}\big)' \in \mathbb{R}^{\frac{D(D-1)}{2}},$
and the model-implied counterpart $\bm{\pi}(\bm{\psi}, \bm{f})
= \left(\pi_{12}(\bm{\psi}, \bm{f}), \pi_{13}(\bm{\psi}, \bm{f}), \ldots, \pi_{D-1,D}(\bm{\psi}, \bm{f})\right)',$
which can be abbreviated as $\bm{\pi}(\bm{\psi})$ when $\bm{f}$ is fixed. Let $\check{\bm{W}}$ be a diagonal weight matrix with diagonal entries $\check w_{ij}\ge 0$ matching the ordering in $\hat{\bm{\pi}}$ (in our case, $\checkw_{ij} = 1/(2-\hat{\pi}_{ij})$). The loss function defined in \eqref{eq:WLS_loss} in the main paper, under a fixed warping, can be written as:
\begin{equation}\label{eq:WLS_loss2}
\ell_{\text{WLS}}(\bm{\psi}) = \left(\bm{\pi}(\bm{\psi}) - \hat{\bm{\pi}}\right)' \check{\bm{W}}\left(\bm{\pi}(\bm{\psi}) - \hat{\bm{\pi}}\right),
\end{equation}
and $\hat{\bm{\psi}}_{\text{WLS}} = \arg\min_{\bm{\psi}}\ell_{\text{WLS}}(\bm{\psi})$ is then a classical minimum distance (CMD) estimator as in \cite{newey1994large}.

\subsection{Identification and consistency} \label{Appendix:WLS_define}

Suppose there exist deterministic limits $\bm{\pi}_0$ and $\check{\bm W}_0$ such that
$\hat{\bm{\pi}}\xrightarrow{p}\bm{\pi}_0$ and $\check{\bm W}\xrightarrow{p}\check{\bm W}_0$.
The population CMD loss is
$$
\ell_{\text{WLS},0}(\bm{\psi}) =
\left(\bm{\pi}(\bm{\psi})-\bm{\pi}_0\right)'
\check{\bm W}_0
\left(\bm{\pi}(\bm{\psi})-\bm{\pi}_0\right).
$$
A sufficient identification condition is the \emph{uniqueness} of the minimizer: there exists a unique $\bm{\psi}_{\text{WLS},0} = \arg\min_{\bm{\psi}\in\bm{\Psi}} \ell_{\text{WLS},0}(\bm{\psi})$ such that
$\ell_{\text{WLS},0}(\bm{\psi}) > \ell_{\text{WLS},0}(\bm{\psi}_{\text{WLS},0})$ for all $\bm{\psi}\ne\bm{\psi}_{\text{WLS},0}$. Under standard extremum-estimator regularity conditions:
(i) the parameter space $\bm{\Psi}$ is compact,
(ii) $\bm{\pi}(\cdot)$ is continuous on $\bm{\Psi}$ (this is naturally satisfied by Brown--Resnick-type extremal processes with power variogram),
and (iii) $\ell_{\text{WLS}}(\bm{\psi})$ converges uniformly to $\ell_{\text{WLS},0}(\bm{\psi})$, i.e., $\sup_{\bm{\psi} \in \bm{\Psi}}\left|\ell_{\text{WLS}}(\bm{\psi})-\ell_{\text{WLS},0}(\bm{\psi})\right| \xrightarrow{p} 0$, one obtains consistency:
${\hat{\bm{\psi}}_{{\text{WLS}}} \xrightarrow{p} \bm{\psi}_{\text{WLS},0}}$.


\subsection{Verification of the regularity conditions for the fixed-warping case}

We verify here the regularity conditions used in Section~\ref{Appendix:WLS_define} for the weighted least squares estimator under a fixed warping architecture $\bm{f}$. Throughout, the number of observed sites $D$ is fixed and the number of temporal replicates satisfies $N\to\infty$.

\paragraph{Consistency of the empirical CEPs and of the weights.}
For each pair $(i,j)$, write the empirical CEP in the generic ratio form $\hat\pi_{ij}=\hat a_{ij}/\hat b_{i,N}$, where
$$
\hat a_{ij}=\frac{1}{N}\sum_{t=1}^N \mathds{1}\{A_{ij,t}\},
\qquad
\hat b_{i,N}=\frac{1}{N}\sum_{t=1}^N \mathds{1}\{B_{i,t}\},
$$
for suitable exceedance events $A_{ij,t}\subseteq B_{i,t}$, where $\mathds{1}\{\cdot\}$ is the indicator function. Assume that, for each fixed pair $(i,j)$, the temporal sequence $\bigl(\mathds{1}\{A_{ij,t}\},\mathds{1}\{B_{i,t}\}\bigr)_{t\ge1}$ is independent and identically distributed, and that $b_{i,0}:=\Pr(B_{i,1})>0$. Then, by the law of large numbers,
$$
\hat a_{ij}\xrightarrow{p} a_{ij,0}:=\Pr(A_{ij,1}),
\qquad
\hat b_{i,N}\xrightarrow{p} b_{i,0},
\qquad \text{as } N\to\infty.
$$
Hence, by the continuous mapping theorem,
$$
\hat\pi_{ij}
=
\frac{\hat a_{ij}}{\hat b_{i,N}}
\xrightarrow{p}
\frac{a_{ij,0}}{b_{i,0}}
=:\pi_{ij,0}.
$$
Since the number of pairs is finite, it follows that $\hat{\bm{\pi}}\xrightarrow{p}\bm{\pi}_0$. Moreover, with diagonal weights defined by $\check w_{ij}=g(\hat\pi_{ij})$ and $g(x)=1/(2-x)$, continuity of $g$ on $[0,1]$ implies
$$
\check w_{ij}\xrightarrow{p}\check w_{ij,0}:=g(\pi_{ij,0}),
\qquad
\check{\bm{W}}\xrightarrow{p}\check{\bm{W}}_0.
$$
If the threshold entering the CEP definition is itself estimated empirically, say $\hat u$, the same conclusion holds provided $\hat u\xrightarrow{p}u_0$ and the relevant distributions are continuous at $u_0$.

\paragraph{Condition (i): localization to a compact subset.}
The original dependence-parameter space
$$
\bm{\Psi}=\{(\varphi,\kappa):\varphi>0,\kappa\in(0,2]\}
$$
is not compact. We therefore assume that there exists a compact subset
$$
\bm{\Psi}'\subset\bm{\Psi}
$$
such that the population criterion $\ell_{\text{WLS},0}(\bm{\psi})$ admits a unique minimizer
$\bm{\psi}_0\in\operatorname{int}(\bm{\Psi}')$ and
$$
\inf_{\bm{\psi}\in\bm{\Psi}\setminus\bm{\Psi}'}
\ell_{\text{WLS},0}(\bm{\psi})
>
\ell_{\text{WLS},0}(\bm{\psi}_0).
$$
Hence the estimation problem can be localized to $\bm{\Psi}'$, on which the remaining regularity conditions are verified.

\paragraph{Condition (ii): continuity of $\bm{\pi}(\bm{\psi})$.}
For fixed warping $\bm{f}$, the pairwise semivariogram between sites $\bm{s}_i$ and $\bm{s}_j$ is
$$
\gamma_{ij}(\bm{\psi},\bm{f})
=
\left(
\frac{\|\bm{f}(\bm{s}_i)-\bm{f}(\bm{s}_j)\|}{\varphi}
\right)^\kappa.
$$
Since $\bm{f}$ is fixed and $\varphi$ is bounded away from zero on $\bm{\Psi}'$, the map $\bm{\psi}\mapsto \gamma_{ij}(\bm{\psi},\bm{f})$ is continuous on $\bm{\Psi}'$. For the Brown--Resnick-type $r$-Pareto model used in the manuscript, $\pi_{ij}(\bm{\psi})$ is a continuous function of $\gamma_{ij}(\bm{\psi},\bm{f})$. Hence each component $\pi_{ij}(\bm{\psi})$ is continuous in $\bm{\psi}$, and so is
$$
\bm{\pi}(\bm{\psi})
=
\bigl(\pi_{12}(\bm{\psi}),\ldots,\pi_{D-1,D}(\bm{\psi})\bigr)'.
$$

\paragraph{Condition (iii): uniform convergence on the localized compact set.}
Define
$$
\ell_{\text{WLS}}(\bm{\psi})
=
\left(\bm{\pi}(\bm{\psi})-\hat{\bm{\pi}}\right)'
\check{\bm{W}}
\left(\bm{\pi}(\bm{\psi})-\hat{\bm{\pi}}\right),
$$
and
$$
\ell_{\text{WLS},0}(\bm{\psi})
=
\left(\bm{\pi}(\bm{\psi})-\bm{\pi}_0\right)'
\check{\bm{W}}_0
\left(\bm{\pi}(\bm{\psi})-\bm{\pi}_0\right).
$$
Since $\bm{\Psi}'$ is compact and $\bm{\pi}(\bm{\psi})$ is continuous on $\bm{\Psi}'$, there exists $M<\infty$ such that
$$
\sup_{\bm{\psi}\in\bm{\Psi}'} \|\bm{\pi}(\bm{\psi})\| \le M.
$$
Moreover, since the number of pairs is finite and $\hat{\bm{\pi}}\xrightarrow{p}\bm{\pi}_0$ together with $\check{\bm{W}}\xrightarrow{p}\check{\bm{W}}_0$, we may write
$$
\ell_{\text{WLS}}(\bm{\psi})-\ell_{\text{WLS},0}(\bm{\psi})
=
\sum_{1\le i<j\le D}
\left[
\check w_{ij}\left(\pi_{ij}(\bm{\psi})-\hat\pi_{ij}\right)^2
-
\check w_{ij,0}\left(\pi_{ij}(\bm{\psi})-\pi_{ij,0}\right)^2
\right].
$$
For each pair $(i,j)$, the difference between the two squared terms is uniformly controlled by
$$
\sup_{\bm{\psi}\in\bm{\Psi}'}
\left|
\left(\pi_{ij}(\bm{\psi})-\hat\pi_{ij}\right)^2
-
\left(\pi_{ij}(\bm{\psi})-\pi_{ij,0}\right)^2
\right|
\le
|\hat\pi_{ij}-\pi_{ij,0}|
\left(2M+|\hat\pi_{ij}|+|\pi_{ij,0}|\right),
$$
while the weight difference satisfies
$$
\sup_{\bm{\psi}\in\bm{\Psi}'}
\left|
(\check w_{ij}-\check w_{ij,0})
\left(\pi_{ij}(\bm{\psi})-\pi_{ij,0}\right)^2
\right|
\le
|\check w_{ij}-\check w_{ij,0}|(M+|\pi_{ij,0}|)^2.
$$
Hence each summand converges to zero uniformly on $\bm{\Psi}'$, and since the number of pairs is finite,
$$
\sup_{\bm{\psi}\in\bm{\Psi}'}
\left|
\ell_{\text{WLS}}(\bm{\psi})-\ell_{\text{WLS},0}(\bm{\psi})
\right|
\xrightarrow{p}0.
$$
Together with condition (i), this shows that the estimation problem can be localized to $\bm{\Psi}'$, so the consistency proof reduces to a standard extremum-estimator argument on a compact parameter set.

\paragraph{Conclusion.}
Under a fixed warping $\bm{f}$, compact restriction of the dependence-parameter space, continuity of the Brown--Resnick-type CEP map, and a law of large numbers for the bounded exceedance indicators used to form the empirical CEPs, conditions (i)--(iv) hold on $\bm{\Psi}'$. Therefore the consistency result in Section~\ref{Appendix:WLS_define} applies.

\subsection{Asymptotic normality derivation} 
Assume a central limit theorem for the pre-estimated CEP vector: 
$$
\sqrt{N}(\hat{\bm{\pi}}-\bm{\pi}_0) \xrightarrow{d} \mathcal{N}(\bm 0,\bm{\Omega}),
$$
and assume that $\bm{\pi}(\bm{\psi})$ is differentiable at $\bm{\psi}_0$ with Jacobian
$$
\bm{A} = \frac{\partial \bm{\pi}(\bm{\psi})}{\partial\bm{\psi}'}\bigg|_{\bm{\psi}=\bm{\psi}_0}\in\mathbb{R}^{\frac{D(D-1)}{2}\times p}.
$$
Then, under standard CMD regularity conditions,
\begin{equation*}
\sqrt{N}\left(\hat{\bm{\psi}}_{\text{WLS}} - \bm{\psi}_0\right) \xrightarrow{d} \mathcal{N}\left( \bm{0}, \bm{H}(\bm{\psi}_0)^{-1} \bm{G}(\bm{\psi}_0) \bm{H}(\bm{\psi}_0)^{-1} \right),
\end{equation*}
as $N\to\infty$, with
\begin{equation*}
\begin{aligned}
\bm{H}(\bm{\psi}) &= \bm{A}' \check{\bm{W}} \bm{A} = \sum_{1\leq i<j\leq D} \checkw_{ij} \frac{\partial \pi_{ij}(\bm{\psi}_0)}{\partial \bm{\psi}} \frac{\partial \pi_{ij}(\bm{\psi}_0)}{\partial \bm{\psi}'} \\
\bm{G}(\bm{\psi}) &= \bm{A}' \check{\bm{W}} \bm{\Omega} \check{\bm{W}} \bm{A} = \sum_{1\leq i<j\leq D}\sum_{1\leq k<l\leq D} \checkw_{ij}\checkw_{kl} \frac{\partial \pi_{ij}(\bm{\psi}_0)}{\partial \bm{\psi}} \Omega_{ij,kl} \frac{\partial \pi_{kl}(\bm{\psi}_0)}{\partial \bm{\psi}'}.
\end{aligned}
\end{equation*}
}

\clearpage

{\color{black}
\section{Supplementary results for the simulation study} \label{Appendix:Simulation}

\subsection{Computational scaling} \label{Appendix:Simulation_computation}

We here quantify the overall computational scaling with respect to $D$ in the simulation study. We examine $D \in \{200, 500, 1000, 2000, 5000, 10000\}$ using both the weighted least squares (WLS) and gradient score matching (GSM) inference procedures. Throughout this experiment, the data are generated under the nonstationary model with Architecture~3 using the risk functional $r_{\text{site}}$, and, for each simulated dataset, we fit both the corresponding nonstationary model (Architecture~3) and the stationary model (i.e., identity warping) in order to compare their computational cost (Tables~\ref{tab:timing_scaling_wls}--\ref{tab:timing_scaling_gsm}). The ``stationary'' timings reported below refer to fitting a stationary model to the same nonstationary datasets. We focus on $r_{\text{site}}$ here for brevity, as the corresponding timing results for the other risk functionals are qualitatively very similar and lead to the same overall conclusions.
For each inference method, we report both the overall wall-clock optimization time (in minutes) and the time required for a single loss evaluation (in seconds); ratios relative to the corresponding smallest time at $D=200$ are shown in parentheses. Note that the optimization-iteration settings differ between the stationary and nonstationary fits, so the reported ratios should be interpreted within each model class rather than compared directly across stationary and nonstationary settings. We use a fixed (pre-specified) number of optimization steps; in this experiment, we set the total number of iterations to 150 for the nonstationary models and 100 for the stationary models. Finally, in the nonstationary case, the loss-evaluation time includes the additional step of computing the warped locations, so it reflects both loss evaluation and warping-update overhead.

Overall, Tables~\ref{tab:timing_scaling_wls}--\ref{tab:timing_scaling_gsm} show that the computational time for the WLS approach scales more gently with $D$, whereas the computational time for GSM becomes increasingly dominated by the cost of loss evaluation (due to the associated dense linear-algebra operations and the possible memory limit) as $D$ grows. In the stationary setting, where no warping is estimated, the impact of loss-evaluation complexity is more apparent in the timing; in the nonstationary setting, the additional warping optimization and the computation of warped locations (included in the reported loss-evaluation time) dampen the apparent scaling of the total runtime, especially for moderate $D$, though GSM still exhibits markedly steeper growth than WLS at larger $D$. Importantly, in both our simulation study and data application we work with $D$ in the few-thousand range, and our empirical results indicate that the associated wall-clock costs remain manageable in practice, even under GSM with its nominal $\mathcal{O}(D^3)$ linear-algebra complexity. Finally, for substantially larger $D$, one could leverage scalable linear-algebra strategies to reduce the cost of the required linear solves, such as sparse modeling ideas discussed by \citet{bolin2025intrinsic}.

\begin{table}[t!]
    \centering
    \caption{Summary of overall optimization time and loss evaluation time for WLS under stationary and nonstationary models. Ratios to the time with $D = 200$ are reported in brackets.}
    \begin{tabular}{l|cc|cc}
    \hline \hline
    \multirow{2}{*}{$D$} &
    \multicolumn{2}{c|}{Stationary} &
    \multicolumn{2}{c}{Nonstationary} \\
    \cline{2-5}
    & Opt. time & Loss eval. time & Overall time & Loss eval. time \\
    & (mins) & (secs) & (mins) & (secs) \\
    \hline
    200   & 0.75 (1.00) & 0.18 (1.00) & 2.45 (1.00)  & 0.41 (1.00) \\
    500   & 0.76 (1.02) & 0.18 (1.01) & 2.49 (1.02)  & 0.42 (1.02) \\
    1000  & 0.80 (1.07) & 0.19 (1.03) & 2.62 (1.07)  & 0.43 (1.04) \\
    2000  & 0.91 (1.22) & 0.20 (1.12) & 2.94 (1.20)  & 0.45 (1.08) \\
    5000  & 2.09 (2.80) & 0.38 (2.12) & 5.82 (2.37)  & 0.64 (1.54) \\
    10000 & 6.31 (8.47) & 0.97 (5.34) & 16.65 (6.80) & 1.25 (3.01) \\
    \hline
    \end{tabular}
    \label{tab:timing_scaling_wls}
\end{table}

\begin{table}[t!]
    \centering
    \caption{Summary of overall optimization time and loss evaluation time for GSM under stationary and nonstationary models. Ratios to the time with $D = 200$ are reported in brackets.}
    \begin{tabular}{l|cc|cc}
    \hline \hline
    \multirow{2}{*}{$D$} &
    \multicolumn{2}{c|}{Stationary} &
    \multicolumn{2}{c}{Nonstationary} \\
    \cline{2-5}
    & Opt. time & Loss eval. time & Overall time & Loss eval. time \\
    & (mins) & (secs) & (mins) & (secs) \\
    \hline
    200   & 1.01 (1.00)    & 0.26 (1.00)   & 2.89 (1.00)    & 0.49 (1.00)   \\
    500   & 1.14 (1.13)    & 0.28 (1.07)   & 3.10 (1.07)    & 0.52 (1.05)   \\
    1000  & 1.60 (1.58)    & 0.38 (1.45)   & 3.89 (1.35)    & 0.62 (1.25)   \\
    2000  & 4.48 (4.44)    & 1.00 (3.85)   & 8.29 (2.87)    & 1.24 (2.51)   \\
    5000  & 46.04 (45.55)  & 10.78 (41.48) & 75.60 (26.16)  & 11.05 (22.33) \\
    10000 & 331.40 (327.82)& 79.46 (305.81)& 533.12 (184.46) & 79.95 (161.60) \\
    \hline
    \end{tabular}
    \label{tab:timing_scaling_gsm}
\end{table}

\subsection{Sensitivity to the ridge penalty $\alpha$}\label{Appendix:alpha_sensitivity}

\begin{figure}[t!]
\centering
\begin{tabular}{c}
\includegraphics[width=0.99\linewidth]{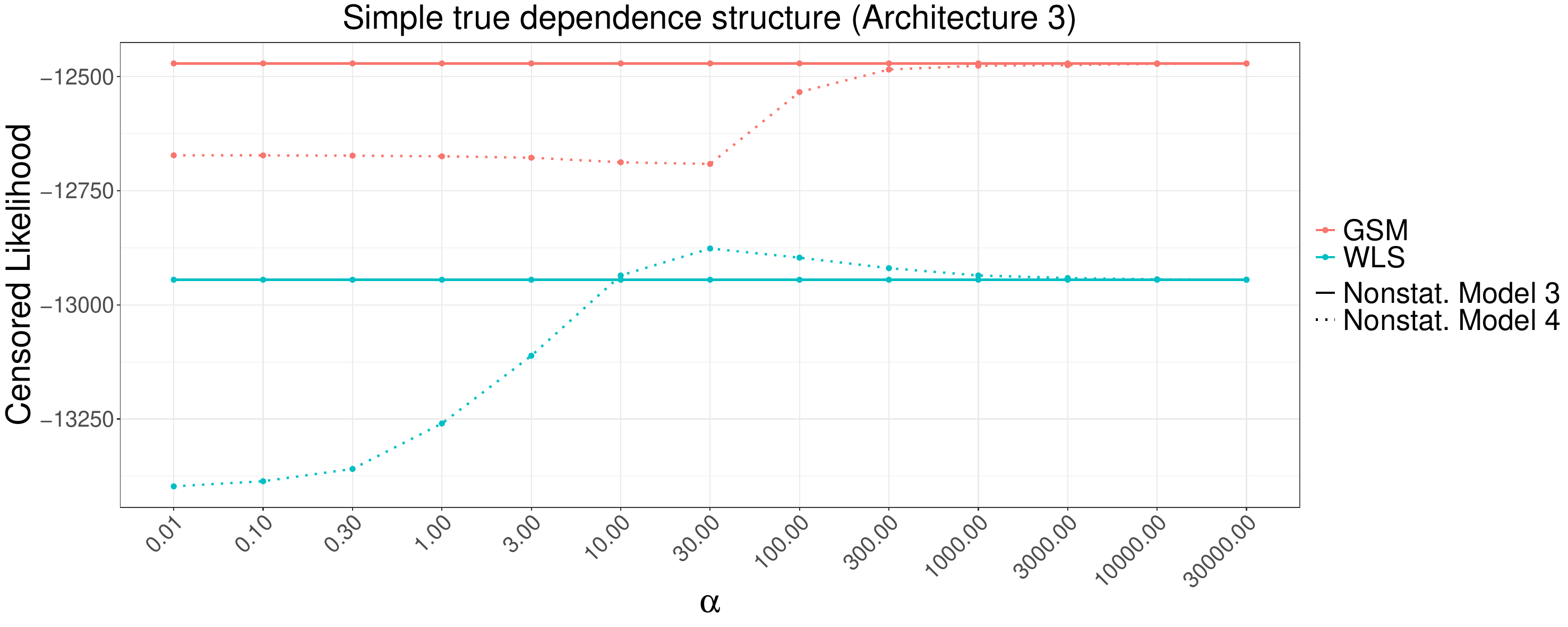}\\
\includegraphics[width=0.99\linewidth]{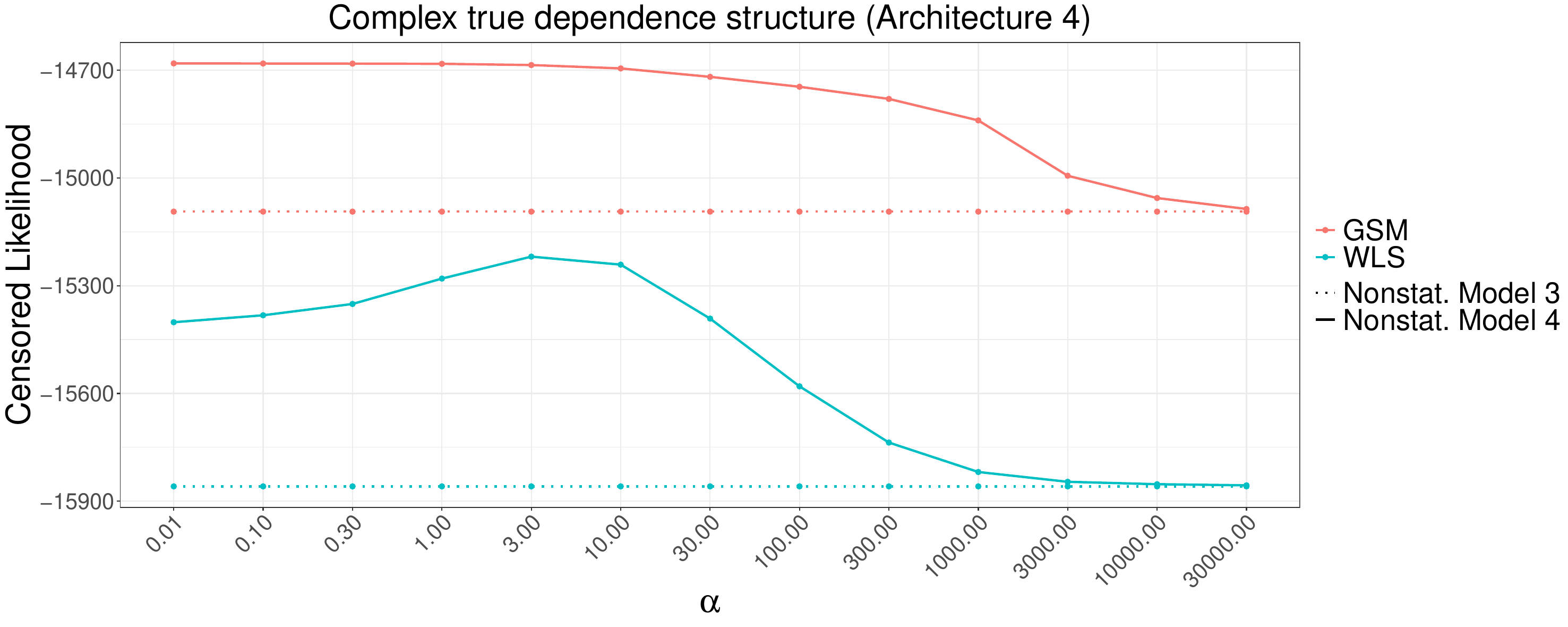}\\
\end{tabular}
\caption{Sensitivity of out-of-sample test censored log-likelihood to the ridge penalty $\alpha$ in \eqref{eq:regularizedLoss} with $D_{\text{train}} = 800$. Top: data generated under Architecture~3; bottom: data generated under Architecture~4. For each dataset and each $\alpha\in\mathcal{A}$, we fit Nonstationary Models~3 and~4 using both GSM and WLS, and evaluate the censored log-likelihood on the test set. In each panel, solid lines correspond to the correctly specified nonstationary model (matching the data-generating architecture), and dotted lines correspond to the misspecified model.}
\label{pic:alpha_sens_800}
\end{figure}

\begin{figure}[t!]
\centering
\begin{tabular}{c}
\includegraphics[width=0.99\linewidth]{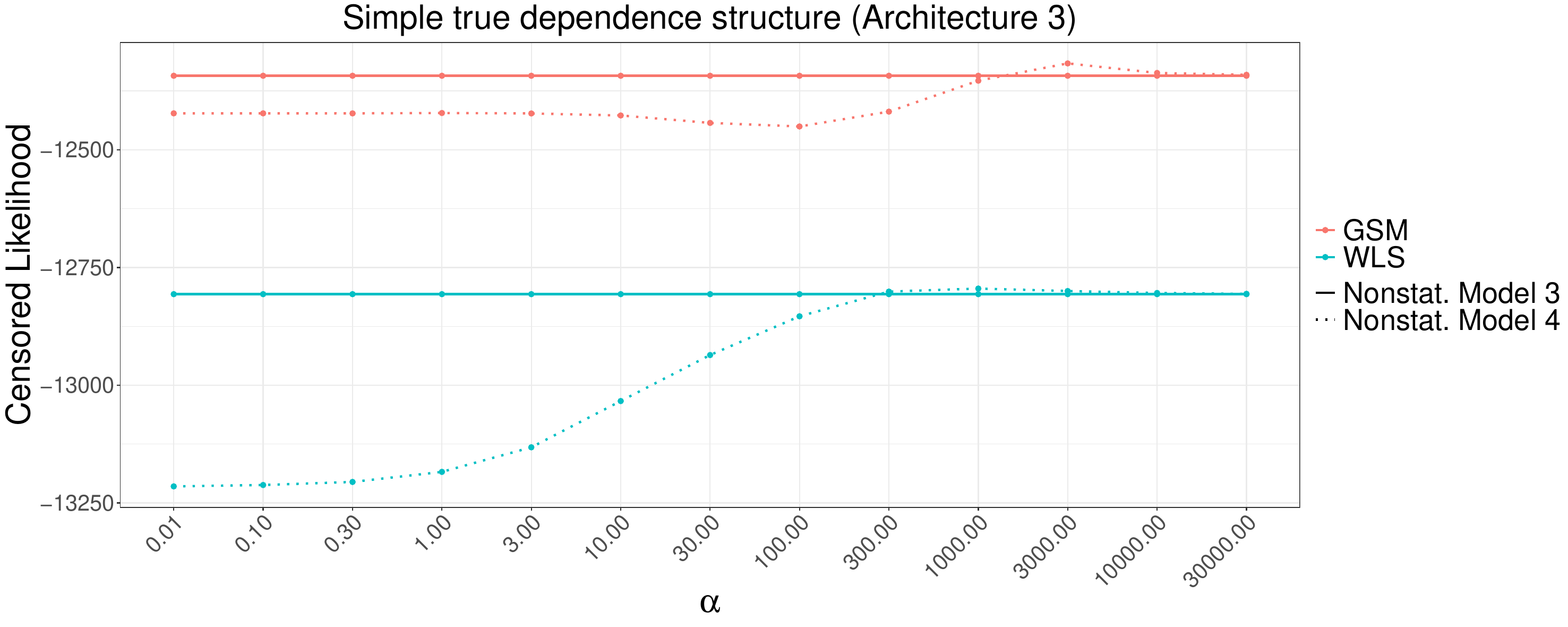}\\
\includegraphics[width=0.99\linewidth]{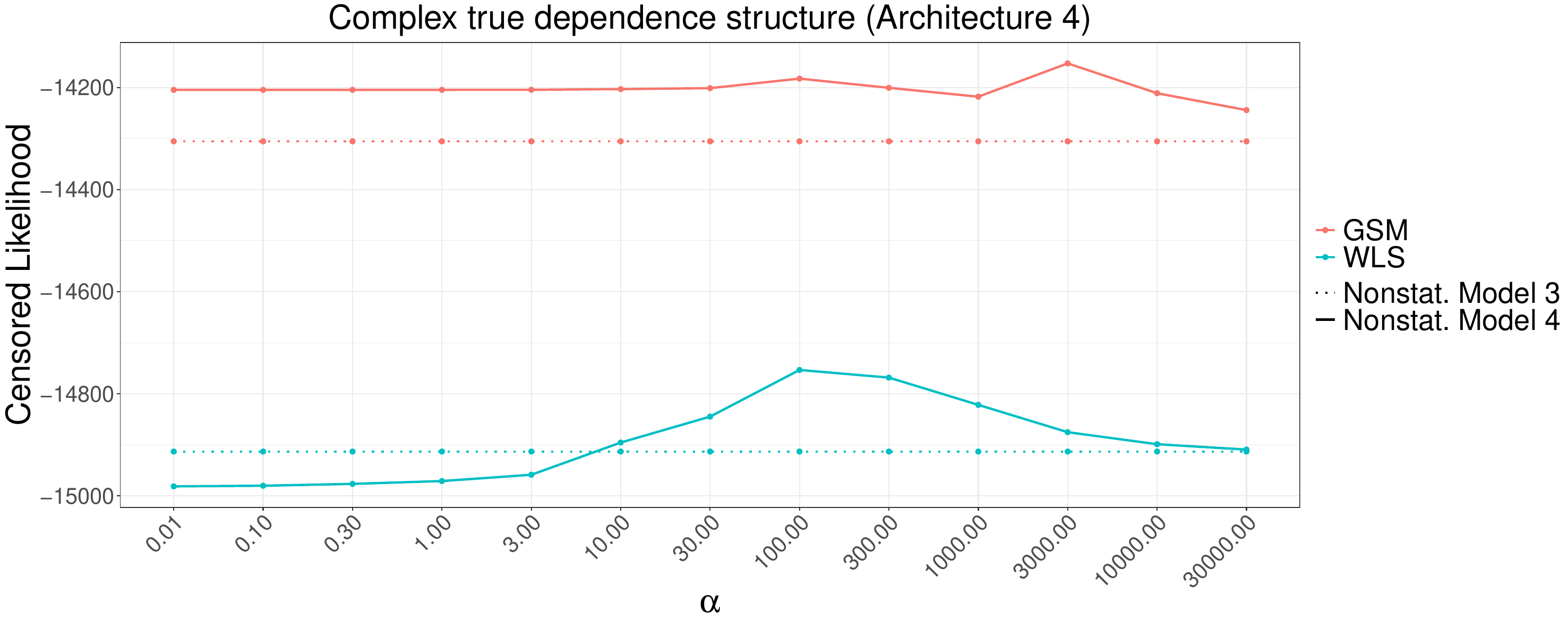}\\
\end{tabular}
\caption{Sensitivity of out-of-sample test censored log-likelihood to the ridge penalty $\alpha$ in \eqref{eq:regularizedLoss} with $D_{\text{train}} = 2000$. Top: data generated under Architecture~3; bottom: data generated under Architecture~4. For each dataset and each $\alpha\in\mathcal{A}$, we fit Nonstationary Models~3 and~4 using both GSM and WLS, and evaluate the censored log-likelihood on the test set. In each panel, solid lines correspond to the correctly specified nonstationary model (matching the data-generating architecture), and dotted lines correspond to the misspecified model.}
\label{pic:alpha_sens_2000}
\end{figure}

We investigate the sensitivity of out-of-sample performance to the choice of ridge penalty $\alpha$ in \eqref{eq:regularizedLoss} of the main paper, which is introduced to regularize the warping and reduce overfitting in the latent warped space. In practice, choosing $\alpha$ involves a practical trade-off: i) when the latent warped space is relatively simple (especially when $D$ is small), a larger $\alpha$ can help prevent overfitting and improve generalization; whereas ii) when the latent warped space is more complex, overly large $\alpha$ may overshrink the fine-scale deformation and lead to lack of fit. For a given model configuration and application, we refit the candidate nonstationary models over a small grid of $\alpha$ values and compare test censored log-likelihood on test locations with $D_{\text{test}}=100$ to select a suitable default $\alpha$ in each setting.

We consider the (logarithmic) grid of $\alpha\in\mathcal{A}$ values, defined as
$$\mathcal{A}=\{0.01,0.1,0.3,1,3,10,30,100,300,1000,3000,10000,30000\},$$ and generate data under Architecture~3 (AWU+SR-RBF(1)) and Architecture~4 (AWU+SR-RBF(1)+SR-RBF(2)). For each dataset and each $\alpha$, we fit Nonstationary Models~3 and~4 using both GSM and WLS, and evaluate the censored log-likelihood on test locations; results are shown in Figures~\ref{pic:alpha_sens_800}--\ref{pic:alpha_sens_2000} (top: Architecture~3; bottom: Architecture~4), for $D_{\text{train}}=800$ and $D_{\text{train}}=2000$, respectively. As the ridge penalty is applied only to SR-RBF$(\ell)$ units with $\ell\ge 2$, Nonstationary Model~3 yields flat curves across $\alpha$.

For GSM (red curves), when the data are generated under Architecture~3 (top panels), larger $\alpha$ improves the performance of Nonstationary Model~4 by curbing overfitting from the extra SR-RBF(2) flexibility. When the data are generated under Architecture~4 (bottom panels), smaller $\alpha$ allows the SR-RBF(2) component to remain sufficiently variable and improves fit. For WLS (blue curves), since it uses only empirical pairwise CEP information (and discards higher-order dependence information), increasing $\alpha$ from small values initially improves generalization for Nonstationary Model~4, while overly large $\alpha$ overshrinks the SR-RBF(2) component so that the performance approaches that of Nonstationary Model~3.

For $D_{\text{train}}=800$, $\alpha=3$ is a reasonable choice, as it provides a good compromise across the two data-generating scenarios and the two inference methods: it already curbs the overfitting seen at very small penalties, while preserving more flexibility than larger values when the richer nonstationary structure is needed. For $D_{\text{train}}=2000$, a somewhat stronger penalty is beneficial, and $\alpha=100$ provides a good compromise across scenarios for the same reason: it is strong enough to control the extra SR-RBF(2) flexibility, but not so strong that it removes the benefit of including it when the deformation is more complex. Although larger values such as $\alpha=1000$ can also perform well, they offer only limited additional improvement while inducing substantially stronger shrinkage. We therefore use $\alpha=3$ for $D_{\text{train}}=800$ and $\alpha=100$ for $D_{\text{train}}=2000$ in the subsequent experiments, viewing these values not as universally optimal choices, but as practical defaults that balance overfitting control and sufficient warping flexibility in the settings considered here. We take the latter choice as practical guidance from an idealized simulation setting with a similar training size for the real data application.

Based on these sensitivity results, we use the grid search mainly as practical guidance from idealized simulation settings, rather than as a full tuning exercise for the real-data application. In principle, for a given model configuration and application, users could follow the same sensitivity-analysis design as above by refitting the candidate nonstationary models over a small grid of $\alpha$ values and comparing out-of-sample censored log-likelihood on test locations (here with $D_{\text{test}}=100$).


\subsection{Robustness to asymptotic-independence misspecification} \label{Appendix:Simulation_misspecification}
\begin{figure}[t!]
\begin{center}
\begin{tabular}{c}
\includegraphics[width=0.24\linewidth]{images_sim/figures-AWU_RBF_2D/true_space_chess.pdf}  \\
\includegraphics[width=0.24\linewidth]{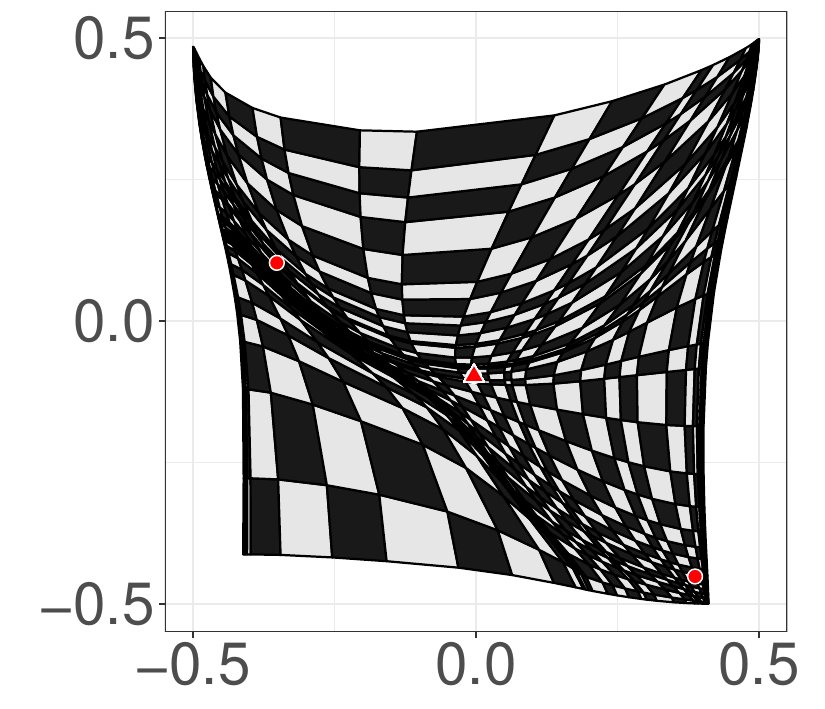} 
\includegraphics[width=0.24\linewidth]{images_sim/figures-AWU_RBF_2D/ind-misspec/miss-GP_max_EC_nonsta_layer3_warped_space_chess.pdf} 
\includegraphics[width=0.24\linewidth]{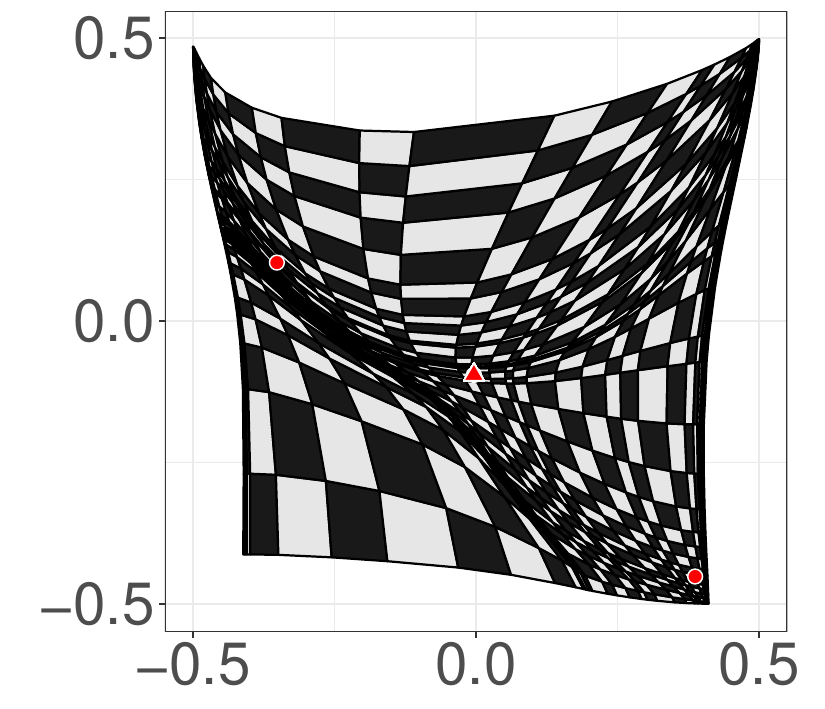} \\
\includegraphics[width=0.24\linewidth]{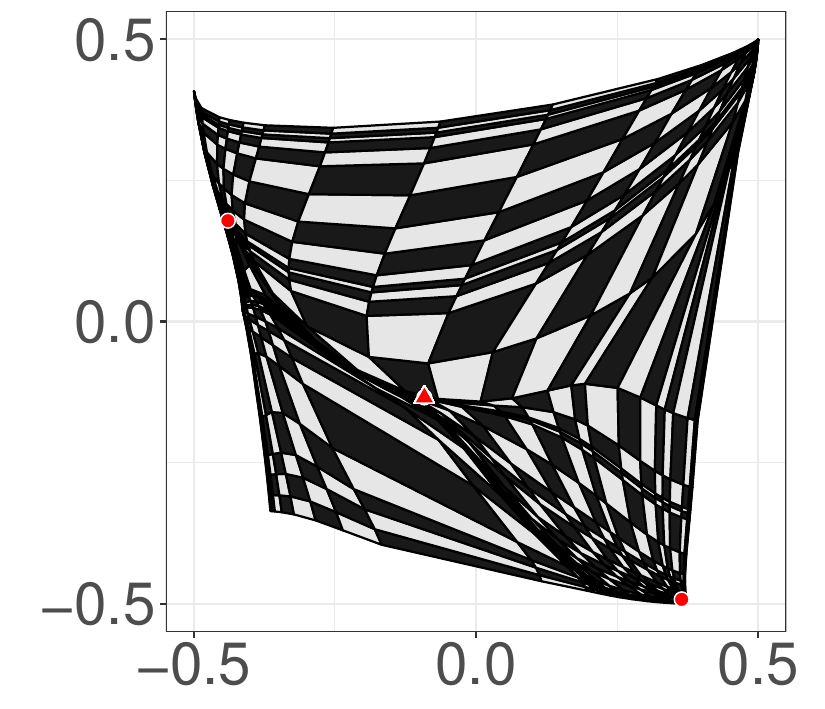} 
\includegraphics[width=0.24\linewidth]{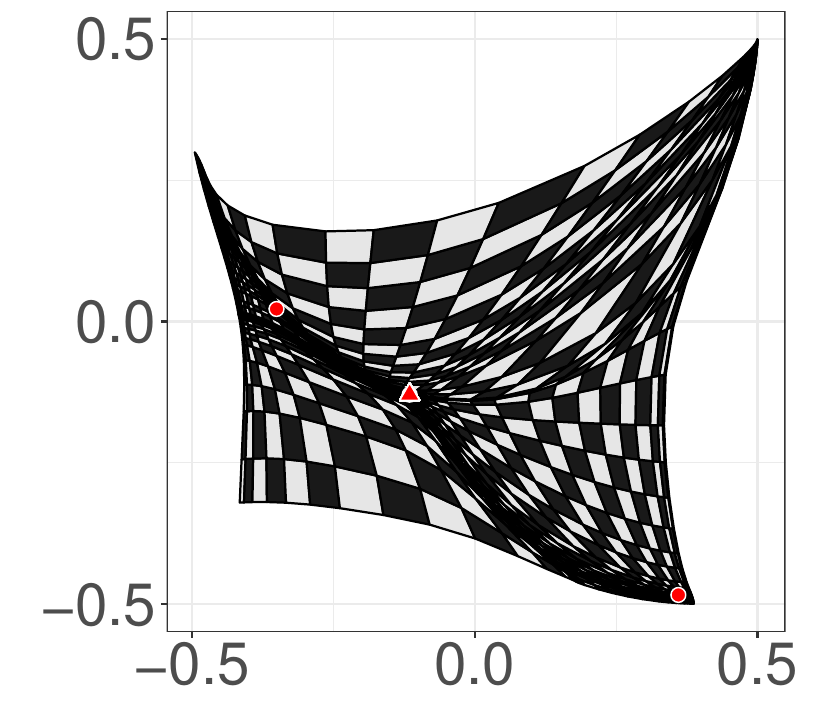} 
\includegraphics[width=0.24\linewidth]{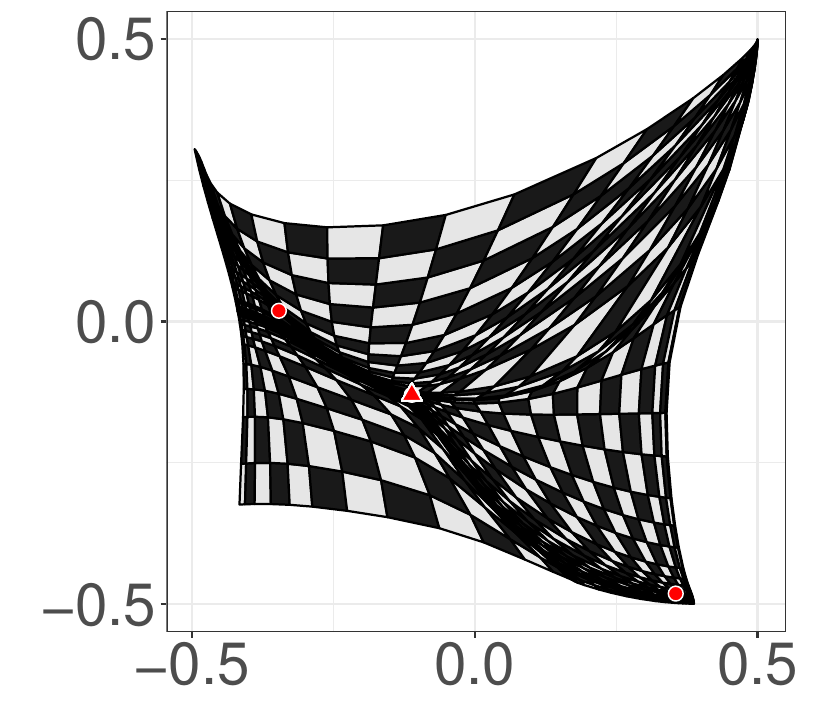} \\
\end{tabular}
\end{center}
\caption{Top row: true warped domain $\mathcal{W}$ generated using Architecture 3. Middle/bottom rows: estimated warped space $\hat{\mathcal{W}}$ in a misspecified setting, where data are generated using an asymptotically independent Gaussian copula with nonstationary model~3, and $\mathcal{W}$ is obtained by fitting an (asymptotically dependent) $r$-Pareto process with the correct nonstationary architecture and with the risk functional $r_{\text{site}}(\cdot)$, $r_{\text{max}}(\cdot)$, and $r_{\text{sum}}(\cdot)$ (column 1--3), respectively. Three reference points are labeled in red.}
\label{pic:misspecification}
\end{figure}

Motivated by the fact that our fitted $r$-Pareto Brown--Resnick model assumes threshold stability and asymptotic dependence, whereas environmental data at observed levels may exhibit weakening tail dependence and thus behave closer to asymptotic independence, we conduct a deliberate misspecification experiment to assess whether learning the spatial deformation is sensitive to the asymptotic regime.
We simulate nonstationary asymptotically independent data using a Gaussian-copula construction with Pareto margins, while inducing nonstationarity through the same warped domain as in our main simulations (Architecture~3). Concretely, letting $\tilde{\bm{s}}_i = \bm{f}_0(\bm s_i)$ denote the true warped coordinates, we generate a Gaussian process $\tilde{X}(\tilde{\bm{s}})$, $\tilde{\bm{s}}\in\mathcal{W}$, with stationary increments (relative to the reference origin $\bm{0}$) on the warped space, with power variogram
\[
\gamma(\tilde{\bm{s}}_i,\tilde{\bm{s}}_j)=\left(\frac{\|\tilde{\bm{s}}_i-\tilde{\bm{s}}_j\|}{\varphi}\right)^{\kappa},
\qquad \varphi>0,\ \kappa\in(0,2],
\]
and covariance (with an anchor location, $\bm 0$) given by the standard variogram identity
$$\mathrm{Cov}\{\tilde{X}(\tilde{\bm{s}}_i),\tilde{X}(\tilde{\bm{s}}_j)\}=\gamma(\tilde{\bm{s}}_i,\bm 0)+\gamma(\tilde{\bm{s}}_j,\bm 0)-\gamma(\tilde{\bm{s}}_i,\tilde{\bm{s}}_j).$$
We then set $U_i=\Phi\{\tilde{X}(\tilde{\bm{s}}_i)\}$ and transform margins to standard Pareto via $X_i=(1-U_i)^{-1}$, yielding asymptotic independence (through the Gaussian copula) but nonstationary dependence in the original space through $\bm{f}_0$.

After standardizing to Pareto margins (as in our pipeline), we fit our asymptotically dependent nonstationary $r$-Pareto model to functional exceedances using the correct architecture (nonstationary model~3), and we repeat this for $r_{\text{site}}(\cdot)$, $r_{\text{max}}(\cdot)$, and $r_{\text{sum}}(\cdot)$, under WLS/GSM inference methods, and $D=500$. As shown in Figure~\ref{pic:misspecification}, the estimated warped domain $\hat{\mathcal W}$ remains very similar to the true warped domain $\mathcal W$ despite the model misspecification. We also repeat the experiment with alternative nonstationary architectures (e.g., nonstationary model~1) and obtain the same qualitative conclusion (results not shown for brevity). 
}

\subsection{Additional results for the simulation study}\label{Appendix:Simulation_add_results}

{\color{black}
\subsubsection{Additional warping-recovery diagnostics for Architecture~3 (Figure~\ref{pic:sim_space})} \label{Appendix:Simulation_recovery}

Figure~\ref{pic:sim_space} in the main paper provides a qualitative visualization of the true and estimated warped spaces under Architecture~3, but explicit cross-panel comparison can be difficult. To localize where different inference methods and architectures under- or over-estimate the deformation, we introduce two location-wise diagnostics that compare the distance profile around each site under the estimated warping $\hat{\bm{f}}$ to that under the oracle warping $\bm{f}_0$ (used in our simulation setting).

Because the fitted extremal dependence relies on the relative geometry of locations in the warped space (i.e., pairwise distances under a stationary and isotropic dependence model), it is natural to assess warping recovery through discrepancies in warped-space distances. The key idea is to compare, for each site $\bm{s}_i$, the entire vector of its distances to all other sites (its distance profile) under $\hat{\bm{f}}$ versus under $\bm{f}_0$, yielding a location-wise diagnostic that directly highlights where the estimated deformation deviates locally from the oracle deformation.

For locations $\{\bm s_i\}_{i=1}^{D}$, define the oracle and fitted warped-space distances
\[
d^{(0)}_{ij}=\|\bm f_0(\bm s_i)-\bm f_0(\bm s_j)\|,
\qquad
\hat d_{ij}=\|\hat{\bm f}(\bm s_i)-\hat{\bm f}(\bm s_j)\|,
\qquad j\neq i.
\]
To account for the range alignment (already considered in our inference), we work with rescaled distances
$\tilde d^{(0)}_{ij}=d^{(0)}_{ij}/\varphi$ and $\tilde{\hat d}_{ij}=\hat d_{ij}/\hat{\varphi}$, where $\varphi$ and $\hat{\varphi}$ denote the oracle and fitted range parameters.
Let $\bar d^{(0)}_i = (D-1)^{-1}\sum_{j\neq i}\tilde d^{(0)}_{ij}$ denote the mean oracle (rescaled) distance from $\bm s_i$ to all other sites.

We then define the (dimensionless) sitewise warping recovery (SWR) score, for location $\bm{s}_i$, as
\[
\mathrm{SWR}(\bm s_i)=
\frac{\sqrt{\frac{1}{D-1}
\sum_{j\neq i}\left(\tilde{\hat d}_{ij}-\tilde d^{(0)}_{ij}\right)^2}}{\bar{d}^{(0)}_i},
\]
which is the root mean squared discrepancy between the rescaled fitted and oracle distance profiles around $\bm s_i$, normalized by $\bar d^{(0)}_i$. Dividing by $\bar d^{(0)}_i$ reduces the influence of spatial position on the score: interior sites typically have smaller average distances to the rest of the domain than boundary sites, and this normalization yields a relative error that is more comparable across sites.

To localize the direction of the discrepancy (over- versus under-estimation), we also report a (normalized) signed bias
\[
\mathrm{SB}(\bm s_i)=
\frac{\frac{1}{D-1}\sum_{j\neq i}\left(\tilde{\hat d}_{ij}-\tilde d^{(0)}_{ij}\right)}{\bar{d}^{(0)}_i},
\]
which is the mean signed discrepancy in the rescaled distance profile around $\bm s_i$, again normalized by $\bar d^{(0)}_i$. Here, $\mathrm{SB}(\bm s_i)>0$ indicates local expansion (overestimation of warped-space distances around $\bm s_i$), whereas $\mathrm{SB}(\bm s_i)<0$ indicates local contraction (underestimation).

\begin{figure}[t!]
\centering
\includegraphics[width=0.95\linewidth]{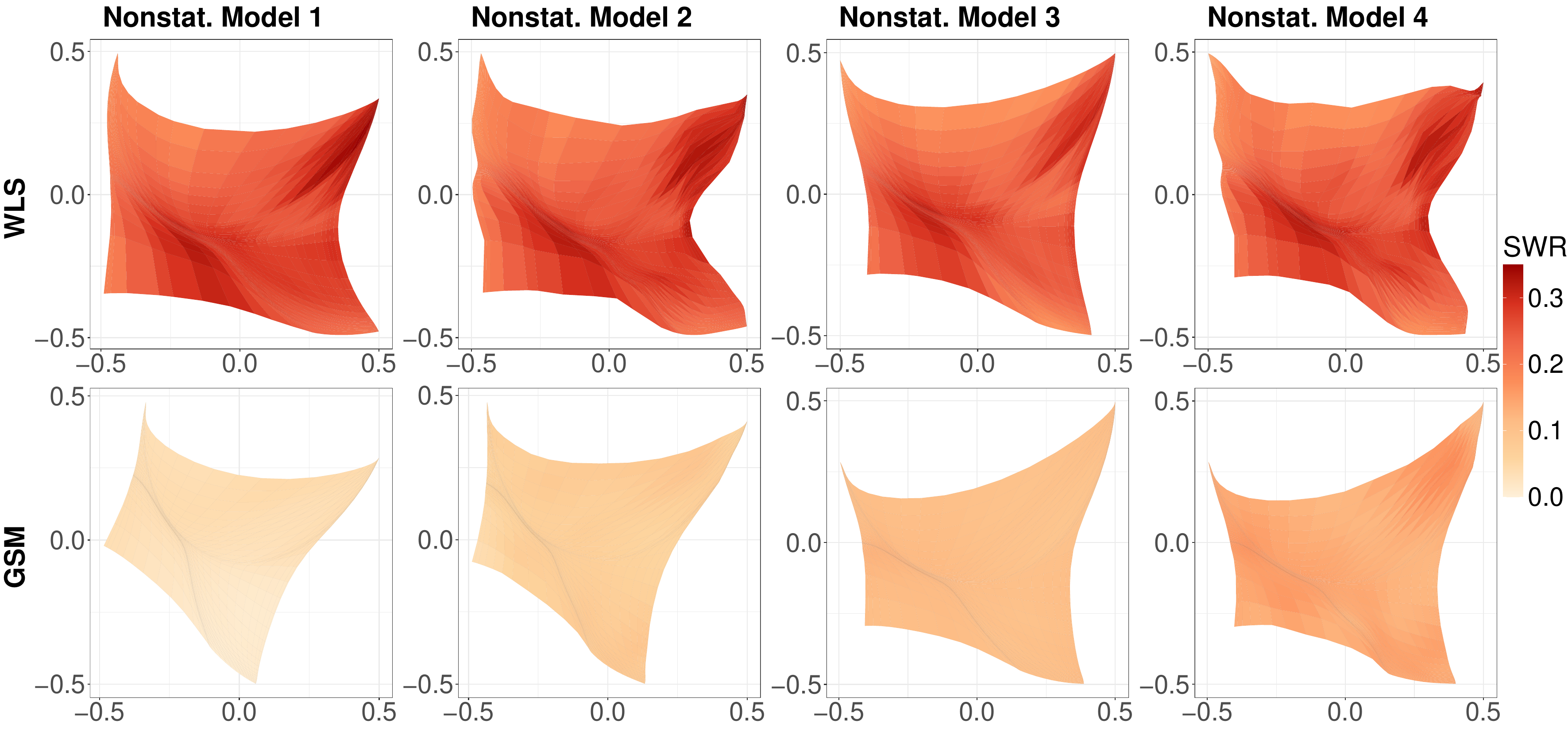}
\caption{Sitewise warping recovery (SWR) diagnostic for the simulation setting of Figure~\ref{pic:sim_space} (data generated under Architecture~3 with risk functional $r_{\max}(\cdot)$). Panels are arranged as in Figure~\ref{pic:sim_space} (columns: fitted nonstationary models 1--4; rows: WLS and GSM). Colors show $\mathrm{SWR}(\bm s_i)$.}
\label{pic:SWR}
\end{figure}

\begin{figure}[t!]
\centering
\includegraphics[width=0.95\linewidth]{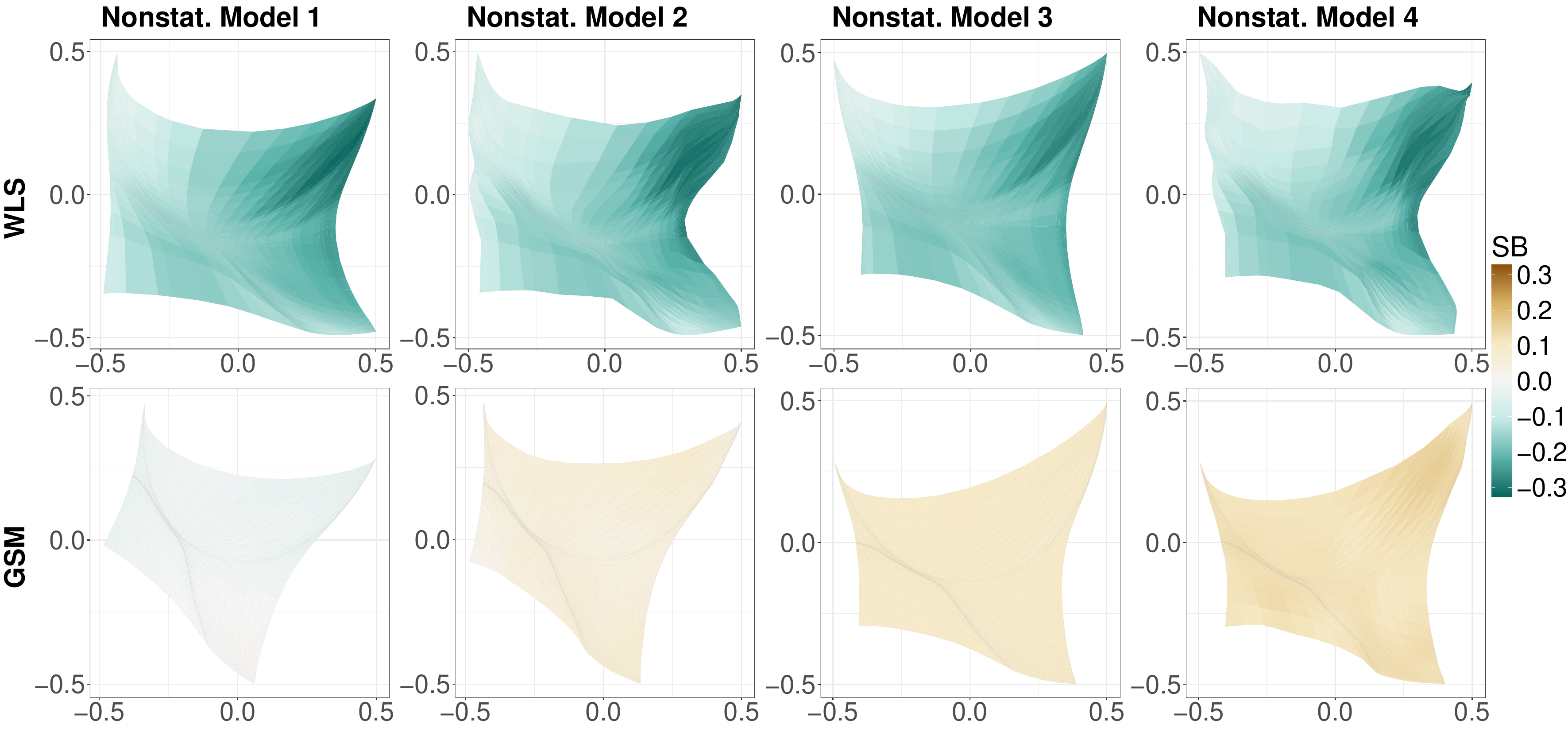}
\caption{Signed bias (SB) diagnostic for the simulation setting of Figure~\ref{pic:sim_space} (data generated under Architecture~3 with risk functional $r_{\max}(\cdot)$). Panels are arranged as in Figure~\ref{pic:sim_space} (columns: fitted nonstationary models 1--4; rows: WLS and GSM). Colors show $\mathrm{SB}(\bm s_i)$, with positive (negative) values indicating local overestimation (underestimation) of warped-space distances around $\bm s_i$.}
\label{pic:SB}
\end{figure}

Figures~\ref{pic:SWR}--\ref{pic:SB} report these diagnostics for the simulation setting of Figure~\ref{pic:sim_space} (data generated under Architecture~3 with risk functional $r_{\max}(\cdot)$), with panels arranged as in Figure~\ref{pic:sim_space}. In Figure~\ref{pic:SWR}, the GSM method (bottom row) yields substantially smaller and more spatially homogeneous SWR values than the WLS method (top row). For WLS, the largest recovery errors concentrate in the upper-right portion of the warped domain, suggesting that the deformation is most difficult to estimate in this region (where the fitted warped mesh exhibits strong local bending/anisotropic stretching). Figure~\ref{pic:SB} further localizes the direction of the discrepancy: WLS shows predominantly negative SB values (local underestimation of warped-space distances, i.e., contraction), again most pronounced near the upper-right corner. In contrast, GSM generally performs well, with SB values much closer to zero; moreover, the Möbius-based Nonstationary Models~1--2 are slightly better on these diagnostics (lower SWR and SB closer to zero) than the models without the Möbius transformation, suggesting that the Möbius transformation provides additional flexibility for warped-space estimation. Together, these diagnostics provide a concise, location-wise summary of where and how WLS/GSM (and different fitted architectures) over/underestimate the deformation, complementing the qualitative estimated warped-space plots in Figure~\ref{pic:sim_space}.

We further note that the extremal dependence measures (i.e., CEP) are not fully determined by the (rescaled) warped-space distances alone: they also depend on the smoothness parameter $\kappa$ through the chosen stationary and isotropic dependence model. In the diagnostics above, we have already accounted for the range alignment by rescaling distances with $\varphi$ and $\hat\varphi$. Thus, $\mathrm{SWR}(\bm{s}_i)$ and $\mathrm{SB}(\bm{s}_i)$ are intended specifically to assess how well the estimated warped-space geometry matches the oracle warped-space geometry in the simulation study; they do not attempt to attribute any remaining CEP discrepancies to differences in $\kappa$ or other dependence-model components.

}

\subsubsection{Results regarding Architectures 1, 2, and 4}\label{Appendix:Simulation_124}

In this section, we report the additional simulation results for data generated under Architectures~1, 2, and~4, complementing the main-text results for Architecture~3 by presenting the corresponding estimated warped spaces, test censored log-likelihoods, and dependence-parameter estimates.

\begin{figure}[hbt!]
\begin{center}
\begin{tabular}{c}
\includegraphics[width=0.24\linewidth]{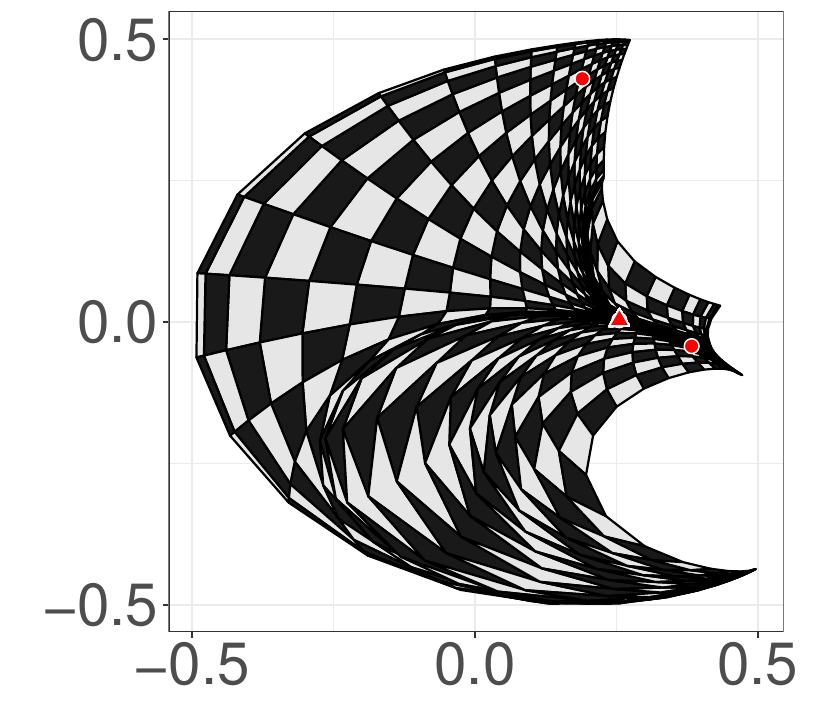}\\
\includegraphics[width=0.24\linewidth]{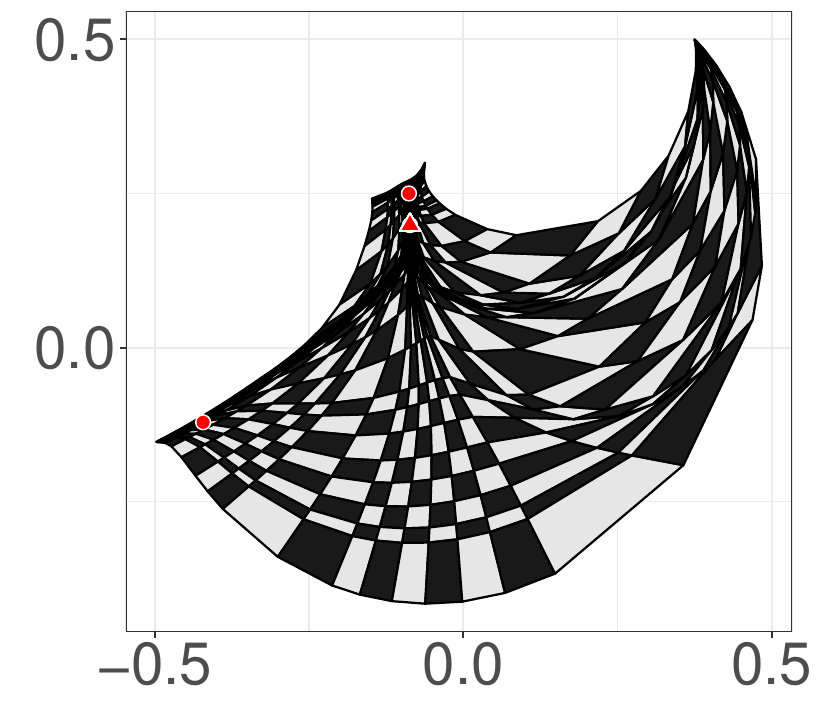} 
\includegraphics[width=0.24\linewidth]{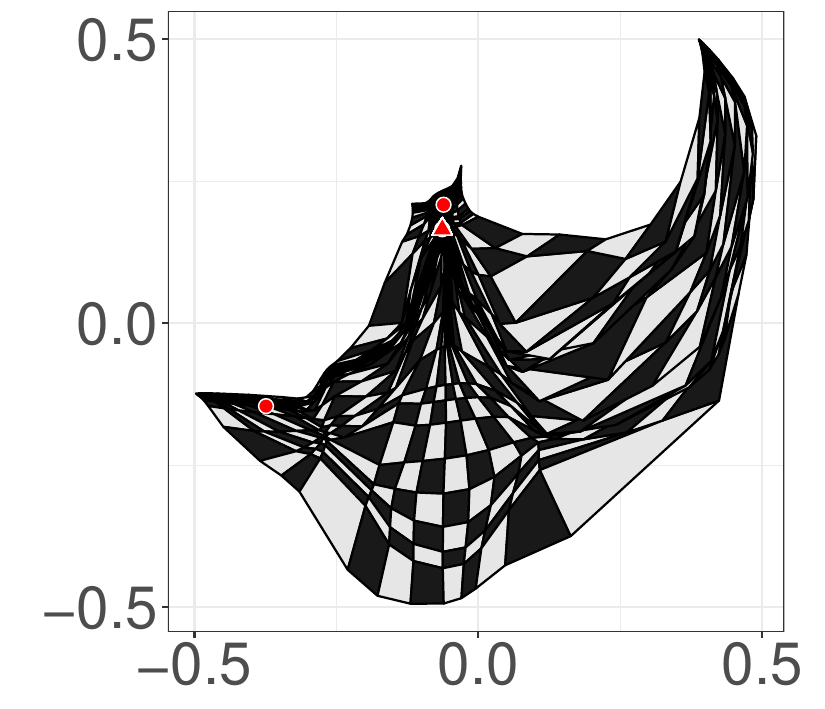}
\includegraphics[width=0.24\linewidth]{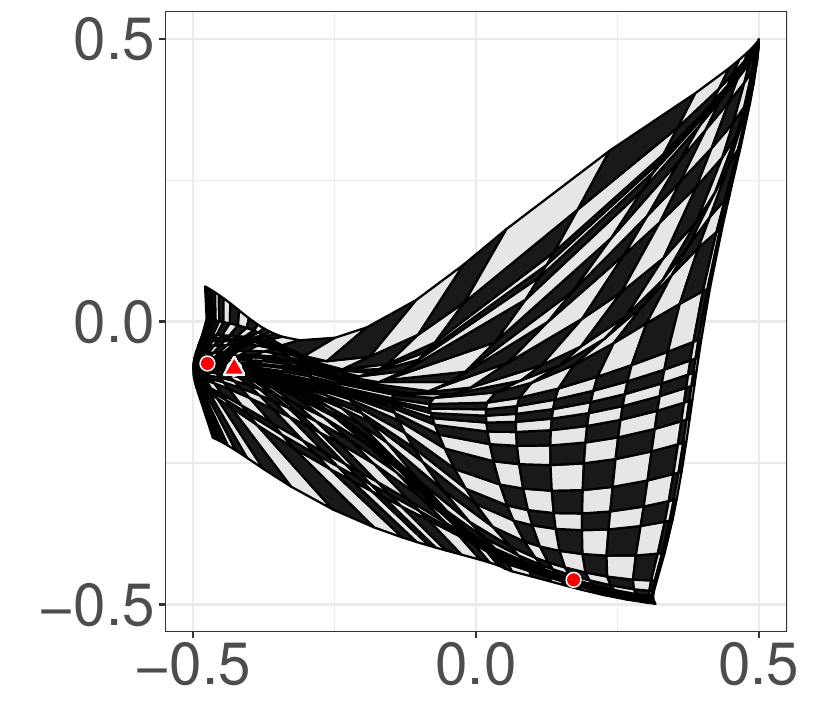}
\includegraphics[width=0.24\linewidth]{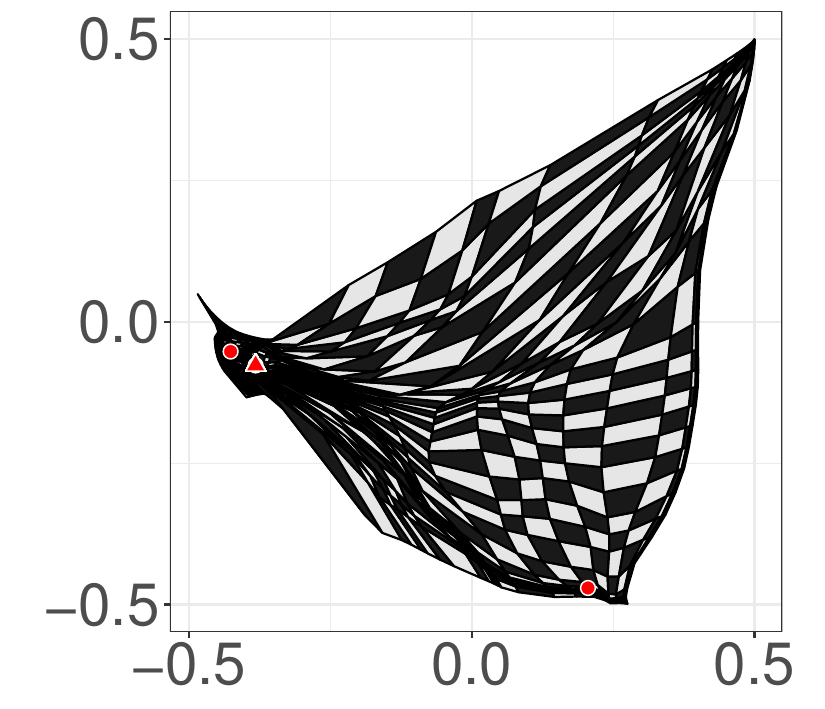} \\
\includegraphics[width=0.24\linewidth]{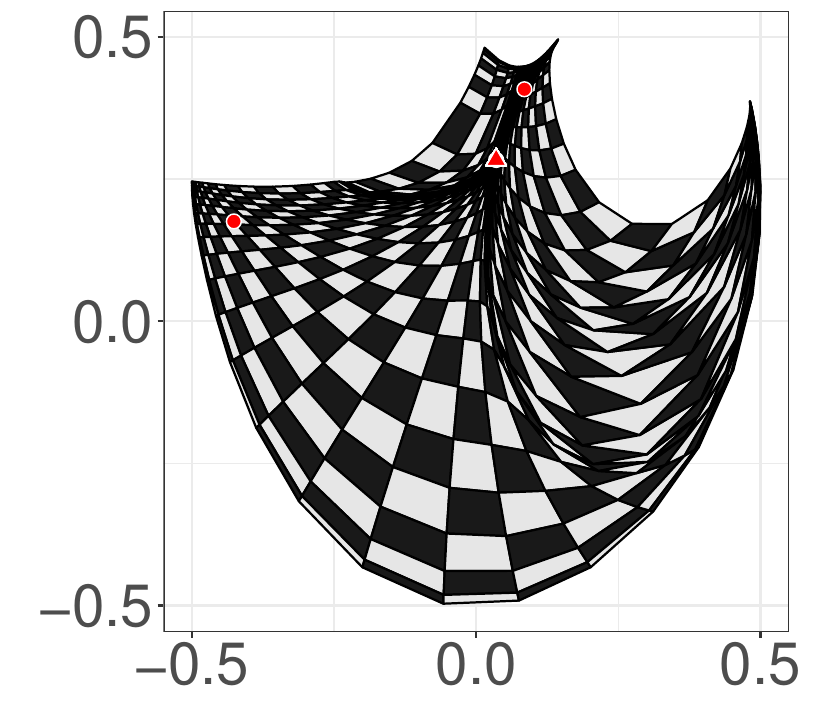} 
\includegraphics[width=0.24\linewidth]{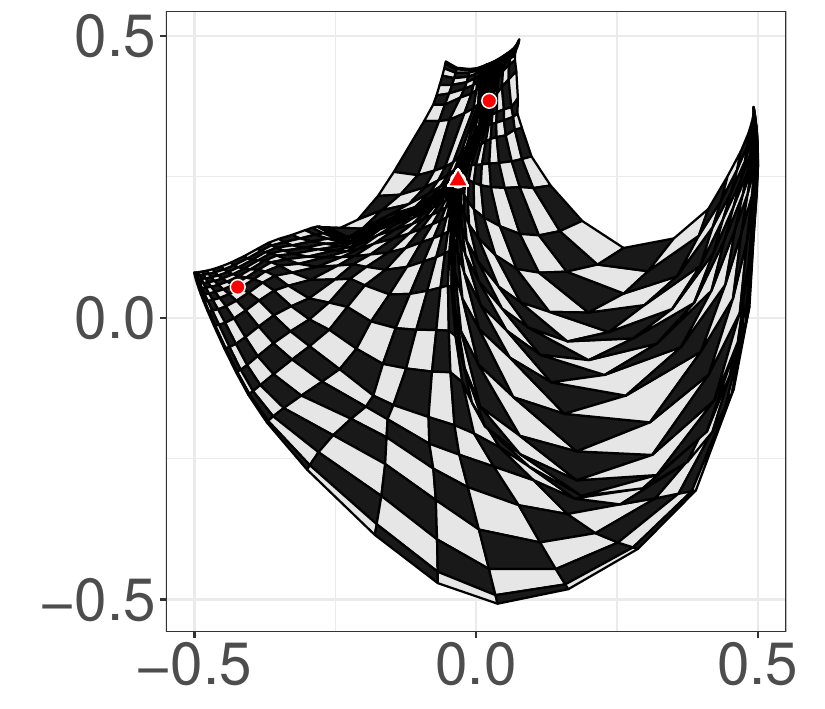}
\includegraphics[width=0.24\linewidth]{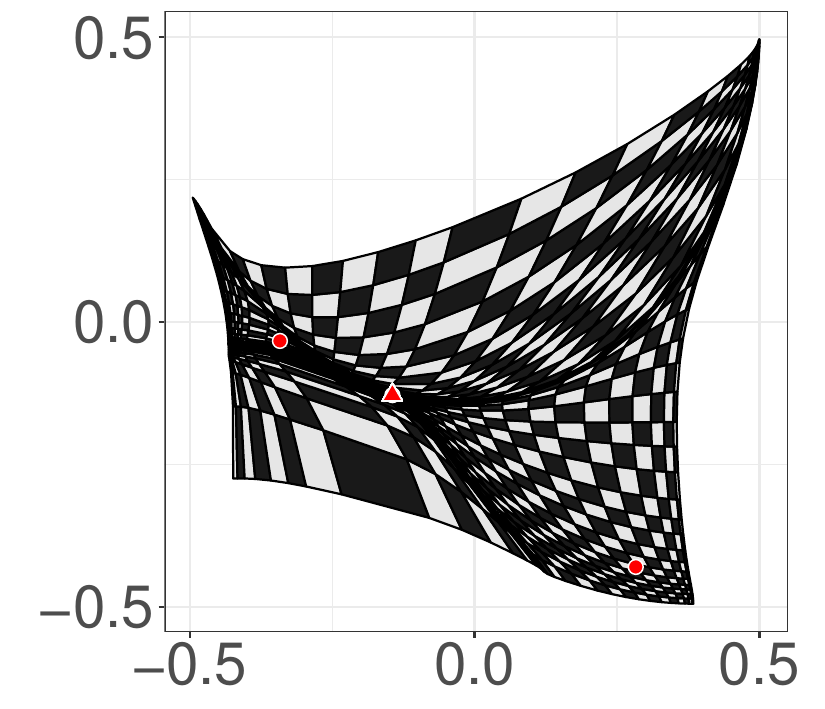}
\includegraphics[width=0.24\linewidth]{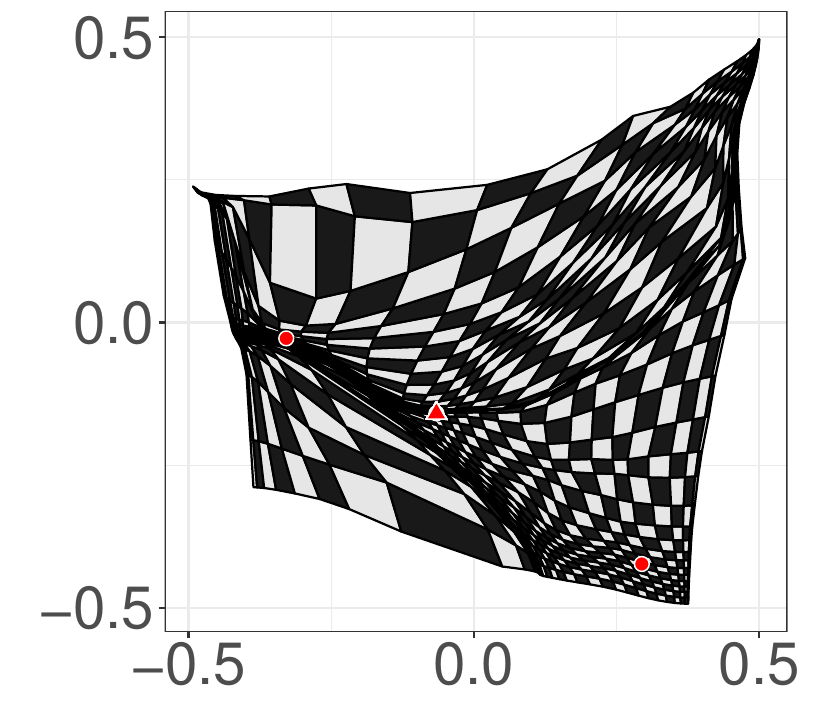}\\
\end{tabular}
\end{center}
\caption{Top row: true warped domain $\mathcal{W}$ generated using Architecture 1. Median and bottom rows: estimated warped space $\hat{\mathcal{W}}$ using nonstationary models 1--4 (column 1--4) with WLS and GSM inference methods, respectively. The data are generated with the risk functional $r_{\text{max}}(\cdot)$, and three reference points are labeled in red.}
\label{pic:sim_space1}
\end{figure} 

\begin{table}[hbt!]
    \centering
    \caption{Test censored log-likelihoods (larger is preferred) evaluated for all fitted models, using both the gradient score matching (GSM) and weighted least squares (WLS) inference methods, and with three different risk functionals. The best models are highlighted in bold for each risk functional and each inference method, and the oracle model based on Architecture~1 is labeled with an asterisk.}
    \begin{tabular}{l|l|c|c|c}
    \hline \hline
    \multirow{2}{*}{Architecture} & Inference & \multicolumn{3}{c}{Risk functional} \\
    \cline{3-5}
    & method & $r_{\text{site}}(\cdot)$ & $r_{\text{max}}(\cdot)$ & $r_{\text{sum}}(\cdot)$ \\
    \hline
Stationary & \multirow{5}{*}{WLS} & $-61327$ & $-16959$ & $20551$ \\ 
Nonstationary 1* &  & $-59495$ & $-15714$ & $23614$ \\ 
Nonstationary 2 &  & $-60690$ & $\mathbf{-15564}$ & $\mathbf{23653}$ \\ 
Nonstationary 3 &  & $\mathbf{-58733}$ & $-16182$ & $22244$ \\ 
Nonstationary 4 &  & $-58831$ & $-16114$ & $22414$ \\ 
\hline 
Stationary & \multirow{5}{*}{GSM} & $-61928$ & $-16961$ & $20236$ \\ 
Nonstationary 1* &  & $-56858$ & $\mathbf{-14793}$ & $\mathbf{25018}$ \\ 
Nonstationary 2 &  & $\mathbf{-56069}$ & $-14906$ & $24843$ \\ 
Nonstationary 3 &  & $-59834$ & $-15181$ & $22666$ \\ 
Nonstationary 4 &  & $-58800$ & $-15516$ & $24268$ \\ 
    \hline
    \end{tabular} 
    \label{tab:sim_compare1}
\end{table}

\begin{table}[hbt!]
    \centering
    \caption{Estimates of the extremal dependence parameters, $\hat{\bm{\psi}}' = (\hat{\varphi}, \hat{\kappa})$, for data generated with Architecture 1, using both the gradient score matching (GSM) and weighted least squares (WLS) inference methods, and three different risk functionals. The true values of the extremal dependence parameters are $\varphi_0 = 0.2$ and $\kappa_0 = 1$. Standard deviations obtained using a nonparametric bootstrap are reported in brackets as subscripts of the corresponding parameter estimate. The oracle model based on Architecture~1 is labeled with an asterisk.}
    \resizebox{\columnwidth}{!}{%
    \begin{tabular}{l|l|c|c|c}
    \hline \hline
    \multirow{2}{*}{Architecture} & Inference & \multicolumn{3}{c}{Risk functional} \\
    \cline{3-5}
    & method & $r_{\text{site}}(\cdot)$ & $r_{\text{max}}(\cdot)$ & $r_{\text{sum}}(\cdot)$ \\
    \hline 
Stationary & \multirow{5}{*}{WLS} & $0.249_{(0.036)}$, $0.809_{(0.075)}$ & $0.481_{(0.160)}$, $0.926_{(0.136)}$ & $0.595_{(0.254)}$, $0.860_{(0.080)}$ \\ 
Nonstationary 1* &  & $0.130_{(0.035)}$, $1.017_{(0.106)}$ & $0.203_{(0.096)}$, $0.889_{(0.158)}$ & $0.290_{(0.151)}$, $0.923_{(0.104)}$ \\ 
Nonstationary 2 &  & $0.132_{(0.041)}$, $0.988_{(0.172)}$ & $0.208_{(0.092)}$, $0.885_{(0.136)}$ & $0.286_{(0.148)}$, $0.886_{(0.107)}$ \\ 
Nonstationary 3 &  & $0.190_{(0.028)}$, $0.835_{(0.084)}$ & $0.297_{(0.100)}$, $0.859_{(0.146)}$ & $0.388_{(0.176)}$, $0.738_{(0.125)}$ \\ 
Nonstationary 4 &  & $0.165_{(0.029)}$, $0.724_{(0.090)}$ & $0.274_{(0.096)}$, $0.824_{(0.144)}$ & $0.369_{(0.177)}$, $0.728_{(0.110)}$ \\ 
\hline 
Stationary & \multirow{5}{*}{GSM} & $0.155_{(0.001)}$, $0.950_{(0.005)}$ & $0.158_{(0.001)}$, $0.959_{(0.004)}$ & $0.157_{(0.001)}$, $0.956_{(0.004)}$ \\ 
Nonstationary 1* &  & $0.173_{(0.005)}$, $0.986_{(0.025)}$ & $0.180_{(0.004)}$, $1.007_{(0.020)}$ & $0.178_{(0.005)}$, $0.996_{(0.028)}$ \\ 
Nonstationary 2 &  & $0.174_{(0.004)}$, $0.988_{(0.019)}$ & $0.178_{(0.004)}$, $1.002_{(0.017)}$ & $0.178_{(0.003)}$, $0.996_{(0.016)}$ \\ 
Nonstationary 3 &  & $0.162_{(0.006)}$, $0.905_{(0.031)}$ & $0.188_{(0.006)}$, $1.061_{(0.034)}$ & $0.173_{(0.006)}$, $0.989_{(0.032)}$ \\ 
Nonstationary 4 &  & $0.170_{(0.004)}$, $0.977_{(0.019)}$ & $0.184_{(0.004)}$, $1.043_{(0.020)}$ & $0.186_{(0.005)}$, $1.053_{(0.029)}$ \\ 
    \hline
    \end{tabular} }
    \label{tab:sim_table1}
\end{table}


\clearpage
\begin{figure}[hbt!]
\begin{center}
\begin{tabular}{c}
\includegraphics[width=0.24\linewidth]{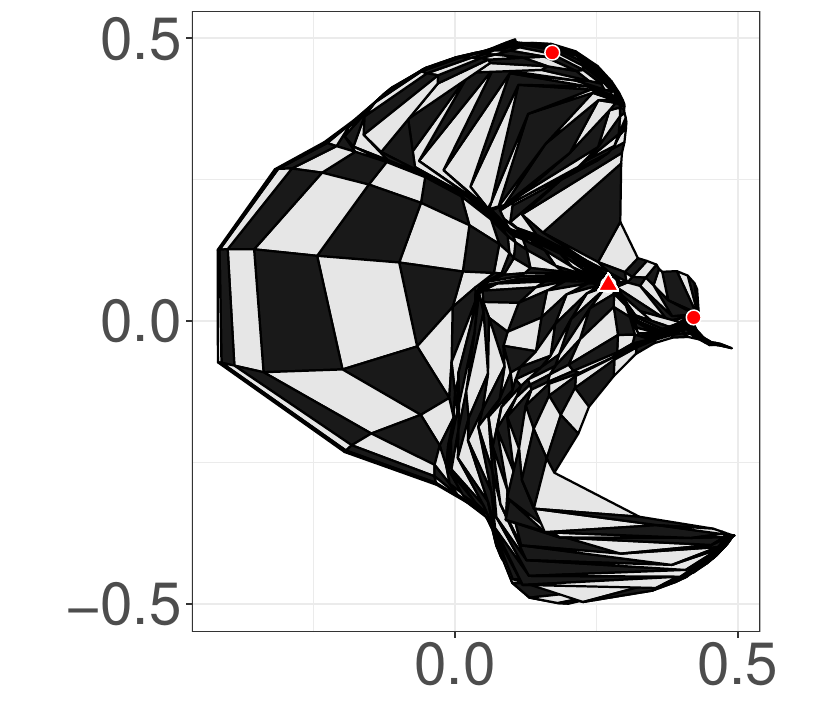}\\
\includegraphics[width=0.24\linewidth]{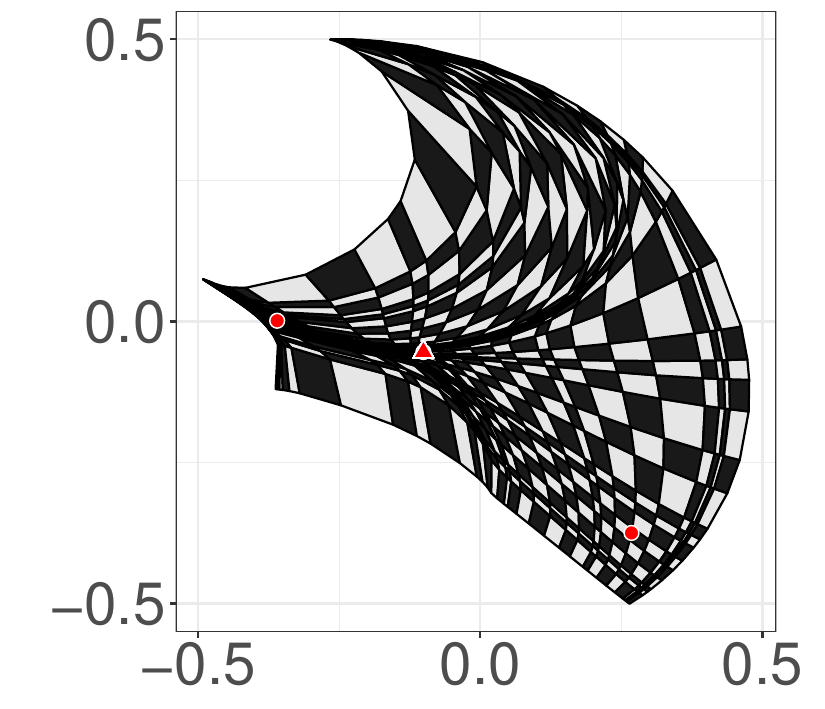} 
\includegraphics[width=0.24\linewidth]{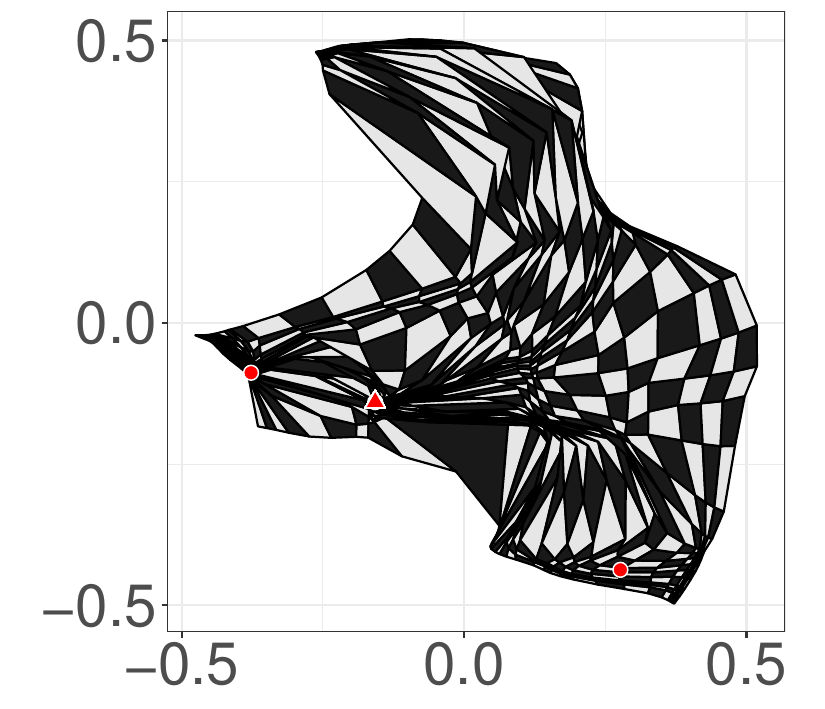}
\includegraphics[width=0.24\linewidth]{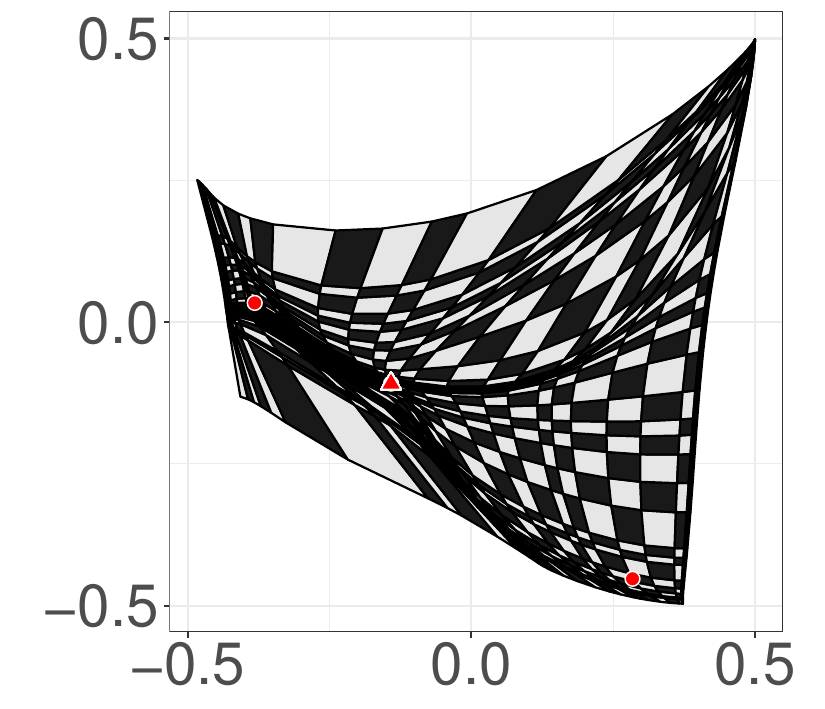}
\includegraphics[width=0.24\linewidth]{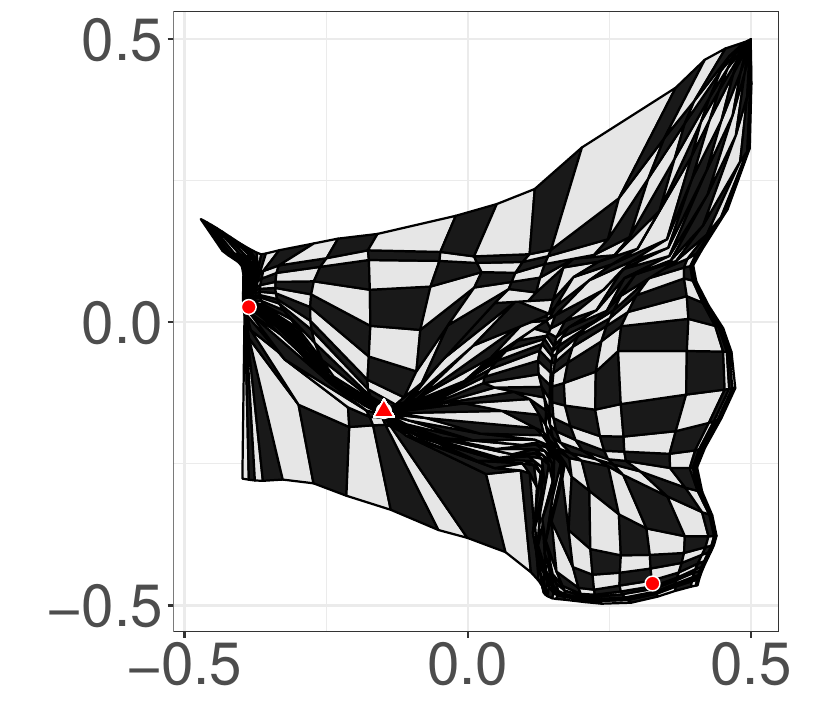} \\
\includegraphics[width=0.24\linewidth]{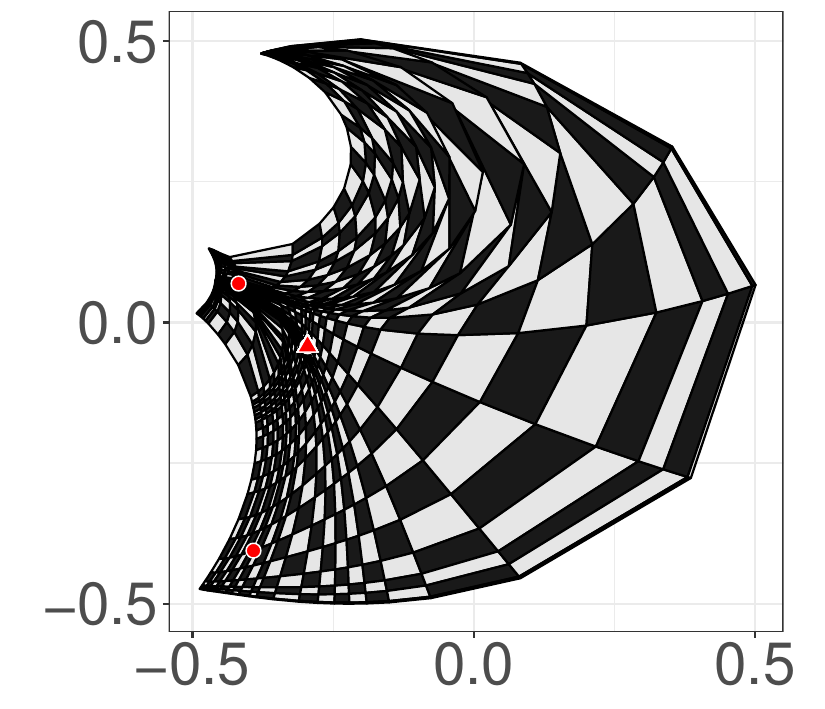} 
\includegraphics[width=0.24\linewidth]{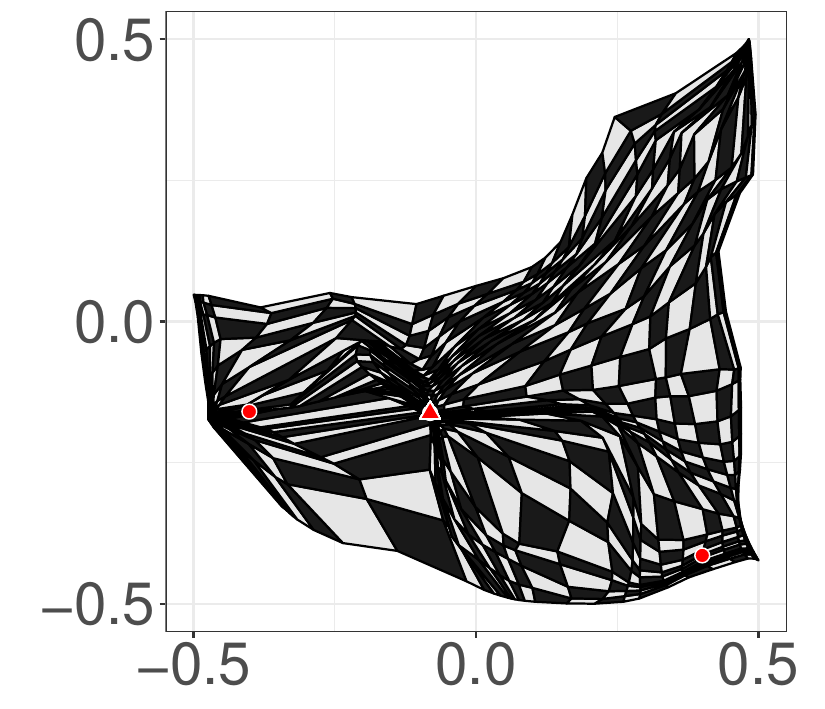}
\includegraphics[width=0.24\linewidth]{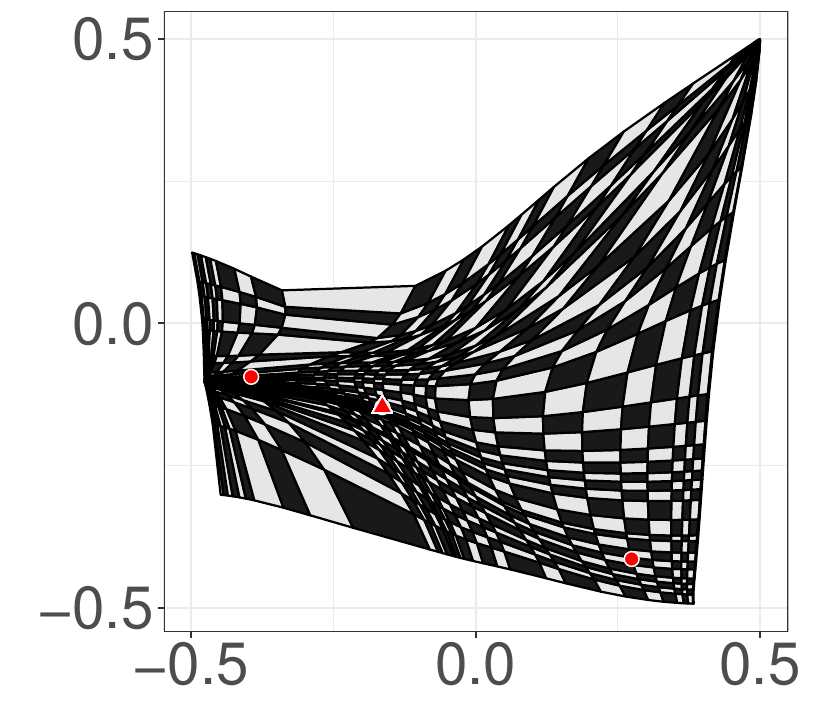}
\includegraphics[width=0.24\linewidth]{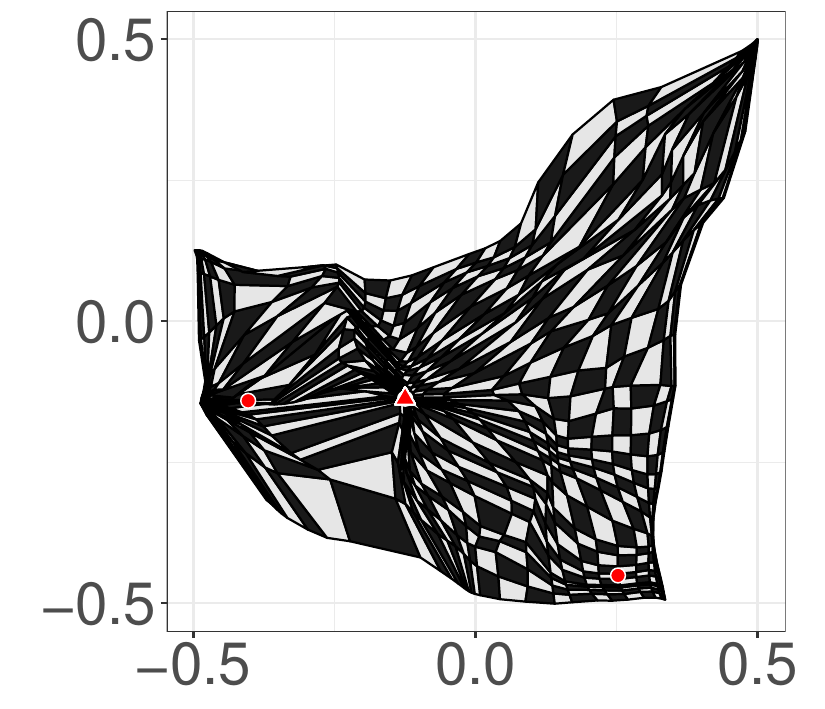}\\
\end{tabular}
\end{center}
\caption{Top row: true warped domain $\mathcal{W}$ generated using Architecture 2. Median and bottom rows: estimated warped space $\hat{\mathcal{W}}$ using nonstationary models 1--4 (column 1--4) with WLS and GSM inference methods, respectively. The data are generated with the risk functional $r_{\text{max}}(\cdot)$, and three reference points are labeled in red.}
\label{pic:sim_space2}
\end{figure} 

\begin{table}[hbt!]
    \centering
    \caption{Test censored log-likelihoods (larger is preferred) evaluated for all fitted models, using both the gradient score matching (GSM) and weighted least squares (WLS) inference methods, and with three different risk functionals. The best models are highlighted in bold for each risk functional and each inference method, and the oracle model based on Architecture~2 is labeled with an asterisk.}
    \begin{tabular}{l|l|c|c|c}
    \hline \hline
    \multirow{2}{*}{Architecture} & Inference & \multicolumn{3}{c}{Risk functional} \\
    \cline{3-5}
    & method & $r_{\text{site}}(\cdot)$ & $r_{\text{max}}(\cdot)$ & $r_{\text{sum}}(\cdot)$ \\
    \hline
Stationary & \multirow{5}{*}{WLS} & $-68255$ & $-14964$ & $17989$ \\ 
Nonstationary 1 &  & $-64719$ & $-13527$ & $20470$ \\ 
Nonstationary 2* &  & $\mathbf{-62131}$ & $\mathbf{-12751}$ & $\mathbf{21282}$ \\ 
Nonstationary 3 &  & $-69555$ & $-13958$ & $19701$ \\ 
Nonstationary 4 &  & $-68469$ & $-12873$ & $20092$ \\ 
\hline 
Stationary & \multirow{5}{*}{GSM} & $-66459$ & $-14927$ & $17505$ \\ 
Nonstationary 1 &  & $-63010$ & $-13755$ & $20150$ \\ 
Nonstationary 2* &  & $\mathbf{-61793}$ & $-13369$ & $\mathbf{20988}$ \\ 
Nonstationary 3 &  & $-63172$ & $-13621$ & $20144$ \\ 
Nonstationary 4 &  & $-62450$ & $\mathbf{-13285}$ & $20968$ \\ 
    \hline
    \end{tabular} 
    \label{tab:sim_compare2}
\end{table}

\begin{table}[hbt!]
    \centering
    \caption{Estimates of the extremal dependence parameters, $\hat{\bm{\psi}}' = (\hat{\varphi}, \hat{\kappa})$, for data generated with Architecture 2, using both the gradient score matching (GSM) and weighted least squares (WLS) inference methods, and three different risk functionals. The true values of the extremal dependence parameters are $\varphi_0 = 0.2$ and $\kappa_0 = 1$. Standard deviations obtained using a nonparametric bootstrap are reported in brackets as subscripts of the corresponding parameter estimate. The oracle model based on Architecture~2 is labeled with an asterisk.}
    \resizebox{\columnwidth}{!}{%
    \begin{tabular}{l|l|c|c|c}
    \hline \hline
    \multirow{2}{*}{Architecture} & Inference & \multicolumn{3}{c}{Risk functional} \\
    \cline{3-5}
    & method & $r_{\text{site}}(\cdot)$ & $r_{\text{max}}(\cdot)$ & $r_{\text{sum}}(\cdot)$ \\
    \hline 
Stationary & \multirow{5}{*}{WLS} & $0.410_{(0.089)}$, $1.037_{(0.070)}$ & $0.293_{(0.056)}$, $1.103_{(0.083)}$ & $0.294_{(0.124)}$, $0.781_{(0.115)}$ \\ 
Nonstationary 1 &  & $0.256_{(0.038)}$, $1.223_{(0.122)}$ & $0.193_{(0.030)}$, $1.134_{(0.103)}$ & $0.135_{(0.035)}$, $0.854_{(0.152)}$ \\ 
Nonstationary 2* &  & $0.255_{(0.038)}$, $1.173_{(0.129)}$ & $0.179_{(0.032)}$, $1.023_{(0.092)}$ & $0.114_{(0.034)}$, $0.702_{(0.134)}$ \\ 
Nonstationary 3 &  & $0.329_{(0.054)}$, $1.121_{(0.113)}$ & $0.219_{(0.037)}$, $1.080_{(0.113)}$ & $0.188_{(0.058)}$, $0.794_{(0.137)}$ \\ 
Nonstationary 4 &  & $0.305_{(0.053)}$, $1.065_{(0.126)}$ & $0.200_{(0.037)}$, $0.995_{(0.099)}$ & $0.160_{(0.060)}$, $0.719_{(0.125)}$ \\ 
\hline 
Stationary & \multirow{5}{*}{GSM} & $0.147_{(0.001)}$, $0.922_{(0.003)}$ & $0.148_{(0.001)}$, $0.927_{(0.004)}$ & $0.150_{(0.001)}$, $0.931_{(0.004)}$ \\ 
Nonstationary 1 &  & $0.153_{(0.017)}$, $0.988_{(0.112)}$ & $0.149_{(0.013)}$, $0.971_{(0.079)}$ & $0.155_{(0.012)}$, $1.004_{(0.090)}$ \\ 
Nonstationary 2* &  & $0.151_{(0.007)}$, $0.947_{(0.045)}$ & $0.153_{(0.008)}$, $0.945_{(0.059)}$ & $0.156_{(0.006)}$, $0.992_{(0.044)}$ \\ 
Nonstationary 3 &  & $0.161_{(0.003)}$, $1.002_{(0.035)}$ & $0.166_{(0.003)}$, $1.033_{(0.016)}$ & $0.163_{(0.002)}$, $1.014_{(0.028)}$ \\ 
Nonstationary 4 &  & $0.163_{(0.003)}$, $0.978_{(0.031)}$ & $0.161_{(0.003)}$, $0.958_{(0.032)}$ & $0.162_{(0.003)}$, $0.994_{(0.035)}$ \\ 
    \hline
    \end{tabular} }
    \label{tab:sim_table2}
\end{table}


\clearpage
\begin{figure}[hbt!]
\begin{center}
\begin{tabular}{c}
\includegraphics[width=0.24\linewidth]{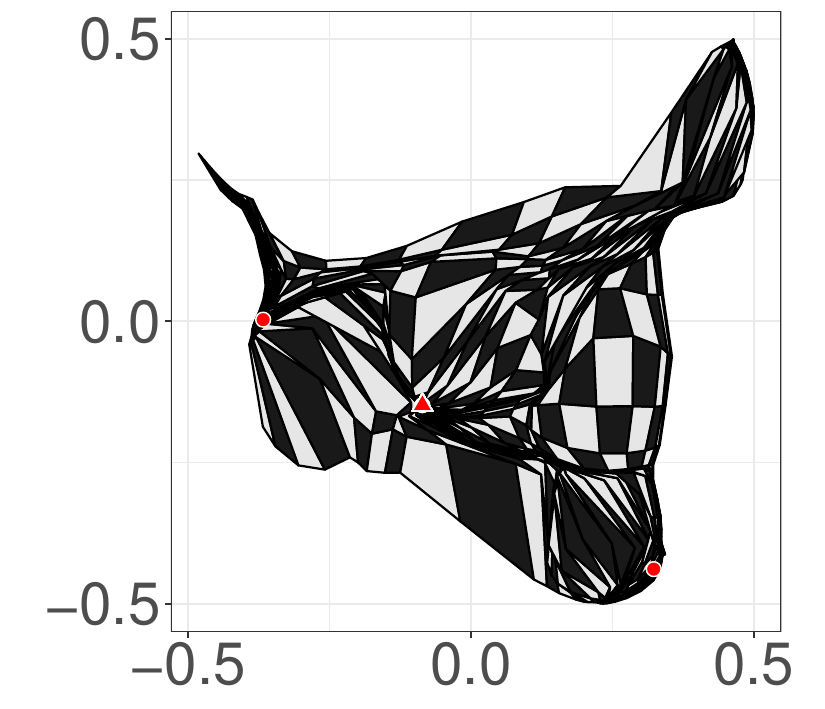}\\
\includegraphics[width=0.24\linewidth]{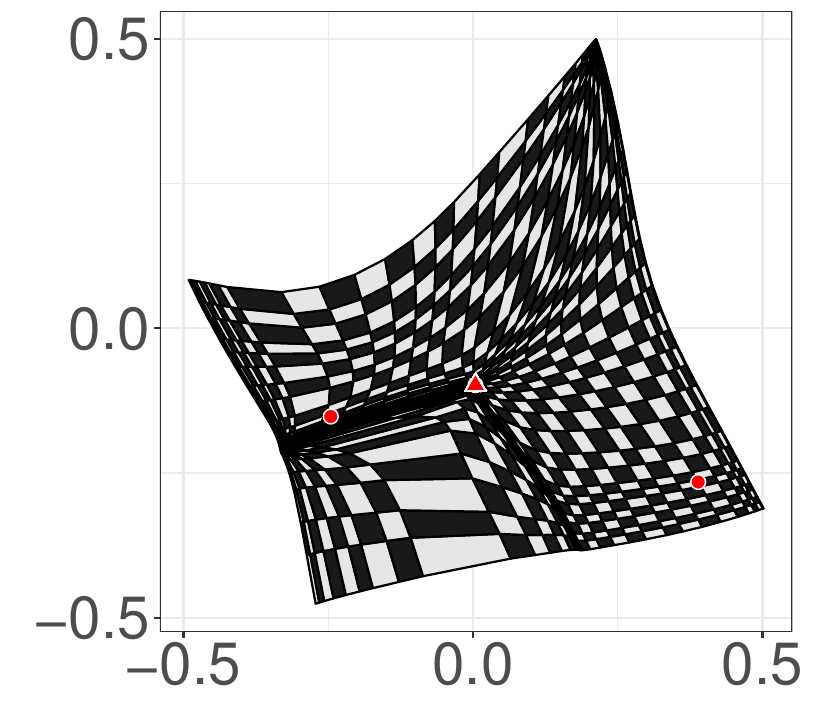} 
\includegraphics[width=0.24\linewidth]{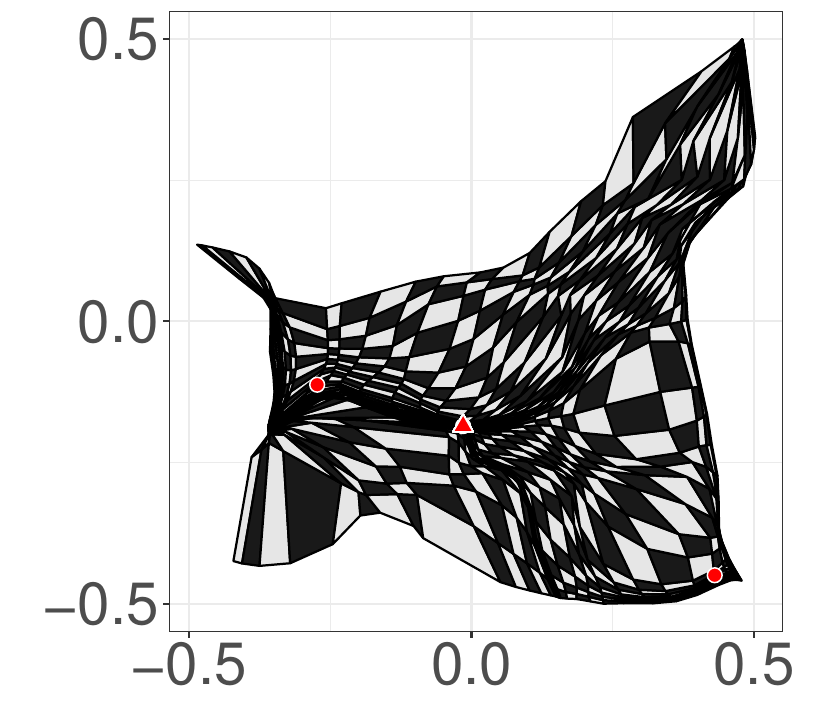}
\includegraphics[width=0.24\linewidth]{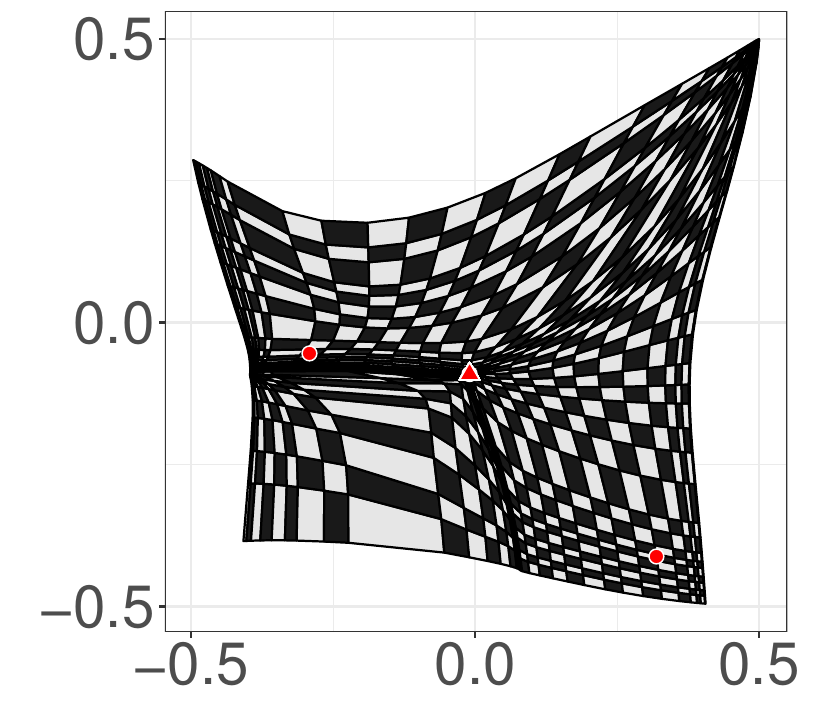}
\includegraphics[width=0.24\linewidth]{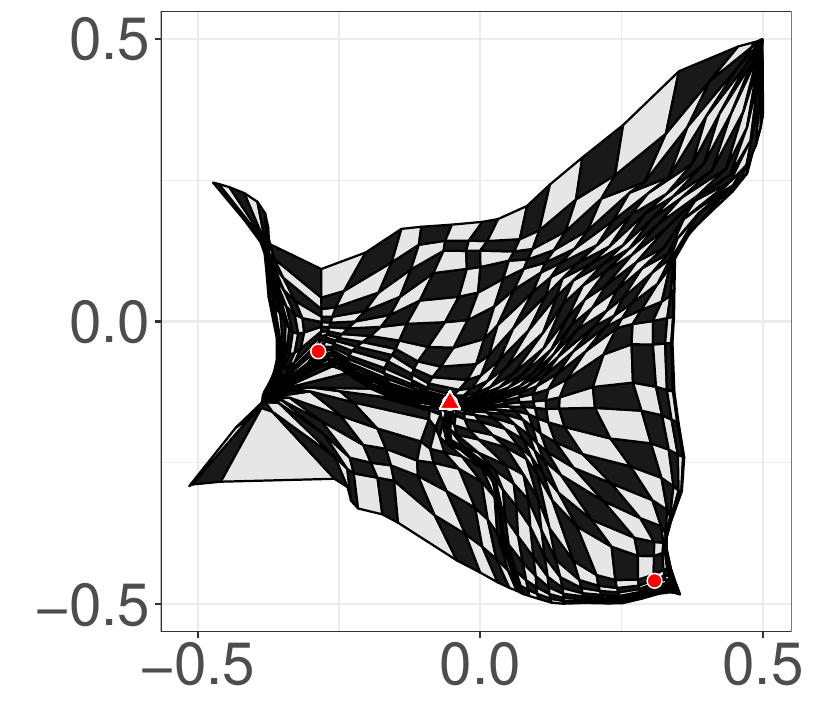}\\
\includegraphics[width=0.24\linewidth]{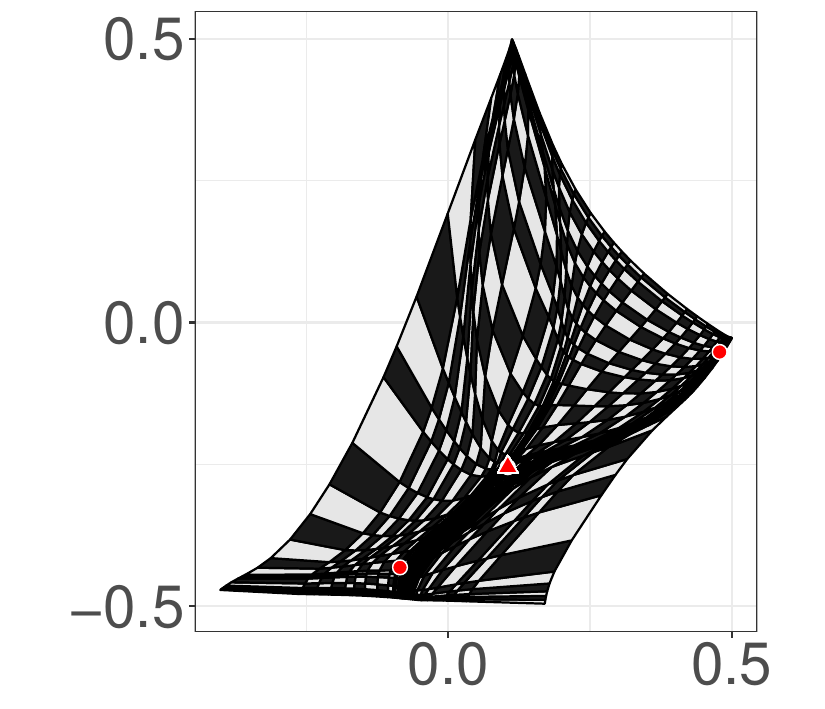} 
\includegraphics[width=0.24\linewidth]{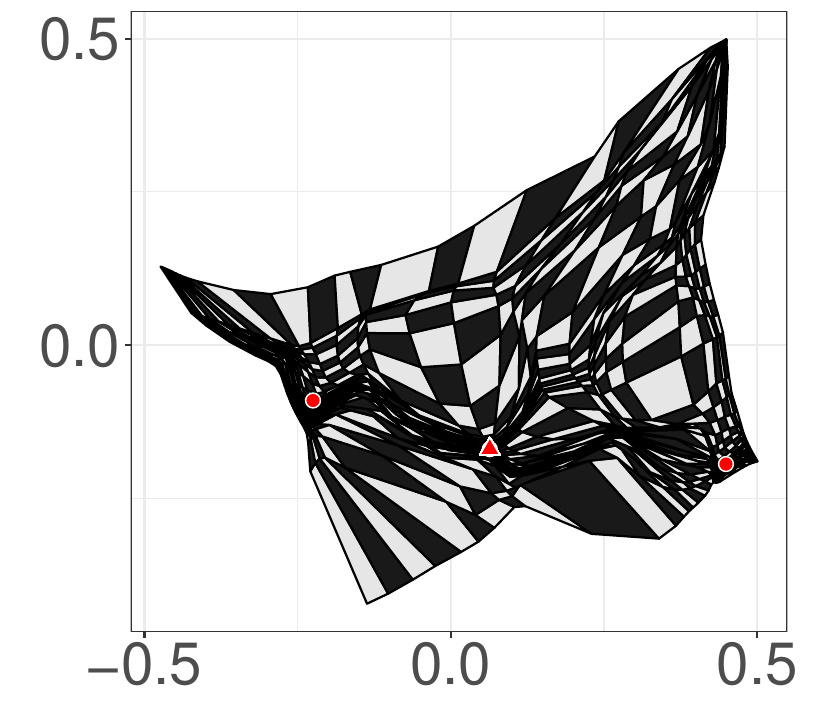}
\includegraphics[width=0.24\linewidth]{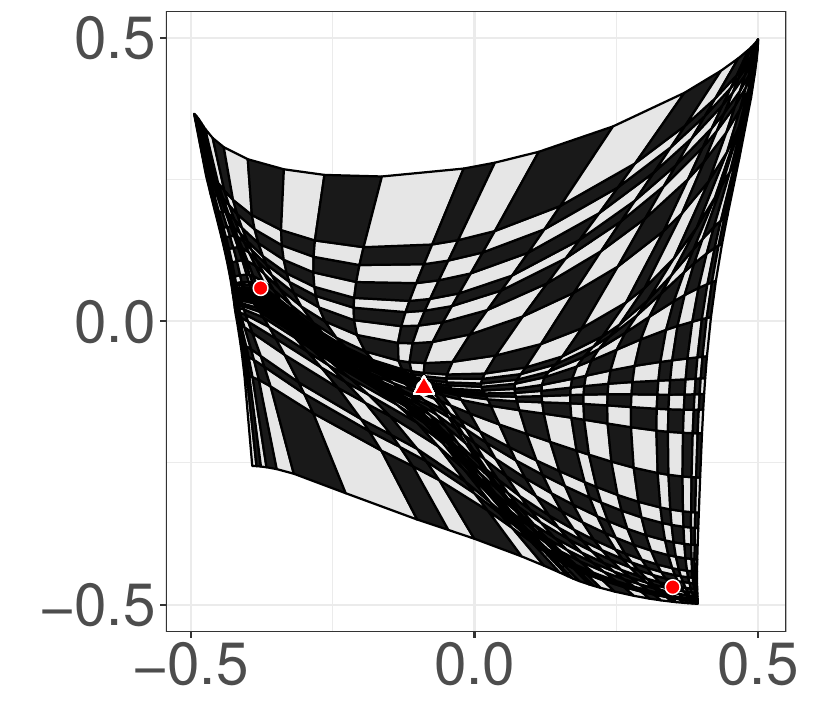}
\includegraphics[width=0.24\linewidth]{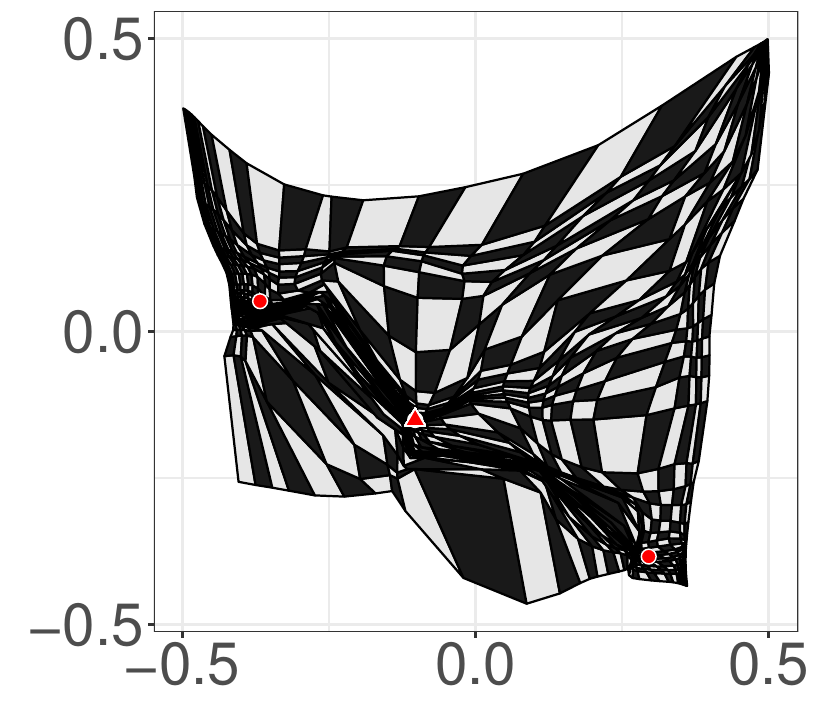} \\
\end{tabular}
\end{center}
\caption{Top row: true warped domain $\mathcal{W}$ generated using Architecture 4. Median and bottom rows: estimated warped space $\hat{\mathcal{W}}$ using nonstationary models 1--4 (column 1--4) with WLS and GSM inference methods, respectively. The data are generated with the risk functional $r_{\text{max}}(\cdot)$, and three reference points are labeled in red.}
\label{pic:sim_space4}
\end{figure} 

\begin{table}[hbt!]
    \centering
    \caption{Test censored log-likelihoods (larger is preferred) evaluated for all fitted models, using both the gradient score matching (GSM) and weighted least squares (WLS) inference methods, and with three different risk functionals. The best models are highlighted in bold for each risk functional and each inference method, and the oracle model based on Architecture~4 is labeled with an asterisk.}
    \begin{tabular}{l|l|c|c|c}
    \hline \hline
    \multirow{2}{*}{Architecture} & Inference & \multicolumn{3}{c}{Risk functional} \\
    \cline{3-5}
    & method & $r_{\text{site}}(\cdot)$ & $r_{\text{max}}(\cdot)$ & $r_{\text{sum}}(\cdot)$ \\
    \hline
    Stationary & \multirow{5}{*}{WLS} & $-56501$ & $-16584$ & $10670$ \\ 
Nonstationary 1 &  & $-58074$ & $-16928$ & $12586$ \\ 
Nonstationary 2 &  & $\mathbf{-55146}$ & $\mathbf{-15111}$ & $\mathbf{13185}$ \\ 
Nonstationary 3 &  & $-57114$ & $-16028$ & $12448$ \\ 
Nonstationary 4* &  & $-58912$ & $-15333$ & $13051$ \\ 
\hline 
Stationary & \multirow{5}{*}{GSM} & $-54593$ & $-16257$ & $10760$ \\ 
Nonstationary 1 &  & $-51740$ & $-14959$ & $13004$ \\ 
Nonstationary 2 &  & $\mathbf{-50596}$ & $\mathbf{-14537}$ & $12985$ \\ 
Nonstationary 3 &  & $-51538$ & $-14923$ & $13027$ \\ 
Nonstationary 4* &  & $-50686$ & $-14577$ & $\mathbf{13177}$ \\
    \hline
    \end{tabular} 
    \label{tab:sim_compare4}
\end{table}

\begin{table}[hbt!]
    \centering
    \caption{Estimates of the extremal dependence parameters, $\hat{\bm{\psi}}' = (\hat{\varphi}, \hat{\kappa})$, for data generated with Architecture 4, using both the gradient score matching (GSM) and weighted least squares (WLS) inference methods, and three different risk functionals. The true values of the extremal dependence parameters are $\varphi_0 = 0.2$ and $\kappa_0 = 1$. Standard deviations obtained using a nonparametric bootstrap are reported in brackets as subscripts of the corresponding parameter estimate. The oracle model based on Architecture~4 is labeled with an asterisk.}
    \resizebox{\columnwidth}{!}{%
    \begin{tabular}{l|l|c|c|c}
    \hline \hline
    \multirow{2}{*}{Architecture} & Inference & \multicolumn{3}{c}{Risk functional} \\
    \cline{3-5}
    & method & $r_{\text{site}}(\cdot)$ & $r_{\text{max}}(\cdot)$ & $r_{\text{sum}}(\cdot)$ \\
    \hline 
Stationary & \multirow{5}{*}{WLS} & $0.274_{(0.029)}$, $1.239_{(0.107)}$ & $0.314_{(0.043)}$, $1.166_{(0.072)}$ & $0.242_{(0.030)}$, $1.202_{(0.080)}$ \\ 
Nonstationary 1 &  & $0.195_{(0.038)}$, $1.201_{(0.130)}$ & $0.208_{(0.035)}$, $1.108_{(0.086)}$ & $0.184_{(0.025)}$, $1.194_{(0.099)}$ \\ 
Nonstationary 2 &  & $0.199_{(0.041)}$, $1.163_{(0.141)}$ & $0.215_{(0.035)}$, $1.147_{(0.097)}$ & $0.171_{(0.026)}$, $1.175_{(0.099)}$ \\ 
Nonstationary 3 &  & $0.230_{(0.032)}$, $1.219_{(0.132)}$ & $0.267_{(0.034)}$, $1.131_{(0.089)}$ & $0.184_{(0.021)}$, $1.200_{(0.099)}$ \\ 
Nonstationary 4* &  & $0.232_{(0.039)}$, $1.197_{(0.148)}$ & $0.247_{(0.037)}$, $1.119_{(0.106)}$ & $0.183_{(0.023)}$, $1.176_{(0.102)}$ \\ 
\hline 
Stationary & \multirow{5}{*}{GSM} & $0.143_{(0.001)}$, $0.908_{(0.002)}$ & $0.143_{(0.001)}$, $0.909_{(0.003)}$ & $0.144_{(0.001)}$, $0.911_{(0.003)}$ \\ 
Nonstationary 1 &  & $0.170_{(0.004)}$, $1.023_{(0.015)}$ & $0.172_{(0.004)}$, $1.065_{(0.017)}$ & $0.172_{(0.004)}$, $1.054_{(0.015)}$ \\ 
Nonstationary 2 &  & $0.166_{(0.006)}$, $1.003_{(0.030)}$ & $0.167_{(0.003)}$, $1.003_{(0.022)}$ & $0.166_{(0.004)}$, $0.999_{(0.026)}$ \\ 
Nonstationary 3 &  & $0.172_{(0.002)}$, $1.015_{(0.016)}$ & $0.178_{(0.003)}$, $1.058_{(0.015)}$ & $0.177_{(0.003)}$, $1.049_{(0.016)}$ \\ 
Nonstationary 4* &  & $0.171_{(0.003)}$, $1.004_{(0.015)}$ & $0.171_{(0.003)}$, $0.999_{(0.019)}$ & $0.172_{(0.003)}$, $1.001_{(0.016)}$ \\ 
    \hline
    \end{tabular} }
    \label{tab:sim_table4}
\end{table}

\clearpage
\section{Supplementary results for the data application} \label{Appendix:Application}

\subsection{Modeling of margins}\label{Appendix:Application_margins}

\begin{figure}[hbt!]
\begin{center}
\begin{tabular}{c}
\includegraphics[width=0.6\linewidth]{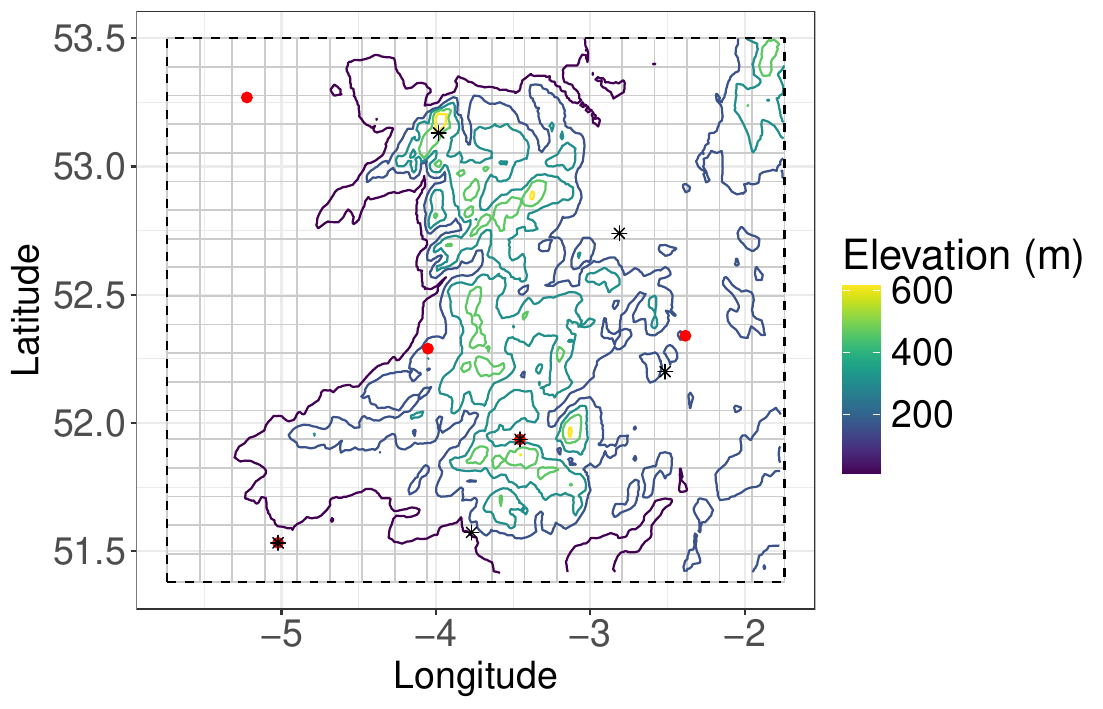}
\end{tabular}
\end{center}
\caption{Geographical map of the study domain, with the position of five sites that violate the null hypothesis (labeled as red dots), namely that the data at these sites do not follow a GPD. We also display six representative sites (labeled as black stars), with two of them corresponding to the sites that violate the null hypothesis, which we use for goodness-of-fit checks.}
\label{pic:UKpr_diag_map}
\end{figure}

\begin{figure}[hbt!]
\begin{center}
\begin{tabular}{c}
\includegraphics[width=0.33\linewidth]{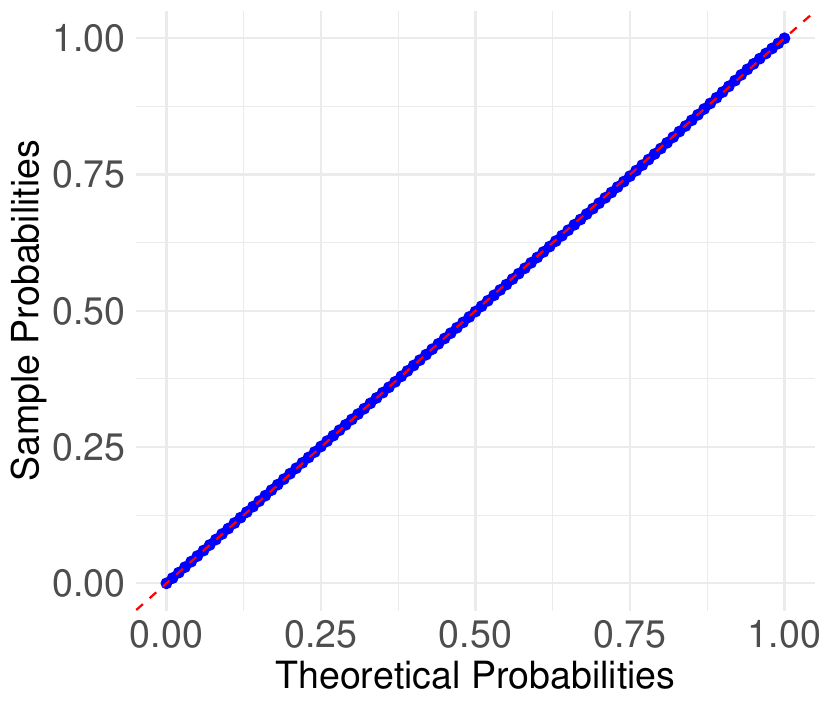} \\
\includegraphics[width=0.99\linewidth]{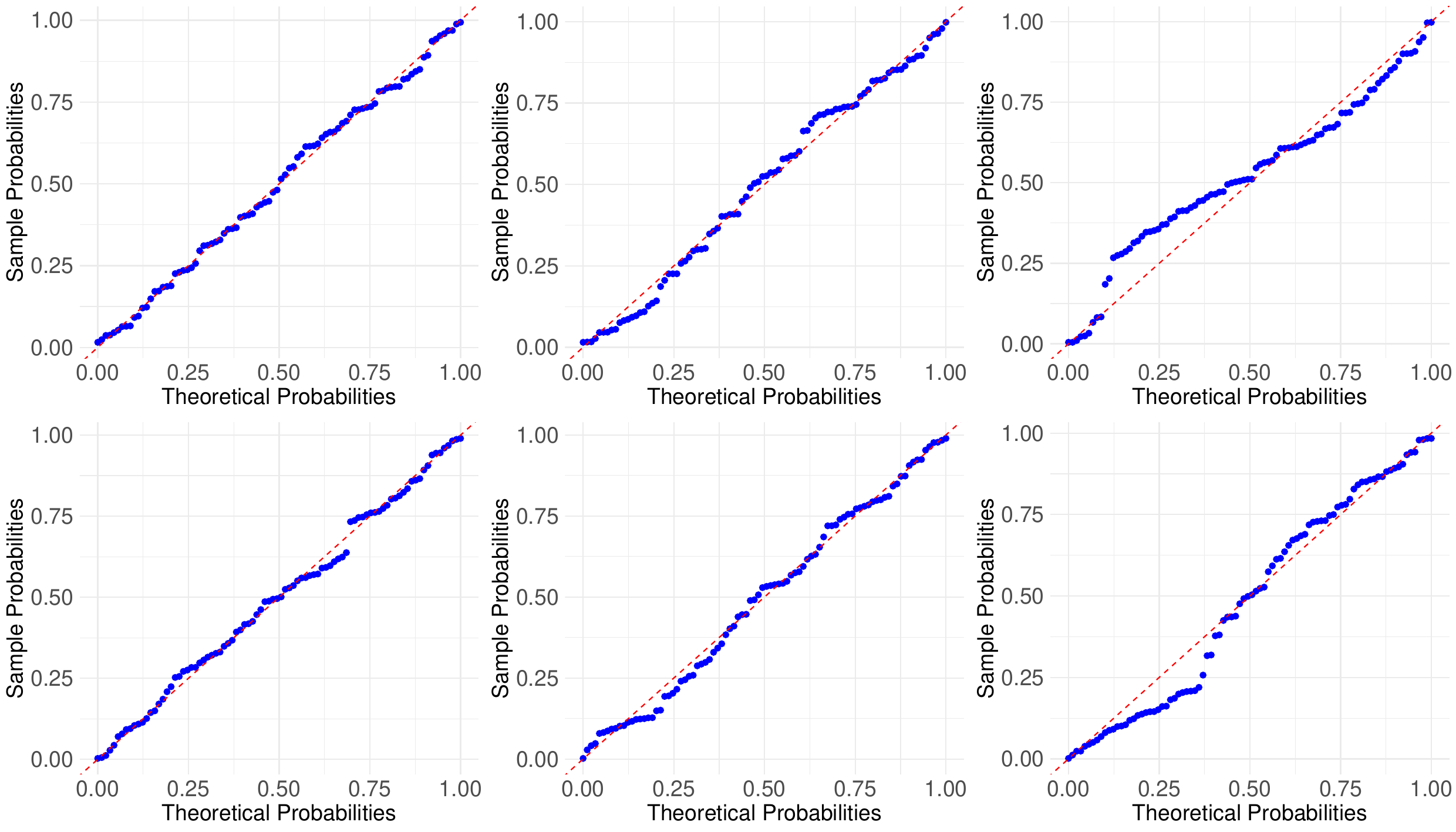}
\end{tabular}
\end{center} 
\caption{The probability-probability (P-P) plot of the marginal excesses for the total domain (top), and at six representative sites (labeled as black stars in Figure~\ref{pic:UKpr_diag_map}), where observations over four of them (the first and the second panels in the second and third rows) do not reject the null hypothesis that the data sampled there follow a GPD, and observations over the other two (the third panel in the second and third rows) which do reject the null hypothesis. } 
\label{pic:UKpr_diag}
\end{figure}

For the marginal distributions, we assume that high quantile exceedances at each site are distributed as a generalized Pareto distribution (GPD) with scale and shape parameters $\tau(\bm{s})$ and $\xi(\bm{s})$, respectively (see Section~\ref{RealDataApplication:UKpr} in the main article for details). To validate marginal fits, we compare the empirical tail distribution from the data and the fitted GPD model using probability-probability (P-P) plots (displayed on the Uniform$(0,1)$ scale) and Kolmogorov--Smirnov (K--S) distances. After fitting the marginal GPD to threshold exceedances, we obtain estimates $\hat{\tau}(\bm{s}_i)$ and $\hat{\xi}(\bm{s}_i)$ for all $i = 1,\ldots,D$. Let $\widehat{G}(\cdot) \equiv G(\cdot~; \hat{\tau}(\bm{s}_i), \hat{\xi}(\bm{s}_i))$ be the estimated tail distribution at site $\bm{s}_i$, with $G(\cdot)$ as in \eqref{eq:GPD} of the main article. If the marginal tail model is correctly-specified, $U_{i} = \widehat{G}\{Y(\bm{s}_i)-u(\bm{s}_i)\}$ (with $u(\bm{s}_i)$ the selected marginal threshold) is approximately $\textrm{Uniform}(0,1)$ distributed, at each site. Thus, using P-P plots and K--S distances, we can check if $U_{i} \sim \textrm{Uniform}(0,1)$.

Out of 12600 sites, only 5 reject the null hypothesis based on the K--S test. P-P plots for six representative sites (see Figure~\ref{pic:UKpr_diag_map}), which include two sites for which the null hypothesis is rejected, are shown in Figure~\ref{pic:UKpr_diag}. The P-P plots generally demonstrate good marginal fits for the data at sites that do not reject the null hypothesis, i.e., the data sampled there follow a GPD, while some slight lack of fit is observed at sites that do reject the null hypothesis. Nevertheless, the fits appear to be quite satisfactory, even at these worst sites. We further display a P-P plot based on all the data from all sites pooled together. The results show that the marginal fit is very good overall.

{\color{black}
\subsection{Threshold choice for functional exceedances}\label{Appendix:Application_threshold}

\begin{figure}[!hbt]
    \centering
    \includegraphics[width=0.8\linewidth]{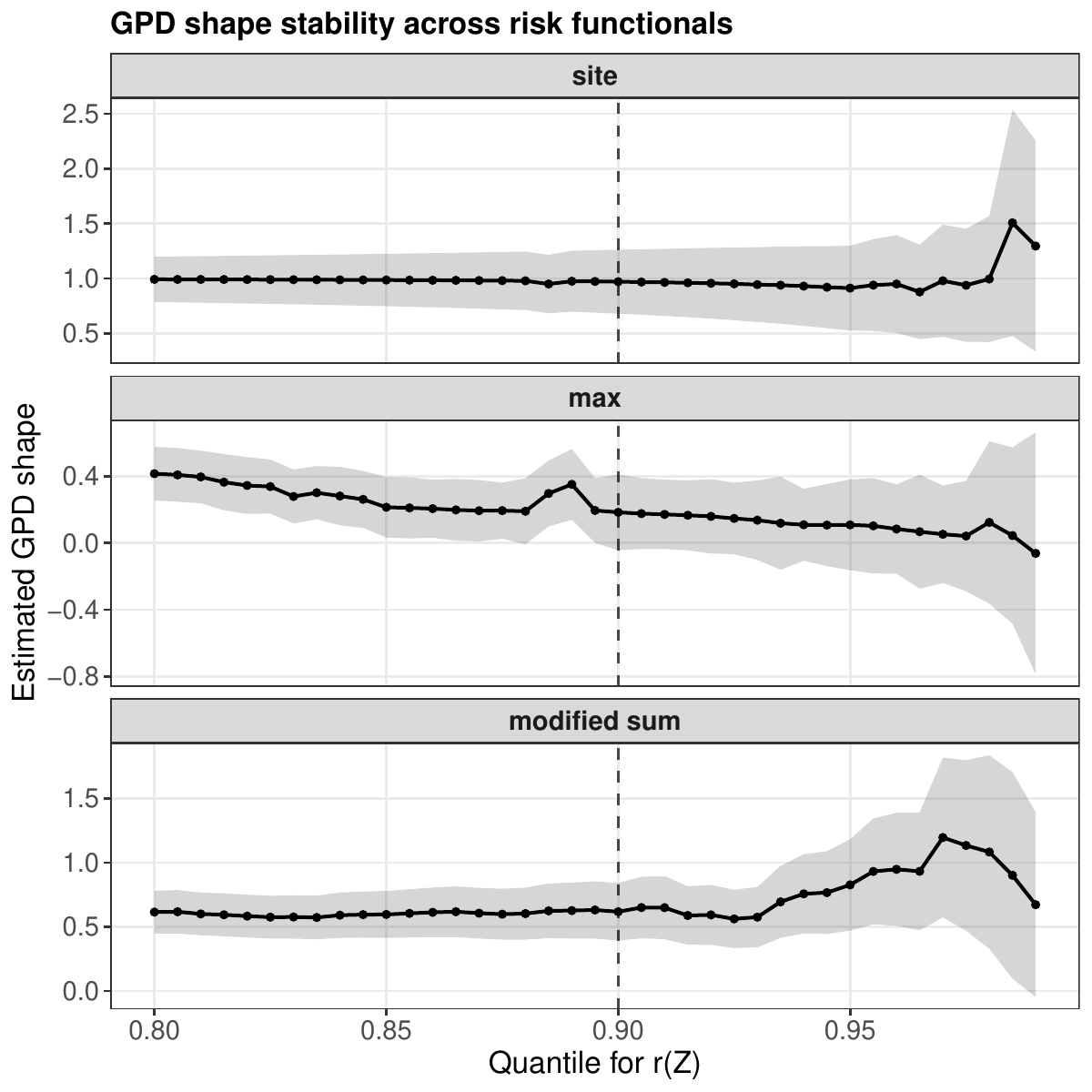}
    \caption{Threshold-stability diagnostic for functional exceedances. For each risk functional $r(\cdot)$, the threshold $u$ is set to the empirical $q$-quantile of $\{r(\bm{x}_t)\}_{t=1}^N$ over a grid $q\in[0.8,0.99]$, and a univariate GPD is fitted to the excesses $\{r(\bm{x}_t)-u: r(\bm{x}_t)>u\}$. The panels report the estimated GPD shape parameter $\hat{\xi}$ as a function of $q$; the vertical dashed line marks the selected quantile $q=0.90$.}
    \label{pic:stability}
\end{figure}

To guide the choice of the functional threshold $u$ for event selection, we follow the standard peaks-over-threshold practice of checking threshold stability. A closely related strategy is used in \cite{de2018high}, who selected events using a very high quantile of the risk functional and support this choice by monitoring the stability of the fitted generalized Pareto distribution (GPD) tail parameter across nearby thresholds.

Motivated by this approach, we compute the empirical risk series $\{r(\bm{x}_t)\}_{t=1}^N$ and, for a grid of candidate quantiles $q\in[0.8,0.99]$, set $u$ to the empirical $q$-quantile of $\{r(\bm{x}_t)\}_{t=1}^N$ and fit a univariate GPD to the corresponding excesses $\{r(\bm{x}_t)-u: r(\bm{x}_t)>u\}$. Figure~\ref{pic:stability} displays the resulting estimates of the GPD shape parameter $\hat{\xi}$ as a function of $q$ for the risk functionals considered in Section~\ref{RealDataApplication} of the main paper.

Across the considered risk functionals, the estimated shape parameter is approximately stable, though at the highest quantiles it becomes more variable and/or shows noticeable drift, consistent with the increased sampling variability caused by too few exceedances. Our choice of $u$, with $q=0.90$, thus appears appropriate for all risk functionals.
}

\clearpage
\subsection{Additional results}

In this section, we present additional results for the UK precipitation application, focusing on uncertainty summaries for the fitted pairwise conditional exceedance probabilities under the fitted nonstationary models.

\begin{figure}[hbt!]
\begin{center}
\begin{tabular}{c}
\includegraphics[width=0.99\linewidth]{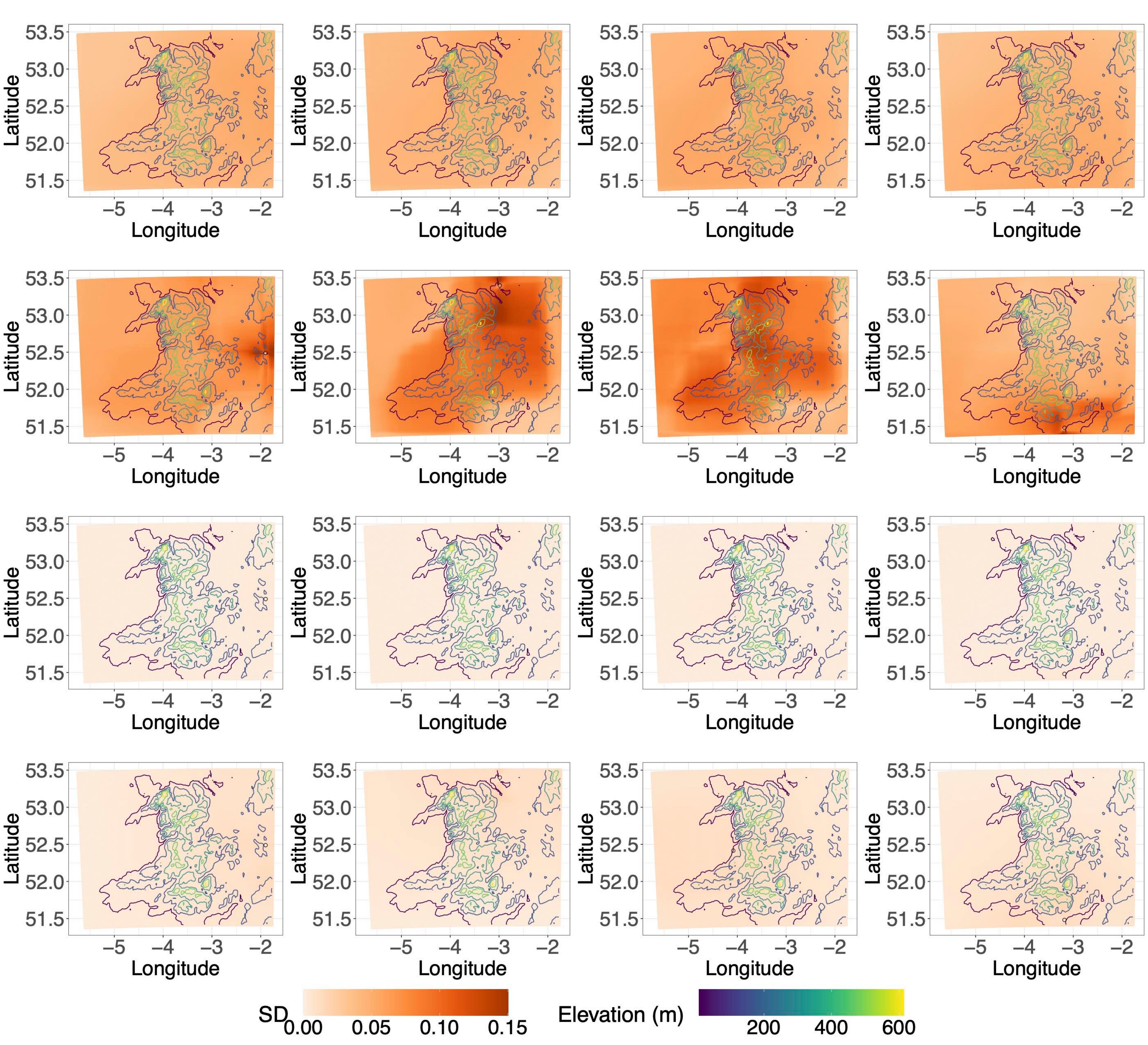} \\
\end{tabular}
\end{center}
\caption{Standard deviation of the fitted pairwise CEPs without (1st and 3rd rows) and with (2nd and 4th rows) the consideration of the deformation for reference cities Birmingham (1st column), Liverpool (2nd column), Aberystwyth (3th column), and Cardiff (4th column) using either the WLS (1st and 2nd rows) or GSM (3rd and 4th rows) inference methods. Elevation contour lines are colored with their magnitude information as references.}
\label{pic:UKpr_unc}
\end{figure}

\clearpage
{\color{black}
\section{Bootstrap design and assessment} \label{Appendix:Bootstrap}

\subsection{Nested bootstrap design and variance decomposition}\label{Appendix:Bootstrap_design}

Deriving theoretical asymptotic distributions that explicitly propagate uncertainty in warping estimation is beyond the scope of this work. To the best of our knowledge, the existing deep compositional spatial model (DCSM) literature in the Gaussian setting does not provide systematic empirical evaluations of bootstrap-based confidence interval coverage (or formal bootstrap-based consistency guarantees), so we cannot directly cite established coverage results in closely matched settings.

More broadly, uncertainty in our modeling pipeline can arise from multiple sources, including: (i) estimation of the spatial warping; (ii) the choice of exceedance threshold $u$ (or the risk-functional quantile used for event selection); (iii) the estimation/standardization of marginal transformations used to obtain approximately common margins; (iv) finite-sample variability from the temporal replicates (or exceedance events) used for inference; (v) sampling variability due to the finite set of observed/training locations; and (vi) stochasticity in numerical optimization (e.g., random initialization and Adam dynamics). Our goal here is therefore to make explicit which components the bootstrap is capturing, to consider reasonable bootstrap variants, and to assess calibration empirically in controlled simulations.

To propagate uncertainty from threshold selection and marginal standardization, each bootstrap run re-estimates the marginal transformations and then re-selects exceedances based on the bootstrapped risk series before refitting the extremal dependence model (and, when applicable, the warping). This ensures that variability induced by the marginal/threshold steps is carried through to the resulting uncertainty quantification.

In addition, our experiments suggest that variability induced by the choice of the finite subset of sampling locations used for training can be non-negligible relative to replicate-level variability. We therefore also consider a location resampling component. Spatial resampling methods have been developed to quantify uncertainty induced by observing a spatial process at a finite set of sites, often via spatial subsampling or block-based resampling, or alternatively via site-level resampling. Both \cite{sherman1996variance} and \cite{lee2002least} have used overlapping spatial subregions to create approximate replicates for uncertainty quantification, and \cite{lahiri2006resampling} developed block-bootstrap procedures for spatial data under stochastic (irregular) sampling designs. However, these block/subsampling approaches are typically justified under stationarity and/or weak-dependence conditions, which are not directly applicable to our setting in the original space where the process is nonstationary. For this reason, we adopt a pragmatic location resampling design to assess site-selection uncertainty in our nonstationary setting.

\paragraph{Location bootstrap design.}
To assess uncertainty induced by the selection of training sites while keeping the number of training locations, $D_{\text{train}}$, fixed, we adopt the following design in the simulation study. We first sample $D_{\text{obs}}=1000$ observed locations to mimic a finite set of available monitoring sites, and then we fit the model using a training set of size $D_{\text{train}}=800$ (i.e., $80\%$ of the observed pool). For the location bootstrap, we repeatedly re-select a training set of size 800 from the same pool of 1000 observed locations, thereby injecting site-selection uncertainty without reducing the effective training sample size. A practical consideration is the fraction $D_{\text{train}}/D_{\text{obs}}$: if the training set is too small relative to the available pool, fitting highly flexible nonstationary models can become unstable (and, in our case, computationally inefficient) due to insufficient spatial coverage. Choosing $80\%$ strikes a pragmatic balance: it keeps the training sample large enough for stable fitting while leaving a nontrivial hold-out pool (here $20\%$) to induce meaningful variability across resampled training-site sets.

We consider a two-level (nested) resampling scheme that jointly accounts for location and temporal variability:
\begin{itemize}
\item {\bf Outer layer (sites):} re-sample a training-location set $\mathcal{D}_{b_1}$ of fixed size $|\mathcal{D}_{b_1}|=D_{\text{train}}=800$ from the observed-location pool $\mathcal{D}_{\text{obs}}$ of size $D_{\text{obs}}=1000$, for ${b_1=1,\ldots,B_1}$.
\item {\bf Inner layer (replicates/events):} conditional on $\mathcal{D}_{b_1}$, re-sample uniformly-at-random temporal replicates $\mathcal{T}_{b_2}$ (with replacement) to construct the bootstrap dataset, i.e., 
${\{x_t(\bm{s}) : t\in \mathcal{T}_{b_2}, \bm{s}\in\mathcal{D}_{b_1}\}}$, for $b_2=1,\ldots,B_2$.
\item {\bf Re-fit:} for each $(b_1,b_2)$, re-run the full inference pipeline on the re-sampled dataset (marginal standardization, exceedances selection, extremal dependence estimation), under two variants:
(i) a ``full re-fit'' bootstrap that re-estimates the warping jointly with the extremal dependence parameters, and
(ii) a ``conditional'' bootstrap that keeps the fitted warping fixed and re-estimates only the extremal dependence parameters.
Denote the resulting estimates by $\hat{\bm{\psi}}_{b_1,b_2}$ and $\hat{\bm f}_{b_1,b_2}$.
\end{itemize}

This nested design aligns with the variance decomposition
$$
\mathrm{Var}(\hat{\bm{\psi}}_{b_1,b_2}) =
\mathrm{Var}\left(\mathbb{E}\{\hat{\bm{\psi}}_{b_1,b_2}\mid\mathcal{D}_{b_1}\}\right) +
\mathbb{E}\left(\mathrm{Var}\{\hat{\bm{\psi}}_{b_1,b_2}\mid\mathcal{D}_{b_1}\}\right),
$$
where the first term reflects site-selection uncertainty and the second reflects replicate/event uncertainty. In practice, we choose $B_1=25$ and $B_2=4$, yielding $B_1B_2=100$ total re-fits. This choice reflects a computational trade-off: $B_1$ must be large enough to stabilize the between-site component $\mathrm{Var}\{\mathbb{E}(\hat{\bm{\psi}}_{b_1,b_2}\mid\mathcal{D}_{b_1})\}$, while a smaller $B_2$ is sufficient to characterize the within-site replicate variability conditional on $\mathcal{D}_{b_1}$. In our experiments, increasing $B_2$ beyond 4 produced negligible changes in the within-site variance estimates relative to the added computational cost.

\begin{table}[t!]
    \centering
    \caption{Proportion of the total variance of the extremal dependence estimates attributable to site-selection uncertainty, defined as
$\mathrm{Var}(\mathbb{E}\{\hat{\bm{\psi}}_{b_1,b_2}\mid\mathcal{D}_{b_1}\})/\mathrm{Var}(\hat{\bm{\psi}}_{b_1,b_2})$,
for data generated with Architecture~3, using both GSM and WLS and three risk functionals. Each table entry reports two values (for $\hat{\varphi}$ and $\hat{\kappa}$, respectively). The oracle model based on Architecture~3 is labeled with an asterisk.}
    \begin{tabular}{l|l|c|c|c}
    \hline \hline
    \multirow{2}{*}{Architecture} & Inference & \multicolumn{3}{c}{Risk functional} \\
    \cline{3-5}
    & method & $r_{\text{site}}(\cdot)$ & $r_{\text{max}}(\cdot)$ & $r_{\text{sum}}(\cdot)$ \\
    \hline 
    Stationary & \multirow{5}{*}{WLS} & $0.066$, $0.042$ & $0.055$, $0.046$ & $0.054$, $0.056$ \\ 
Nonstationary 1 &  & $0.060$, $0.043$ & $0.059$, $0.042$ & $0.053$, $0.057$ \\ 
Nonstationary 2 &  & $0.060$, $0.046$ & $0.062$, $0.043$ & $0.058$, $0.056$ \\ 
Nonstationary 3* &  & $0.062$, $0.047$ & $0.057$, $0.039$ & $0.057$, $0.055$ \\ 
Nonstationary 4 &  & $0.062$, $0.049$ & $0.056$, $0.041$ & $0.058$, $0.054$ \\ 
\hline 
Stationary & \multirow{5}{*}{GSM} & $0.011$, $0.009$ & $0.010$, $0.010$ & $0.010$, $0.009$ \\ 
Nonstationary 1 &  & $0.056$, $0.047$ & $0.054$, $0.049$ & $0.061$, $0.055$ \\ 
Nonstationary 2 &  & $0.063$, $0.062$ & $0.055$, $0.057$ & $0.052$, $0.048$ \\ 
Nonstationary 3* &  & $0.067$, $0.063$ & $0.062$, $0.061$ & $0.058$, $0.059$ \\ 
Nonstationary 4 &  & $0.058$, $0.056$ & $0.055$, $0.058$ & $0.061$, $0.058$ \\ 
    \hline
    \end{tabular}
    \label{tab:sim_table_sdratio}
\end{table}

Table~\ref{tab:sim_table_sdratio} indicates that, for the Architecture~3 setting considered here, site-selection uncertainty contributes a relatively small but non-negligible fraction of the total variability in $\hat{\bm{\psi}}_{b_1,b_2}$, with a broadly similar magnitude under WLS and GSM. In the application, the fitted model is based on one representative training subset of size $D_{\text{train}}=2000$, selected from an assumed observed set of $D_{\text{obs}}=2500$ sites. We do not separately decompose the contribution of this site-selection uncertainty in the application itself; rather, the nested resampling results reported here are intended to show, in a controlled simulation setting, that this additional source of variability is present but secondary, while the main-text application remains focused on the fitted extremal dependence structure for one representative training subset.

For WLS, the site-selection component is typically around $5\%$--$7\%$ for $\hat{\varphi}$ and about $4\%$--$6\%$ for $\hat{\kappa}$ across architectures and risk functionals. For GSM, the ratios are of comparable order in the nonstationary settings (roughly $5\%$--$7\%$ for both $\hat{\varphi}$ and $\hat{\kappa}$), whereas in the stationary case they are closer to $1\%$. Across both methods, the ratios for $\hat{\varphi}$ and $\hat{\kappa}$ are generally similar in magnitude, suggesting that site selection affects range and smoothness estimation to a comparable extent in this experiment. Overall, these results suggest that replicate/event resampling remains the dominant contributor to uncertainty under Architecture~3, but that including a location-resampling layer still provides a principled way to quantify the additional variability induced by finite site selection.

\subsection{Empirical coverage assessment}\label{Appendix:Bootstrap_coverage}

To directly assess calibration, we evaluate empirical coverage of bootstrap-based 95\% confidence intervals for summaries of the recovered nonstationary extremal dependence structure, with the pairwise CEP summaries as our primary focus, and for the individual dependence parameters as a secondary reference point, in our simulation study. Since the main uncertainty question of interest here is whether uncertainty in the estimated deformation needs to be propagated into interval estimation, we focus on comparing two bootstrap variants:
\begin{itemize}
\item {\bf Full re-fit:} each bootstrap sample re-estimates the warping $\bm{f}$ jointly parameters $\bm{\psi}=(\varphi,\kappa)'$;
\item {\bf Conditional on the warping:} the estimated warping is kept fixed, and only $\bm{\psi}$ is re-estimated in each bootstrap sample.
\end{itemize}

We do not include site re-sampling as an additional factor in this coverage experiment, since its effect has already been examined separately through the nested variance decomposition in Section~\ref{Appendix:Bootstrap_design}. Likewise, we do not consider a separate exceedance-level resampling comparison here, in order to keep the calibration study targeted and computationally manageable.

The coverage study is conducted for data generated under Architecture~3. For each simulated dataset, we fit Nonstationary Models~1 and~3 using both the GSM and WLS inference methods. This gives four fitted model/inference combinations in total. For each combination, we generate bootstrap replicates under the two variants detailed above, construct percentile-based 95\% confidence intervals for both the individual extremal dependence parameters and the pairwise CEP summaries, and record whether the true values or corresponding oracle summaries are contained in the corresponding intervals. Repeating this over 500 independently generated datasets yields empirical coverage percentages for these quantities of interest.

\begin{table}[t!]
\centering
\caption{Empirical coverage (\%) of 95\% confidence intervals for pairwise CEPs at the three reference sites shown in Figure~\ref{pic:sim_pairCEPs}, based on 500 Monte Carlo datasets generated under Architecture~3. Results are reported for the fully re-estimated warping bootstrap only, where the warping is re-estimated in each bootstrap sample. For each reference site, the two numbers in each entry correspond to the mean/median summaries of pairwise CEPs with respect to the chosen reference site. These are the more practically relevant summaries for assessing recovery of the nonstationary extremal dependence structure.}
\begin{tabular}{l|c|c|c}
\hline \hline
\multirow{2}{*}{Model (inference method)} & \multicolumn{3}{c}{Coverage for pairwise CEPs} \\
\cline{2-4}
& site 1 & site 2 & site 3 \\
\hline
Nonstationary Model 1 (WLS) &  96.1 / 96.8  &  86.7 / 89.2  &  84.4 / 82.2 \\
Nonstationary Model 3 (WLS) &  94.3 / 95.4  &  88.3 / 87.8  &  90.7 / 92 \\
Nonstationary Model 1 (GSM) &  80.5 / 80.8  &  77.7 / 78.2  &  78.5 / 78.4 \\
Nonstationary Model 3 (GSM) &  76.6 / 77.2  &  73.2 / 73.5  &  74.9 / 74.9 \\
\hline
\end{tabular} 
\label{tab:sim_table_coverage_CEP}
\end{table}

\begin{table}[t!]
\centering
\caption{Empirical coverage (\%) of 95\% confidence intervals for the (individual) dependence parameters $(\varphi,\kappa)$, based on 500 Monte Carlo datasets generated under Architecture~3. We compare intervals obtained by fully re-estimating the warping in each bootstrap sample versus intervals obtained conditionally on the estimated warping.}
\resizebox{\columnwidth}{!}{
\begin{tabular}{l|cc|cc}
\hline \hline
\multirow{2}{*}{Model (inference method)} & \multicolumn{2}{c|}{Full refit} & \multicolumn{2}{c}{Conditional on warping} \\
\cline{2-5}
& Coverage for $\varphi$ & Coverage for $\kappa$ & Coverage for $\varphi$ & Coverage for $\kappa$ \\
\hline
Nonstationary Model 1 (WLS) &  76.0  &  64.6  &  78.6  &  61.8 \\
Nonstationary Model 3 (WLS) & 60.4  &  59.2  &  72.0  &  59.2 \\
Nonstationary Model 1 (GSM) &  32.06  &  58.32  &  18.24  &  19.04 \\
Nonstationary Model 3 (GSM) &  30.46  &  41.88  &  15.43  &  16.83 \\
\hline
\end{tabular}}
\label{tab:sim_table_coverage}
\end{table}

As our main inferential target is the recovery of the nonstationary extremal dependence structure itself, Table~\ref{tab:sim_table_coverage_CEP} is the more practically relevant calibration result. It reports empirical coverage for pairwise CEP summaries (mean/median of all pairwise CEPs with respect to the three reference sites shown in Figure~\ref{pic:sim_pairCEPs} of the main paper), based on the full re-fit bootstrap. The results are encouraging overall. Under WLS, coverage is close to nominal for site~1 and somewhat lower but still reasonable for sites~2 and~3. This suggests that the difficulty of recovering the extremal dependence structure (with calibrated uncertainty quantification) varies across regions of the warped space. This spatial heterogeneity is plausible, as some regions may be harder to estimate because of stronger local distortion or differences in the effective density of sites after warping. Under GSM, coverage is more systematically below nominal across all three reference sites, which is consistent with our empirical observation that GSM tends to produce more ``stable'' estimates, both for the dependence parameters and for the warped geometry, and hence yields relatively narrow bootstrap intervals. If some finite-sample bias remains, such narrower intervals are more likely to miss the truth, leading to lower empirical coverage. 
As a secondary reference point, Table~\ref{tab:sim_table_coverage} summarizes the corresponding coverage results for the individual dependence parameters. As the range parameter $\varphi$ is only identifiable up to an overall scaling in the warped space, its empirical coverage has limited practical interpretability, and we therefore focus primarily on the smoothness parameter $\kappa$. The coverage for $\kappa$ is below the nominal 95\% level in all cases, but improves when the warping is re-estimated, especially for GSM. These results suggest that propagating warping-estimation uncertainty is important, even though interval calibration for the individual parameters remains imperfect in this challenging nonstationary setting.

Overall, although some undercoverage remains, especially for GSM, the nonparametric bootstrap appears substantially more calibrated for the practically relevant CEP summaries than for the individual dependence parameters.

}


\end{document}